\documentclass[12pt]{article}

\usepackage{gensymb}
\usepackage{url}
\usepackage{amssymb,amsmath, amsfonts}
\usepackage{cases}
\usepackage{latexsym}
\usepackage{graphicx}
\usepackage{relsize}
\usepackage{natbib}
\usepackage{rotating}
\usepackage[includeheadfoot,left=1.in,right=1.in,top=1.in,bottom=1.in,bindingoffset=0.in,nohead]{geometry}
\usepackage[onehalfspacing]{setspace}
\usepackage{booktabs} 
\usepackage{appendix}
\usepackage[pdfborder={0 0 0}]{hyperref}
\usepackage{float}
\usepackage{xcolor}
\usepackage{caption}
\usepackage{subcaption}
\usepackage{minibox}  
\usepackage{floatpag}
\usepackage[most]{tcolorbox}

\usepackage{tikz}
\usetikzlibrary{matrix,chains,positioning,decorations.pathreplacing,arrows}

\begin{document}
	
	\title{\vspace{-1.5cm} Assessing and Comparing \\Fixed-Target Forecasts  of   Arctic  Sea Ice:\\
	Glide Charts for Feature-Engineered \\Linear Regression and Machine Learning Models}

	\author{Francis X. Diebold\\University of Pennsylvania \and Maximilian G\"obel\\University of Lisbon \and Philippe Goulet Coulombe\\Université du Québec à Montréal \\$~$}
	
	\maketitle

\smallskip

	\begin{spacing}{1}
		
		\noindent \textbf{Abstract}:   We  use ``glide charts" (plots of sequences of root mean squared forecast errors as the target date is approached) to evaluate and compare  fixed-target forecasts of Arctic sea ice.  We first use them  to evaluate  the simple  feature-engineered linear regression (FELR) forecasts of \cite{DieboldGoebel2021}, and to compare FELR forecasts to  naive pure-trend benchmark forecasts. Then we introduce a much more sophisticated feature-engineered \textit{machine learning} (FEML) model, and we use glide charts to evaluate FEML forecasts  and compare them to a FELR benchmark. Our substantive results include the frequent appearance of predictability thresholds, which differ across months, meaning that accuracy initially fails to improve as the target date is approached but then increases progressively once a threshold lead time is crossed.  Also,  we find that FEML can improve appreciably over FELR when forecasting ``turning point" months in the annual cycle at horizons of one to three months ahead.

		\thispagestyle{empty}

\bigskip
\bigskip

\bigskip

		\noindent {\bf Key words}: Seasonal  climate forecasting, forecast evaluation and comparison, prediction

		\bigskip

		\noindent  {\bf JEL codes}: Q54, C22, C52, C53

		\bigskip

		\noindent {\bf Contact}:  goulet\_coulombe.philippe@uqam.ca

	\end{spacing}

	\clearpage
	
	\setcounter{page}{1}
	\thispagestyle{empty}

\section{Introduction} 

Arctic sea ice is melting very quickly as the planet warms (e.g., \cite{DRice} and the many references therein), which brings both major economic opportunities and major risks. Opportunities/benefits include new accessibility for extracting deposits of natural gas, petroleum, and other natural resources, as well as the emergence of trans-Arctic shipping lanes, which will enhance international trade by reducing both transportation costs and piracy-riddled chokepoints on other routes.  Risks/costs include increased emissions and environmental damage due to discharges, spills, and soot deposits \citep{Bekkers2016, Petrick2017}. Finally and more broadly, melting sea ice will have important geopolitical consequences for Arctic sea-lane control \citep{ebinger2009}.

For all of the above reasons, the temporal path and pattern of Arctic sea ice diminution are of particular interest, and Arctic sea ice forecasting has received significant attention \citep{SeaIceBookCh4}.  From a real-time online perspective, there are two key approaches.  The first is fixed-horizon forecasting, where, for example, each month we forecast one month ahead, month after month, ongoing, as in \cite{DRice}.  The second is  fixed-target forecasting, where each month we forecast a fixed future target date, month after month, ending when we arrive at the target date, as in \cite{DieboldGoebel2021}.  In this paper we consider the fixed-target scenario, which has generated  substantial interest in highlighting Arctic sea ice diminution both within years (as September 30 is approached, say) and across years (comparing the sequence of Septembers, say).\footnote{For example, each summer since 2008 the Sea Ice Prediction Network (SIPN)  has sponsored the Sea Ice Outlook  (SIO) competition for predicting September average daily Arctic sea ice extent. {See \url{https://www.arcus.org/sipn} for SIPN, and see \url{https://www.arcus.org/sipn/sea-ice-outlook} for SIO.} September extent forecasts are produced by many research groups mid-month in June, July, and August, and evaluated once September ends and the outcome is known.  {Insightful post-season SIO assessments have been   produced annually (the most recent is \cite{SIO_postseason2022}), and similarly-insightful multi-year retrospective SIO assessments have been  produced occasionally \citep{StroeveEtAl2014,HamiltonStroeve2016,Hamilton2020}.}}

Forecast accuracy naturally increases  as information accumulates and the target date is approached.   A key question is how to quantify that accuracy, and how quickly, and with what pattern, it improves as the target date is approached.  In this paper we use glide charts (plots of sequences of root mean squared forecast errors as the target date is approached) to address those questions in the contexts of two  models for Arctic sea ice forecasting, the feature-engineered linear regression (FELR) model of \cite{DieboldGoebel2021}, and a new and potentially-superior feature-engineered machine learning (FEML) model developed in this paper, \textcolor{black}{building} on the tree-based ``macro random forest" of \cite{MRF}.

We proceed as follows.  In section \ref{felr} we  review  the FELR model, display and discuss its glide charts for each month of the year, and compare them to those of a naive (pure trend) benchmark.    In section \ref{feml} we introduce   the FEML model, display and discuss its glide charts for each month of the year, and compare them to those of a different and more sophisticated benchmark,   FELR.  Hence FELR appears throughout, but in different roles.  It appears first in section \ref{felr} as a candidate model to be compared to a naive benchmark, and then in section \ref{feml} as a more sophisticated benchmark against which a potentially even \textit{more}  sophisticated candidate model is compared.  We conclude in section \ref{concl}.

\section{ Glide Charts for \\ \hspace*{0.08em} Feature-Engineered Linear Regression (FELR)} \label{felr} 

Here we review \textcolor{black}{the} FELR model of \cite{DieboldGoebel2021} and display its glide charts for Arctic sea ice forecasting.  In particular, we treat FELR as a simple but hopefully-sophisticated model -- in the tradition of the  ``KISS Principle" of forecasting: ``Keep it sophisticatedly simple"  \citep{Zellner1992} -- and assess its fixed-target forecasting performance relative to a naive benchmark forecast, a simple linear trend.  We do so in part to illustrate the construction and interpretation of glide charts, and in part because we are interested in FELR and the improvements it may deliver relative to more naive models. Later, in section \ref{feml},  we turn the tables and use {FELR} as the {benchmark} when assessing a more sophisticated non-parametric nonlinear feature-engineered machine learning model.

\subsection{Feature-Engineered Linear Regression} 

To understand FELR, one must understand the real-time fixed-target forecasting exercise in which it is embedded.  In our subsequent empirical work, we will consider fixed-target forecasting for a selected target month (the monthly average of daily observations), conditioning on the expanding daily historical sample as the end of the target month is approached, performing 120 daily estimations and making 120 corresponding fixed-target forecasts, starting 120 days before the last day of the target month and continuing to the last day of the target month.\footnote{{\color{black} We focus on the monthly aggregate rather than raw daily readings, because the monthly aggregate is the object of interest in many climate studies (see \cite{VARCTIC} and references therein), and also in the \textcolor{black}{SIO annual forecasting competition \citep{SIO_postseason2022}}.  Furthermore, raw daily readings are likely to include undesirable high-frequency noise from satellite measurements and post-processing  \citep{iceplus}.}}  Many variations and extensions (e.g., forecasting a particular target day rather than a target monthly average) can be implemented. Although our framework is applicable to fixed-target forecasting of any variable, our subsequent empirical work will  focus  on Arctic sea ice extent ($SIE$) \textcolor{black}{and we have specialized the notation below to this particular exercise.} {\color{black}  Given the importance of seasonality not only in intercepts,  but also in trends and dynamics \citep{DRice},  we run regressions for each month separately.}

Fully general notation gets tedious, so we take a specific example. Consider fixed-target forecasting for  September average daily sea ice extent, $SIE_{9}$, conditioning on the expanding historical sample as we move from June through the end of September.  In FELR, September extent is regressed on an intercept, a linear trend term, and three additional covariates:
\begin{equation}
	\label{full}
	SIE_{9} ~{\rightarrow} ~c, ~ Time,~ SIE_{LastMonth}, ~SIE_{Last30Days}, ~ SIE_{Today},
\end{equation}
where $SIE_9$ denotes September average daily extent, ``$\rightarrow$" denotes ``is regressed on", and the rest of the notation is obvious.\footnote{\cite{DieboldGoebel2021} use the term ``benchmark predictive model" (BPM) rather than FELR.}

As a concrete illustration, and approximately following the SIO forecasting competition \citep{SIO_postseason2022}, consider the  $SIE_{9}$ forecasts on four days: 6/10, 7/10, 8/10 and 9/10. Immediately,  the 6/10 regression used to produce the June forecast of September is
$$
SIE_{9} ~{\rightarrow}~ c, ~ Time,~  SIE_{5}, ~ SIE_{5/\textcolor{black}{12}\_ 6/10} ,~ SIE_{6/10},
$$
the 7/10 regression used to produce the July forecast  of September is
$$
SIE_{9} ~{\rightarrow}~ c, ~ Time,~  SIE_{6}, ~ SIE_{6/11\_ 7/10} ,~ SIE_{7/10},
$$
the 8/10 regression  used to produce the August forecast  of September is
$$
SIE_{9} ~{\rightarrow}~ c, ~ Time,~  SIE_{7}, ~ SIE_{7/\textcolor{black}{12}\_ 8/10} ,~ SIE_{8/10},
$$
and the 9/10 regression  used to produce the September forecast  of September is
$$
SIE_{9} ~{\rightarrow}~ c, ~ Time,~  SIE_{8}, ~ SIE_{8/\textcolor{black}{12}\_ 9/10} ,~ SIE_{9/10}.
$$
Of course the four days above were chosen just as an illustration, conforming approximately  with SIO forecast dates.  In reality we can produce a forecast on any of the days before  the last day of September.

Perhaps surprisingly given their simplicity, the  FELR  forecasts  are quite sophisticated in certain respects of   relevance for sea ice forecasting.   First, they capture low-frequency linear trend dynamics via conditioning on $Time$.  Second, they capture medium-frequency inertial (autoregressive) dynamics around  trend by  conditioning on  $SIE_{LastMonth}$.  Finally, they capture high-frequency dynamics by augmenting the conditioning on historical monthly information (via $SIE_{LastMonth}$) with potentially-invaluable recent daily  information, via $SIE_{Last30Days}$ and $SIE_{Today}$. 

{\color{black}  Empirical results validate such modeling choices,  as FELR is a more-than-adequate benchmark, surpassing the SIO median (the median of all submitted forecasts) for September and the three horizons for which the latter is available.  This can be seen in the September subplot of Figure \ref{MSEOOS_FELR_FEML}.  While the linear trend is widely used as a generic reference point,  the SIO median,  very much like the mean of the Survey of Professional Forecasters in macroeconomics,  is a tenacious contender \textcolor{black}{ against which to benchmark   new approaches} \citep{andersson2021seasonal}.  The crucial practical advantage of FELR over the SIO median is obviously that we can generate its forecasts for more than three arbitrarily-fixed dates and a single target month,} producing direct (rather than iterated) FELR forecasts  day-by-day, using model parameter estimates optimized to the remaining predictive horizon, thanks to the trivial simplicity and speed of FELR estimation by linear least-squares regression.\footnote{One makes a  multi-period ``direct" forecast with a horizon-specific multi-period-ahead estimated model.   In contrast, one makes a multi-period ``iterated" forecast with a one-period-ahead estimated model, iterated forward for the desired number of periods.  Direct projections are theoretically superior under model misspecification (which is always the relevant case), because they directly minimize the relevant multi-step predictive loss, as per  \cite{Ing2003}, Theorem 4 and Corollary 3.}  We exploit this fact below  to make and examine 120 daily fixed-target Arctic sea ice forecasts from June through September.

\subsection{Glide Charts}

We measure   forecast performance, and its evolution as the target date is approached,  with \textit{RMSFE glide charts} (RGCs).  An RGC is simply a plot of the sequence of root mean squared forecast errors (RMSFEs) from the ordered sequence of 120 regressions with the conditioning information expanding as the target date is approached, where $RMSFE = \sqrt{\frac{e'e}{T}}$, for regression residual vector $e$ and sample size $T$.\footnote{ We are of course not the first to work with glide charts or similar constructs (whatever the name) for sea ice forecasts, whether in absolute terms or relative to a benchmark.  Key recent references include \cite{ChevallierEtAl2013}, \cite{DayEtAl2014}, \cite{HawkinsEtAl2016}, and \cite{BushukEtAl2019}.}

\begin{figure}[tp]  
	\caption{Glide Charts: Feature-Engineered Linear Regression}
	\begin{center}
		\includegraphics[trim={0mm 0mm 0mm 0mm},clip,width=.3\textwidth]{{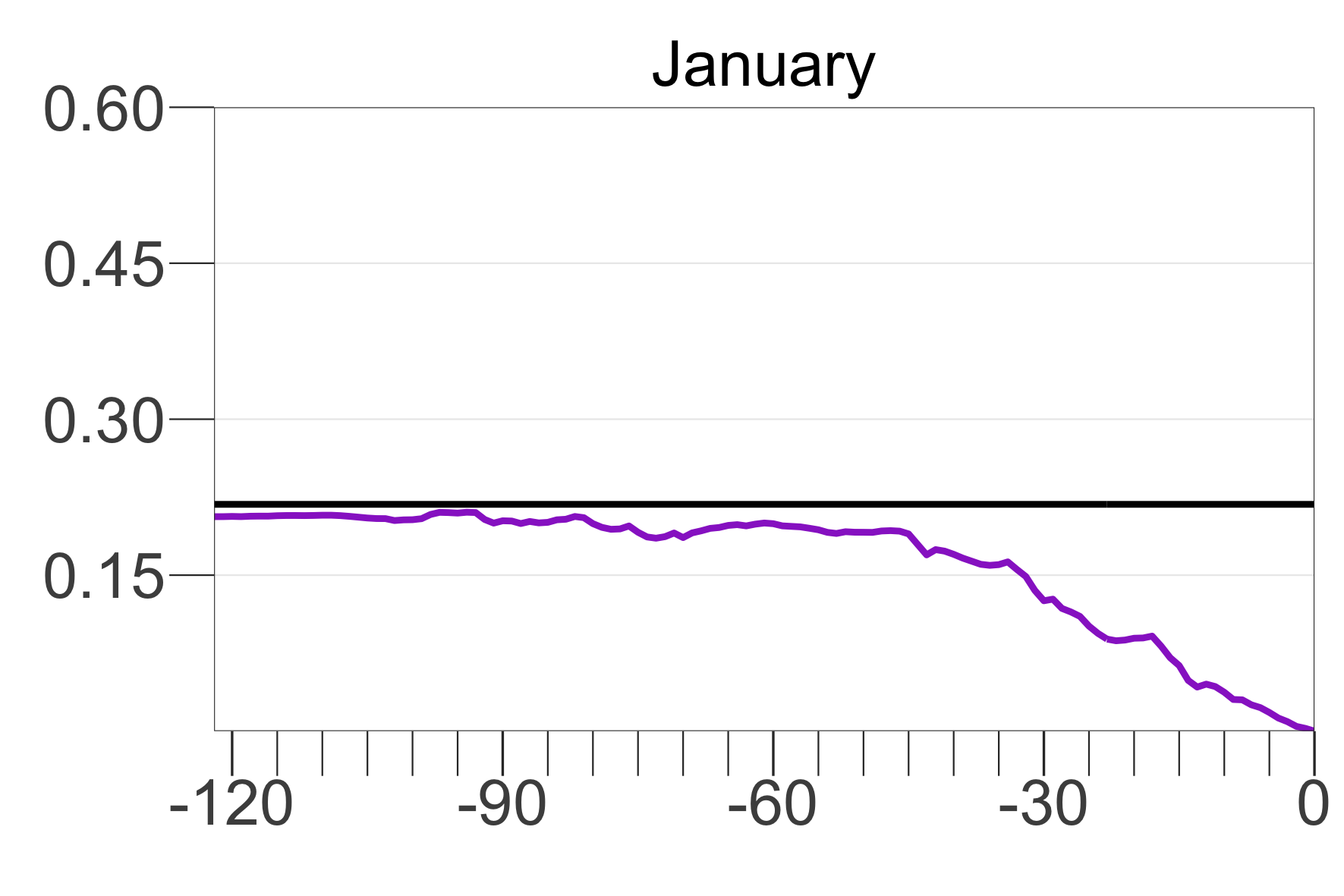}}
		\includegraphics[trim={0mm 0mm 0mm 0mm},clip,width=.3\textwidth]{{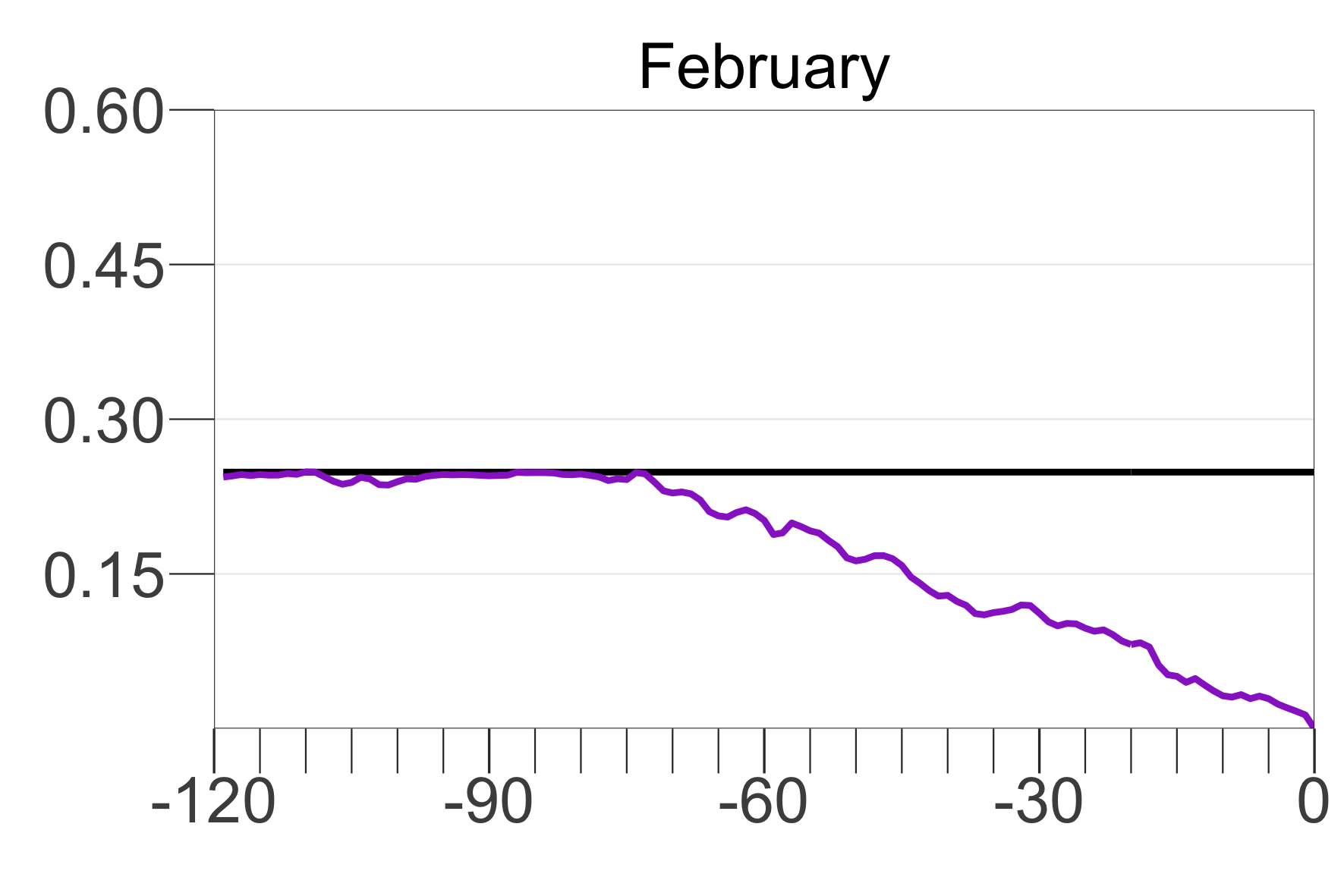}}
		\includegraphics[trim={0mm 0mm 0mm 0mm},clip,width=.3\textwidth]{{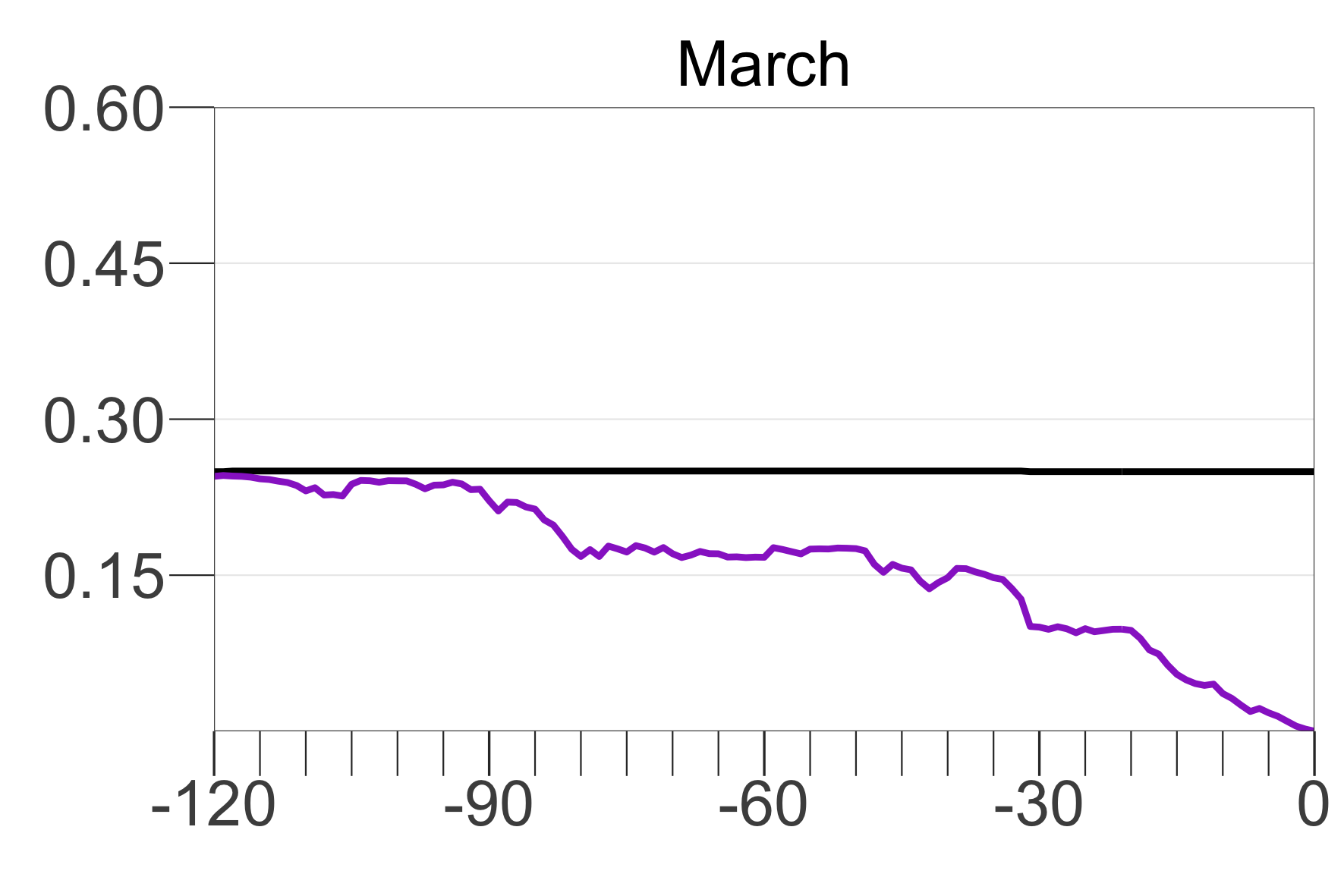}}
		\includegraphics[trim={0mm 0mm 0mm 0mm},clip,width=.3\textwidth]{{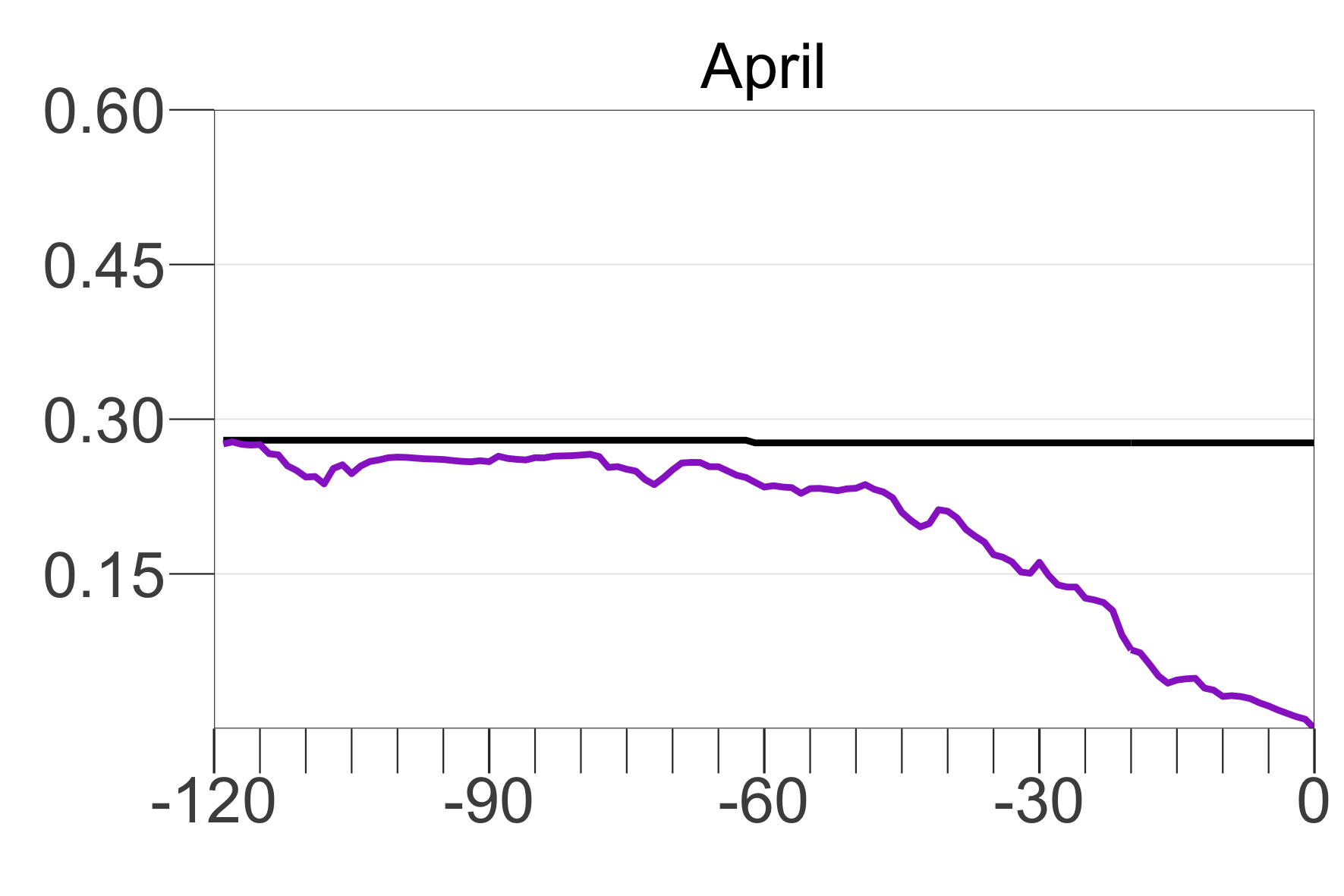}}
		\includegraphics[trim={0mm 0mm 0mm 0mm},clip,width=.3\textwidth]{{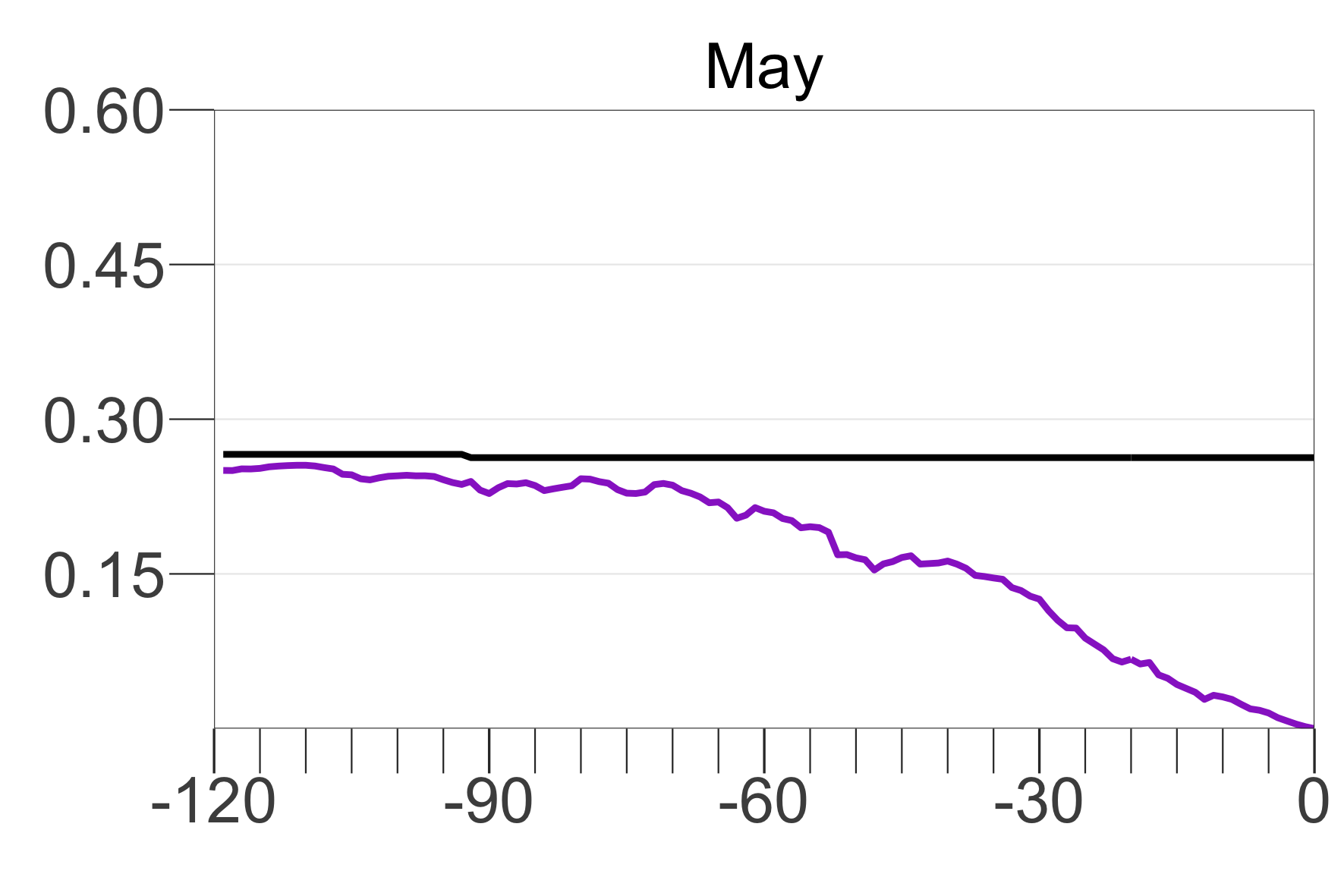}}
		\includegraphics[trim={0mm 0mm 0mm 0mm},clip,width=.3\textwidth]{{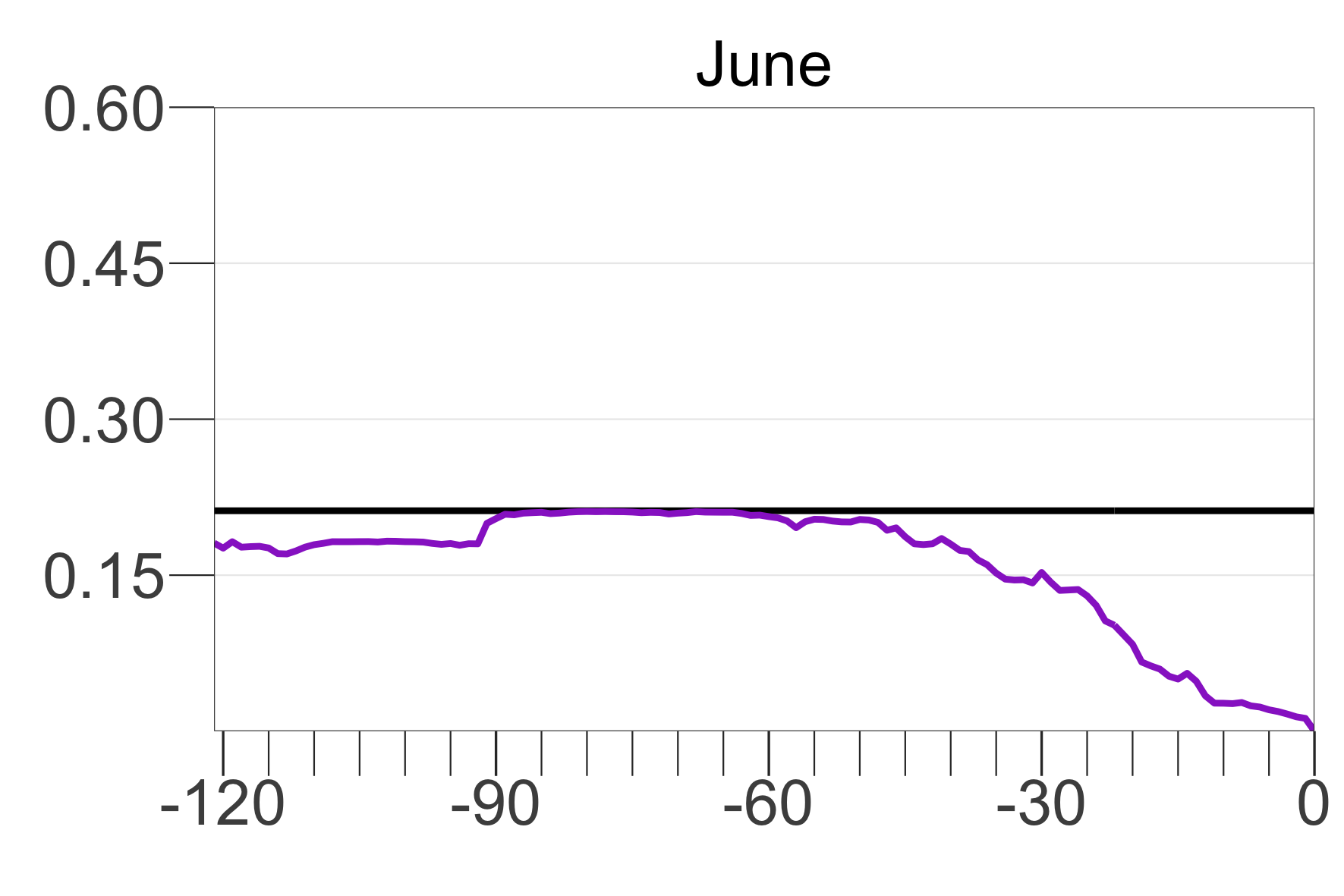}}
		\includegraphics[trim={0mm 0mm 0mm 0mm},clip,width=.3\textwidth]{{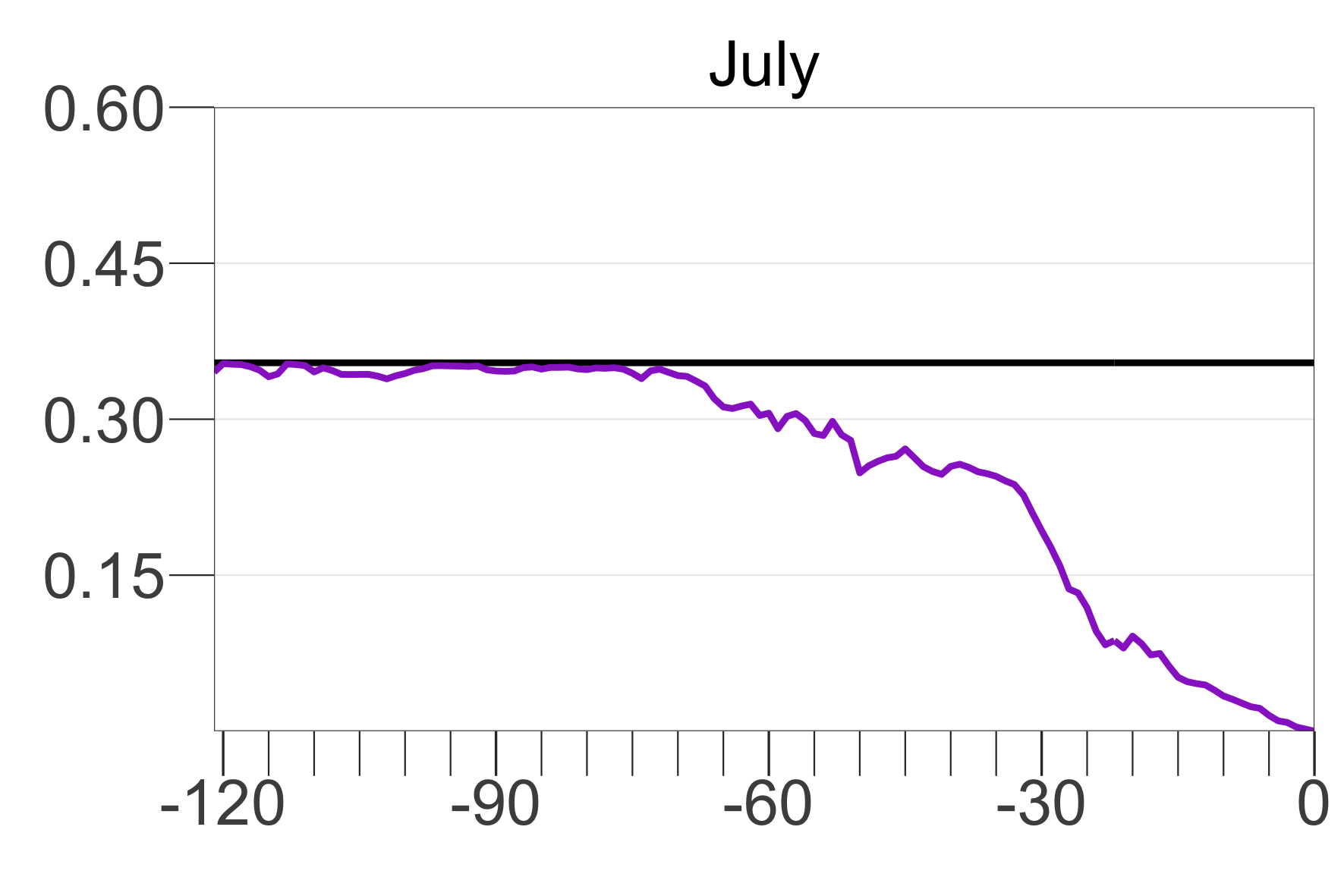}}
		\includegraphics[trim={0mm 0mm 0mm 0mm},clip,width=.3\textwidth]{{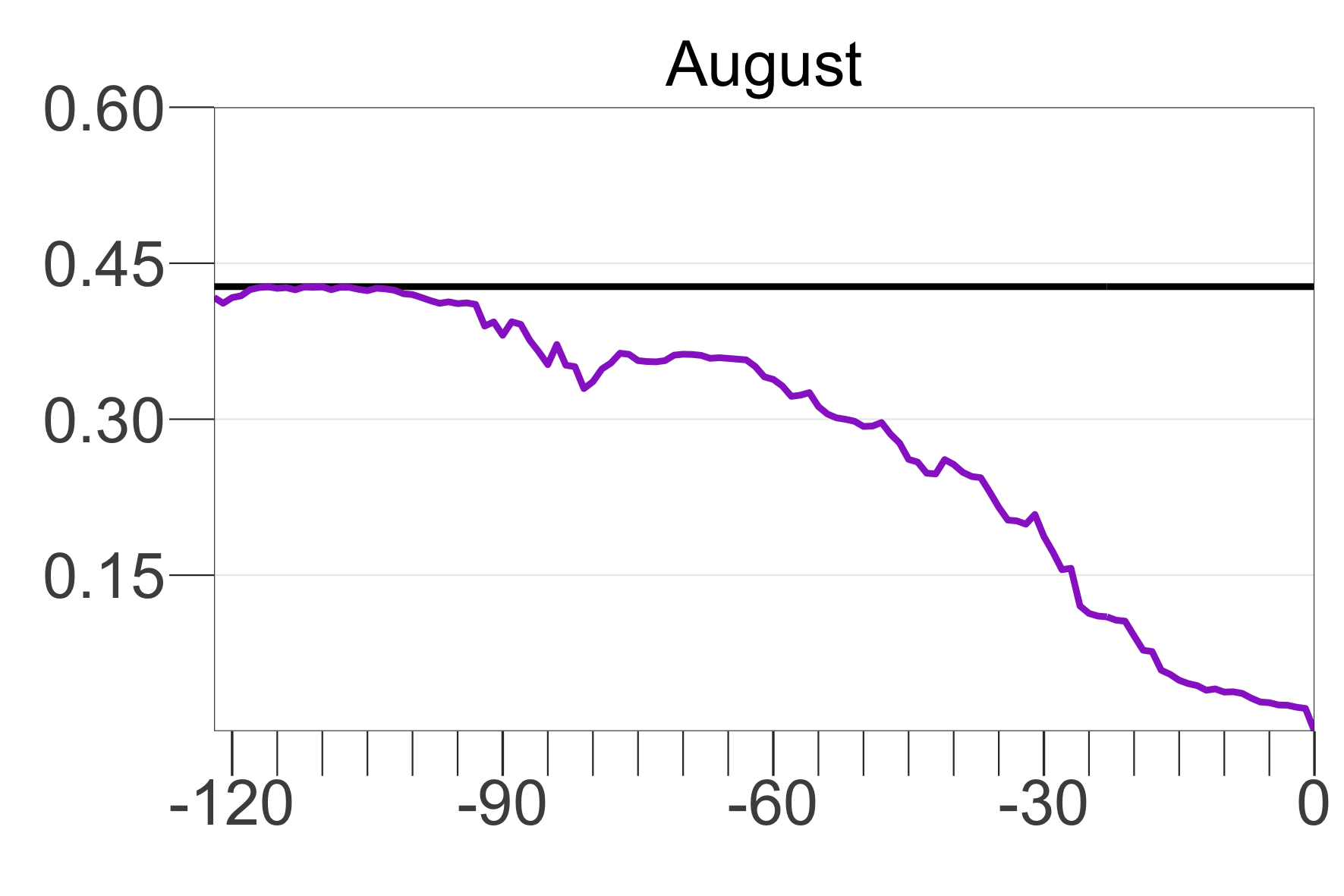}}
		\includegraphics[trim={0mm 0mm 0mm 0mm},clip,width=.3\textwidth]{{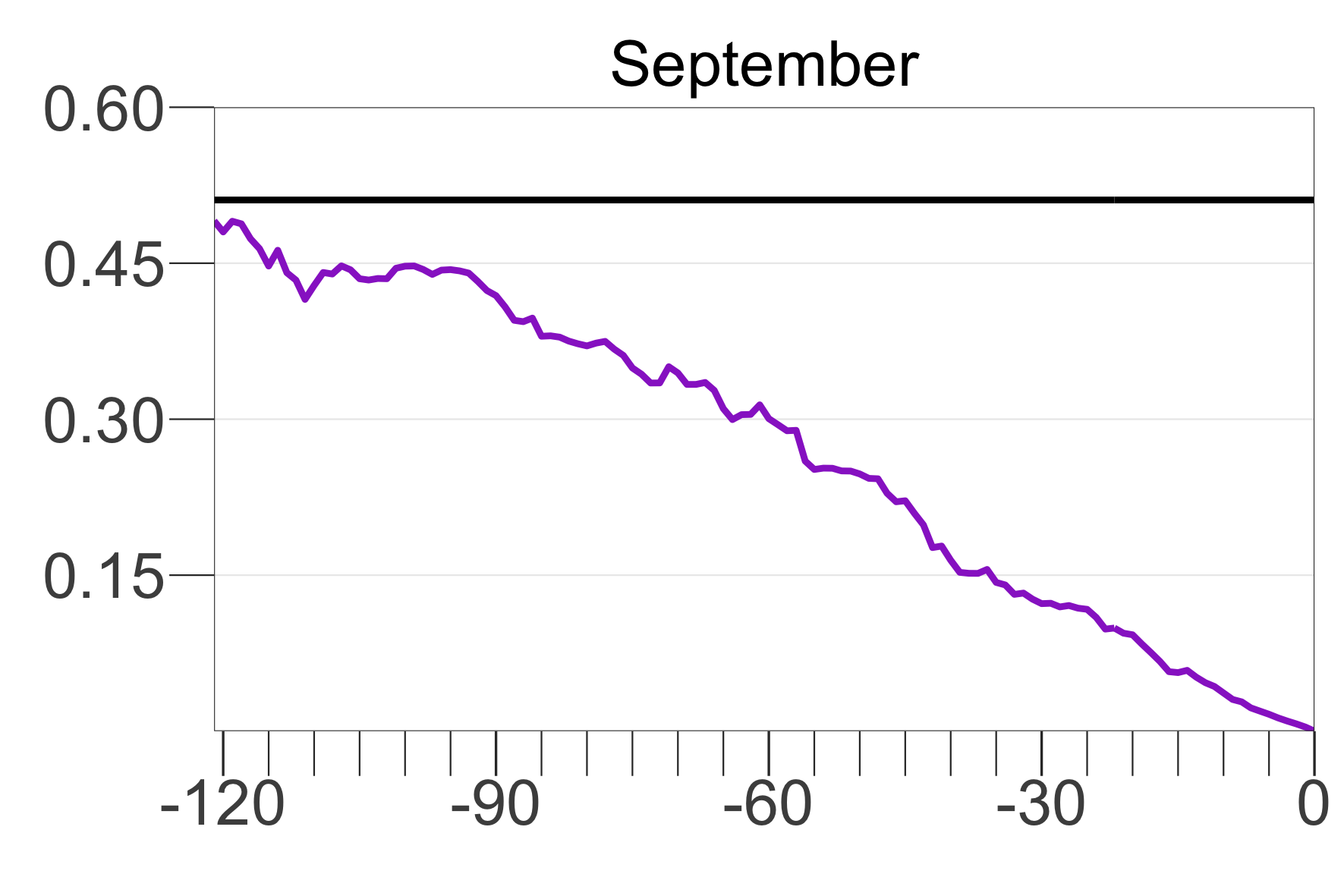}}
		\includegraphics[trim={0mm 0mm 0mm 0mm},clip,width=.3\textwidth]{{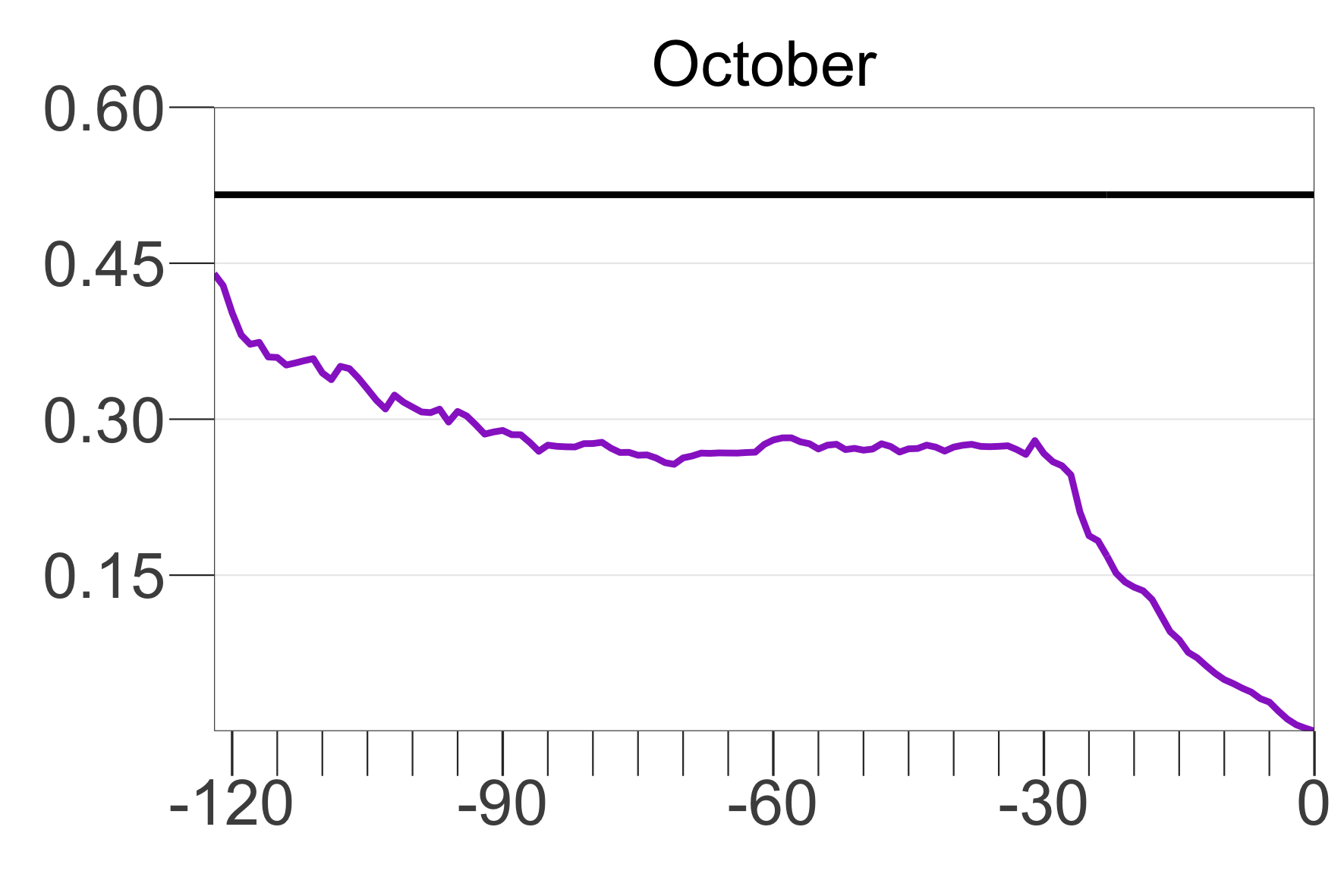}}
		\includegraphics[trim={0mm 0mm 0mm 0mm},clip,width=.3\textwidth]{{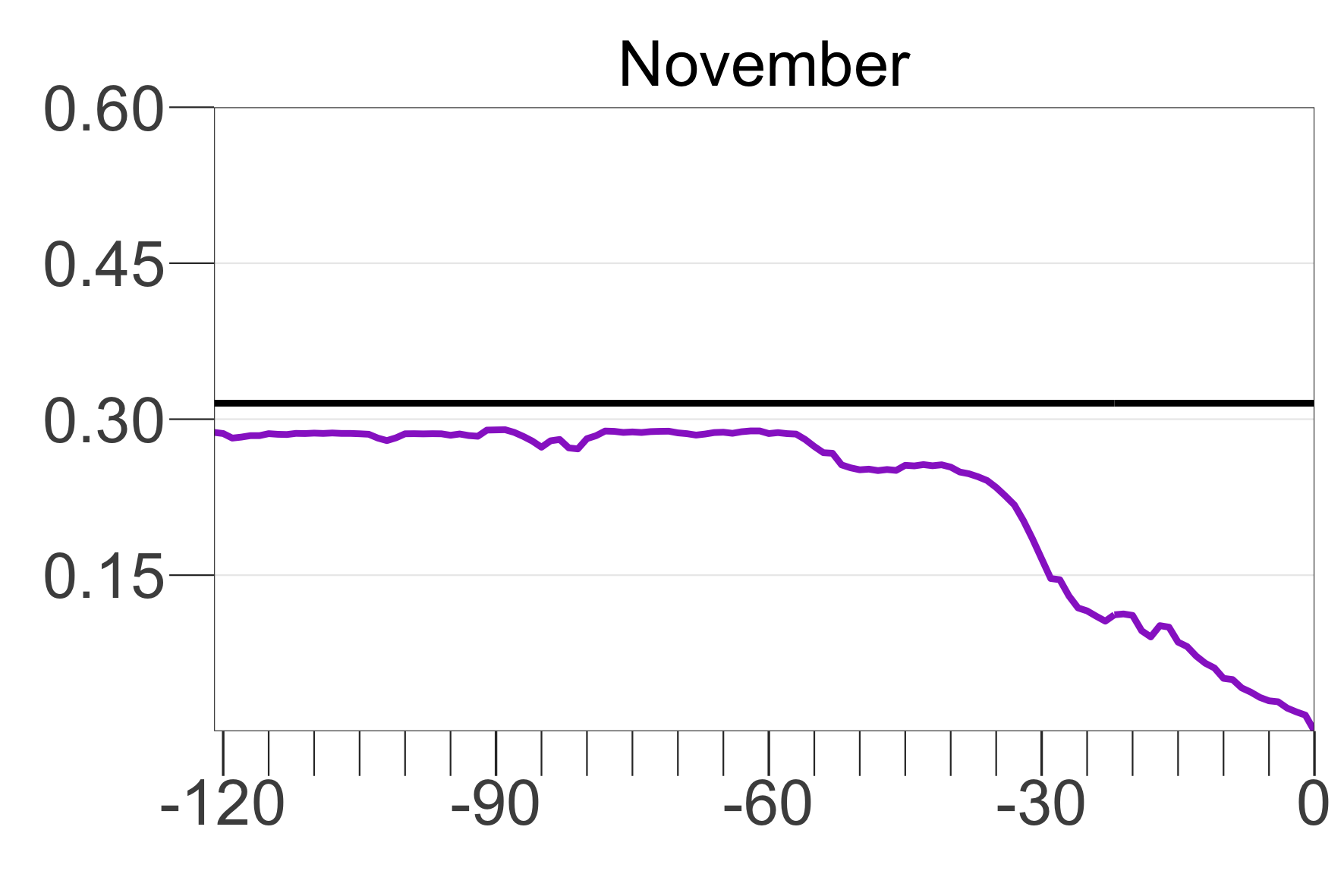}}
		\includegraphics[trim={0mm 0mm 0mm 0mm},clip,width=.3\textwidth]{{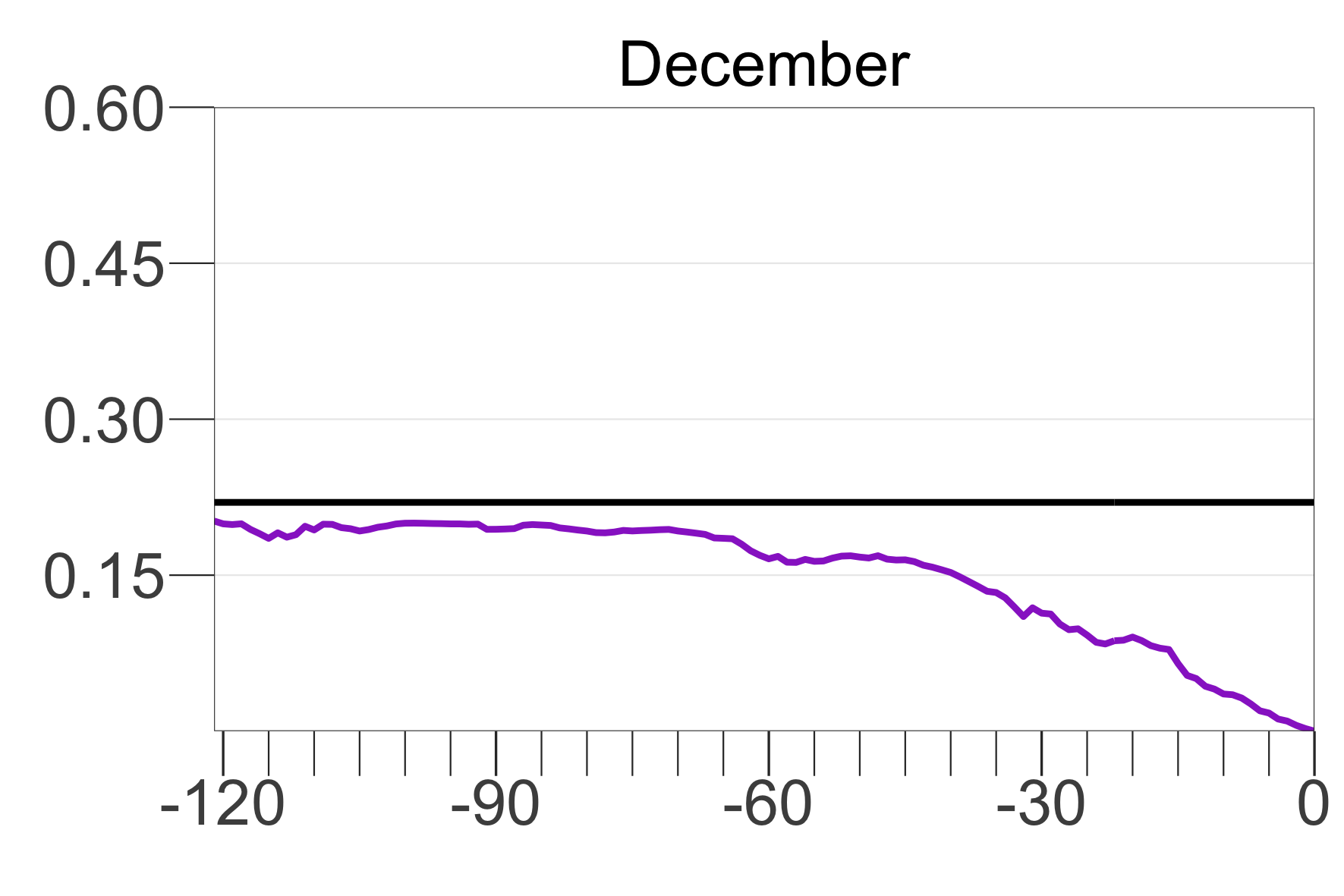}}
	\end{center}
	\label{glideRMSE}
	\begin{spacing}{1.0}  \noindent \footnotesize  Notes: We show RMSFE from linear trend models and  feature-engineered linear regression models:
		$$
		SIE_{m} ~{\rightarrow} ~c, ~ Time ~~(black)
		$$
		$$
		SIE_{m} ~{\rightarrow} ~c, ~ Time,  ~SIE_{LastMonth}, ~SIE_{Last30Days}, ~ SIE_{Today} ~~(orchid),
		$$
 with the estimation sample expanding over 120 days. The horizontal axes show  the number of days until the end of the target month $m$.  In some instances $SIE_{LastMonth} = SIE_{Last30Days}$, so that some models would suffer from perfect multicollinearity. In these cases, we drop $SIE_{Last30Days}$.  See text for details. 
\normalsize
	\end{spacing}
\end{figure}

We use $SIE$  observational data from 1979 to 2020 from the National Snow and Ice Data Center (NSIDC), which  uses the NASA team algorithm to convert microwave brightness readings into ice coverage \citep{Fettereretal2017}.\footnote{See Appendix \ref{data_appendix} for details.}    We start with a 4-month lead (120 days).  Once the 120 FELR regressions are run, we construct RGCs. RGCs naturally differ across months, so we examine the shapes for each month separately.    

  In Figure \ref{glideRMSE} we show the twelve monthly FELR RGCs, together with the twelve monthly benchmark linear trend  RGCs for comparison.   First consider the linear trend RGCs, which are flat, for all months and horizons.  This is expected, because they capture only extremely low-frequency dynamics and hence take almost no account of evolving conditions as the target date is approached. 
  
 Now consider the  FELR RGCs, which  are very different.  Very early on, near date $T-120$, they are little different from those of the linear trend benchmark, but accuracy eventually improves  (so the RGCs drop), achieving perfection by the target date $T$ (so the RGCs are 0).  Moreover, the precise glide chart paths differ across months.   
 
 Most months show FELR  predictability thresholds, meaning that accuracy initially fails to improve as the target date is approached, but then increases progressively once a predictability threshold lead time is crossed. For example, the peak-summer (July) sea ice forecast shows no increase in accuracy until roughly $T-60$, after which accuracy rapidly improves.  This corroborates the results of \cite{DayEtAl2014}, who find a spring ``predictability barrier" for summer pan-Arctic SIE predictions, but contrasts with those of \cite{BushukEtAl2019}. 

Interestingly, the predictability thresholds are earliest for the low-ice months of August, September (when Arctic sea ice achieves its minimum), and October. The August threshold is around $T-90$, the September threshold is evidently around $T-120$, and the October threshold is even earlier -- literally off the chart!

\section{  Glide Charts for  \\ \hspace*{0.08em} Feature-Engineered  Machine Learning (FEML)   } \label{feml}

Although FELR clearly captures salient features in $SIE_m$ at various forecasting horizons, it remains a linear model.  Simultaneously,  there are ample plausible sources of nonlinear $SIE_m$ dynamics, including tipping points and feedback loops \citep{maslanik2007younger}.  \textcolor{black}{With the} shape of those nonlinearities being unknown, we turn to flexible nonlinear machine learning (ML) approaches to estimate them.  A significant roadblock to that enterprise, however, is that completely nonparametric ML methods simply will not work on a sample of size $T\textcolor{black}{\approx}40$.  We confront this situation by using a  feature-engineered  machine learning  (FEML) approach, which, as the name suggests,  retains the feature engineering that made FELR  a successful benchmark, but with substantial generalization.  In particular, the ML model we consider is Macro Random Forest \citep{MRF}, which builds nonlinearities around FELR rather than  modeling  everything nonparametrically.  Hence FEML continues to capture linear signals precisely as with FELR, but it can also capture additional nonlinear signals, while continuing to economize on degrees of freedom.

To make such points more clear, we first review the basics of Macro Random Forest (MRF) and then describe the various FEML specifications that will be used in our subsequent forecasting exercise.

\subsection{Feature-Engineered Machine Learning}

Here we introduce a  flexible tree-based class of nonparametric nonlinear feature-engineered models for fixed-target sea ice forecasting.

\subsubsection{Macro Random Forest}

\cite{MRF} proposes a new form of random forest (RF,  \citealt{breiman2001random}) better suited for time series,  especially macroeconomic data where the available series are typically of short length.   The model is
\begin{align*}
y_{t}& ={X}_{t}\beta _{t}+\epsilon _{t} \\
\beta _{t}& =\mathcal{F}(S_{t})
\end{align*}%
where $S_{t}$ are the state variables governing time variation and $\mathcal{%
F}$ a forest. $S_{t}$ can be a large data
set, beyond what is included in FELR.  ${X}$ determines the \textit{linear} model that we want to be time-varying.  By design, it is preferable that ${X}\subset S$ be parsimonious and a priori important compared to the larger $S$.   For instance,  one can use lags of $y_{t}$ for ${X}_{t}$ when an appreciable degree of persistence is suspected.  In this paper, ${X}_{t}$ will be the features of FELR.  

While an advantage of the method is its potential for interpretation via the generalized time-varying parameters $\beta_t$,  of greater interest here are its predictive advantages in an environment with scarce data and a strong linear signal.  Indeed, it is easy to see that, if  $\mathcal{F}$ ends up hardly nonlinear -- or seen differently,  mostly time-invariant --  FEML collapses to FELR.  In contrast, a plain RF that learns nothing, collapses to the unconditional mean.  Thus,  FEML constructs the conditional mean economically by starting with FELR and incorporating nonlinearities (as much as one can afford with \textcolor{black}{$T\approx40$})  around it.  In contrast,  a plain RF would struggle to capture linear autoregressive signals effectively using hard-thresholding functions (the trees) and would be left with little or no degrees of freedom for ``real" nonlinearities.  For much more on this point, see \cite{MRF}.

The estimation is carried \textcolor{black}{out} through a greedy algorithm, which, in its most basic form,  is to run  
\begin{equation}
\begin{aligned} \min _{j \in \mathcal{J^-}, \enskip c \in \rm I\!R} \Bigg[ &
\min _{\beta_{1}}\sum_{t \in l_1(j,c)} 
\left(y_{t}-{X}_{t}\beta_{1}\right)^{2} + & \min _{\beta_{2}} \sum_{t \in l_2(j,c)} \left(y_{t}-{X}_{t} \beta_{2}\right)^{2} \Bigg]. \end{aligned}  \label{mrf_algo}
\end{equation}%
recursively to construct trees. In words, at each potential tree split, we obtain the optimal variable $S_{j}$ (i.e., the best $j$ out of the random subset of predictors indexes $\mathcal{J}^{-}$) with which to split our sample, \textcolor{black}{and $c$, i.e. the value at which we should split $j$.}   
The resulting outputs $j^{\ast }$ and $c^{\ast }$ are used to split $l$ (the parent "leaf") into two children leaves, $l_{1}$ and $l_{2}$.  Splitting things in halves, and those halves in other small halves eventually \textcolor{black}{leads to obtaining} leaves of size  1 (or a small number) yielding  $\beta_t$,   a coefficient at each point in time.

As in RF,  the core sources of regularization in MRF are (1) averaging over a diversified ensemble of trees generated by Bagging, and (2) random eligibility of predictors for splits  $\mathcal{J}^{-}\subset \mathcal{J}$.\footnote{\color{black} This, and the fact that $\beta_t$ comes from very small leaves obtained from running \eqref{mrf_algo}  recursively, is precisely why we get a different $\beta_t$ for each $t$. } Nonetheless, $\beta_t$'s (and corresponding predictions) may benefit from additional regularization---this is particularly true of short time series where the potency of Bagging is more limited.  Time smoothness is made operational by taking the
``rolling-window view" of time-varying parameters. That is, the tree solves
many weighted least squares problems including close-by
observations.  To keep computational demand low,  \cite{MRF} suggests to use a kernel $w(t;\zeta)$ designed as a symmetric 5-step Olympic podium.  Informally, the kernel puts a weight of 1 on observation $t$, a weight of $\zeta<1$ \textcolor{black}{on} observations $t-1$ and $t+1$
and a weight of $\zeta^2$ \textcolor{black}{on} observations $t-2$ and $t+2$.  Finally,  a
small Ridge penalty is added for matrices to invert nicely even
in very small leaves,  which will be inevitably prevalent in our application.  With those additions,  \eqref{mrf_algo} \textcolor{black}{turns into} the more sophisticated
\begin{equation}
\begin{aligned} \min _{j \in \mathcal{J^-}, \enskip c \in \rm I\!R} \Bigg[ &
\min _{\beta_{1}}\sum_{t \in l_1^{RW}(j,c)} w(t;\zeta)
\left(y_{t}-{X}_{t}\beta_{1}\right)^{2}+\lambda \Vert \beta_1 \Vert_2
\\ + & \min _{\beta_{2}} \sum_{t \in l_2^{RW}(j,c)}
w(t;\zeta)\left(y_{t}-{X}_{t} \beta_{2}\right)^{2}+\lambda \Vert
\beta_2 \Vert_2 \Bigg] ,\end{aligned}  \label{mrf_algo_rw}
\end{equation}%
where $l_1^{*}(j,c)$ and $l_2^{*}(j,c)$ denote the expanded leaves incorporating the aforementioned neighboring observations in time space.

To put things in perspective,  a standard RF is a restricted version of MRF where ${X}_{t}=\iota $,  $\lambda =0$, $\zeta =0$ and the block size for Bagging is 1.  Said differently,  the sole regressor is an intercept,  there is no within-leaf shrinkage, and Bagging
is carried \textcolor{black}{out} as-if we were working with a cross-section.  As discussed earlier,  by design,  MRF will have an edge over RF whenever linear signals included in ${X}_{t}$ are strong
and the number of training observations (or signal-to-noise ratio) is low.  Clearly,  all those boxes are checked in this paper's application.

\subsubsection{Two Specifications}

We consider two MRF specifications corresponding to different configurations of $S_t$ and $X_t$.  First,  the FEML model has a linear part $X_t$ comprising the very same features of the FELR model, i.e., an intercept ($c$), a linear time trend ($Time$), SIE of the previous month ($SIE_{LastMonth}$), the average SIE over the last 30 days ($SIE_{Last30Days}$), and today's  measurement of the Arctic's sea ice extent ($SIE_{Today}$).  State variables $S_t$ feature a larger set of potentially informative climate variables, akin to \cite{VARCTIC}'s VARCTIC,  designed to proxy the current state of the Arctic.  In particular, we include (1) all features entering $X_t$, (2) daily SIE measurements of the previous 14 days, (3) daily Sea Ice Thickness (SIT) measurements of the latest 14 days available, (4) the average SIT over the latest 30 days of available measurements, (5) lags of monthly measurements of SIE, SIT, CO2 and Air Temperature (AT), and (6) the first five principal components of the feature set described in (1)-(5), which may help in summarizing  the key variation in the relatively large $S_t$.\footnote{Daily SIT measurements are published at the end of the following month, i.e. on April 25$^{\text{th}}$ the time series covers data only through the end of February. The data for March is not released until May 1$^{\text{st}}$.  We use data that is publicly available at the time of forecast.  Consequently, depending on the exact day at which one is making a prediction, the SIT enters $S_t$ with a lag of one to two months.}$^,$\footnote{The number of lags differs by variable, but all have a common starting point.  For example,  when making a  prediction on January 20$^{\text{th}}$ of year $t$, the monthly lags for SIE, SIT, CO2 and AT start with a measurement for January of year $t-1$, and end with the latest month on which complete information is available. Thus, for SIE this boils down to 12 monthly lags from January of $t-1$ until December of $t-1$.  For SIT, we only have information until and including the whole month of November $t-1$. Finally, for CO2, the monthly lags run from January $t-1$ through October $t-1$, and AT enters with monthly estimates for January $t-1$ through September $t-1$.} Details on the provenance of the various data series appear in Appendix \ref{data_appendix}.

Second, the ``Pocket FEML" model has the full $S_t$ of {FEML} but $X_t$ is a subset of FELR's regressors, namely  $c$,  $Time$, and $SIE_{Today}$.  The motivation for a restricted FELR as a linear part is that the size of $X_t$ ultimately reduces the potential depth of trees in the forest for very small datasets.  This is due to the fact that the algorithm needs to run small ridge regressions in each leaf, and the larger that regression gets, the larger the minimal leaf size must be to accommodate that operation.  In short, it  limits the expressivity of the trees by restricting their depth.  Thus,  the potential benefits of a more condensed  $X_t$ is to discard partially redundant information,  avoid near-singularity problems in small terminal nodes,  and ultimately leave more room for nonlinearities in $\mathcal{F}$.  The cost is that, obviously,  with respect to the {MRF} specification, we lose the linear signals from the less noisy  $SIE_{LastMonth}$  and  $SIE_{Last30Days}$ (although they are included in $S_t$).  While the  necessity of those is uncertain prior to 30-day-ahead forecasts,  they are mechanically essential for short-run forecasts.  This will be clearly visible in \textcolor{black}{the} empirical results.  Obviously,  this is known ex ante and a forecaster can simply switch to FELR or FEML past that threshold.  

Regarding tuning parameters,  we set values that are a priori more adequate in an environment with very little data.  The sampling rate for features in $S_t$ is $\frac{1}{3}$,  which is standard.  The subsampling rate of data rows is $\frac{9}{10}$ which is rather high and limits the potential for \textcolor{black}{tree-diversification} coming from that source.  The upside is that it allows for slightly deeper trees to be grown,  which is much needed when faced with a small $T$.  $\zeta$ is set to 0.25 which reflects the view that we expect little persistence \textcolor{black}{in} the underlying $\beta_t$'s at the yearly frequency.  $\lambda$ is 1 (higher than what is used in typical macroeconomic specifications),  and brings helpful regularization when both the data subsampling rate is high and $\zeta$ is low.  The prior mean for the ridge shrinkage is switched from 0 to \textcolor{black}{values of} OLS coefficients,  which reflects the view,  {\color{black} like in the choice of a higher $\lambda$ and the specific $X_t$'s},  that if there is an improved model to be found, it should not be excessively far from FELR.  {\color{black} Our main out-of-sample results are robust to non-trivial deviations in both $\lambda$ and $\zeta$. }

\subsection{Glide Charts}

Here we display  glide charts for our two FEML versions (FEML, Pocket FEML) and compare them to glide charts for our two  FELR verions (FELR, Pocket FELR).\footnote{In addition, to disentangle whether Pocket FEML's performance differential comes from nonlinearities versus a sparser inherent linear equation,  we also include Pocket FELR in our set of benchmarks.}  We also distinguish between in-sample and out-of-sample versions.

\subsubsection{An In-Sample Analysis}

\begin{figure}[tp]  
	\caption{Glide Charts: FELR and FEML}
	\begin{center}
		\scalebox{1}[1.25]{\includegraphics[trim={0mm 10mm 0mm 10mm},clip,width=.325\textwidth]{{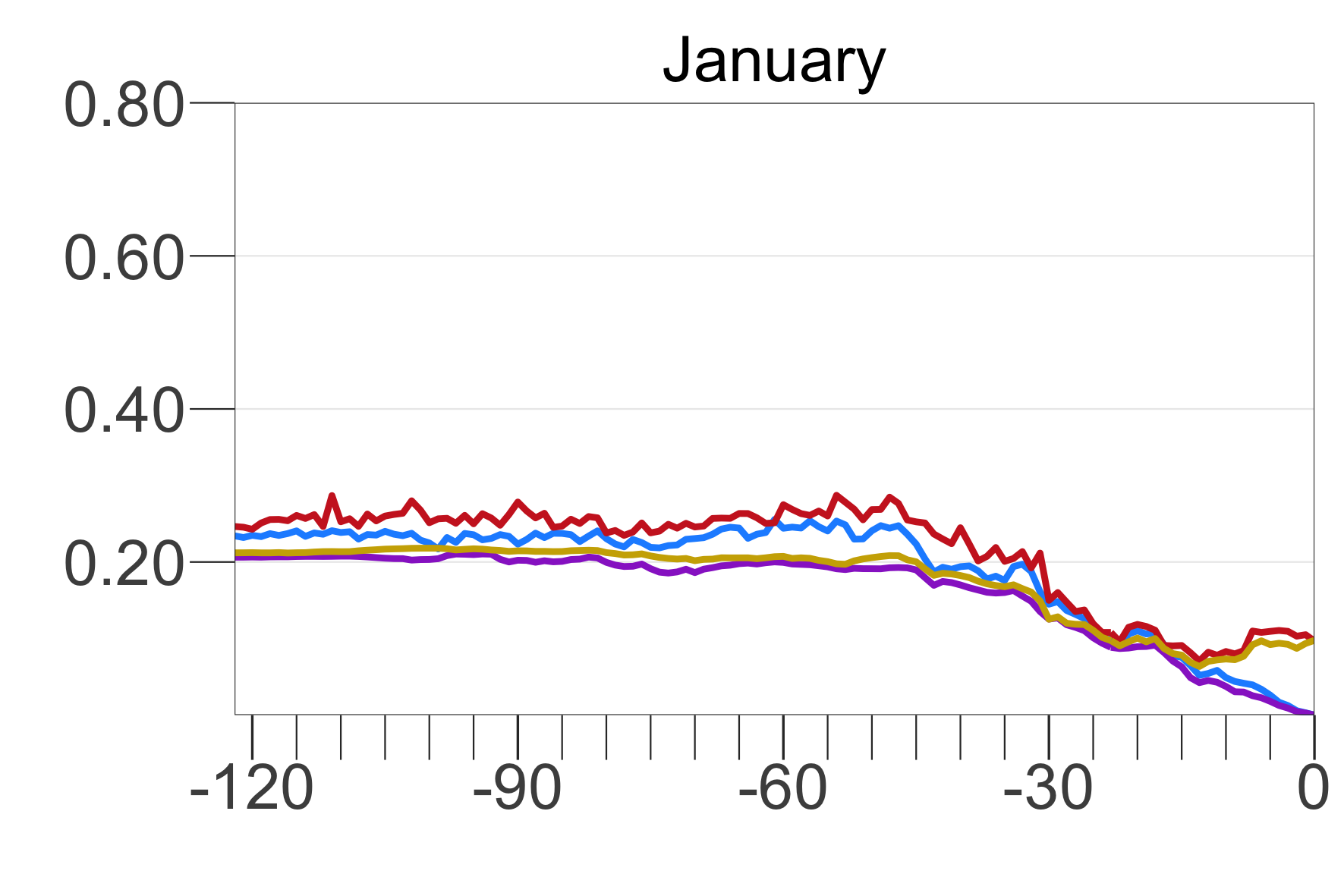}}}
		\scalebox{1}[1.25]{\includegraphics[trim={0mm 10mm 0mm 10mm},clip,width=.325\textwidth]{{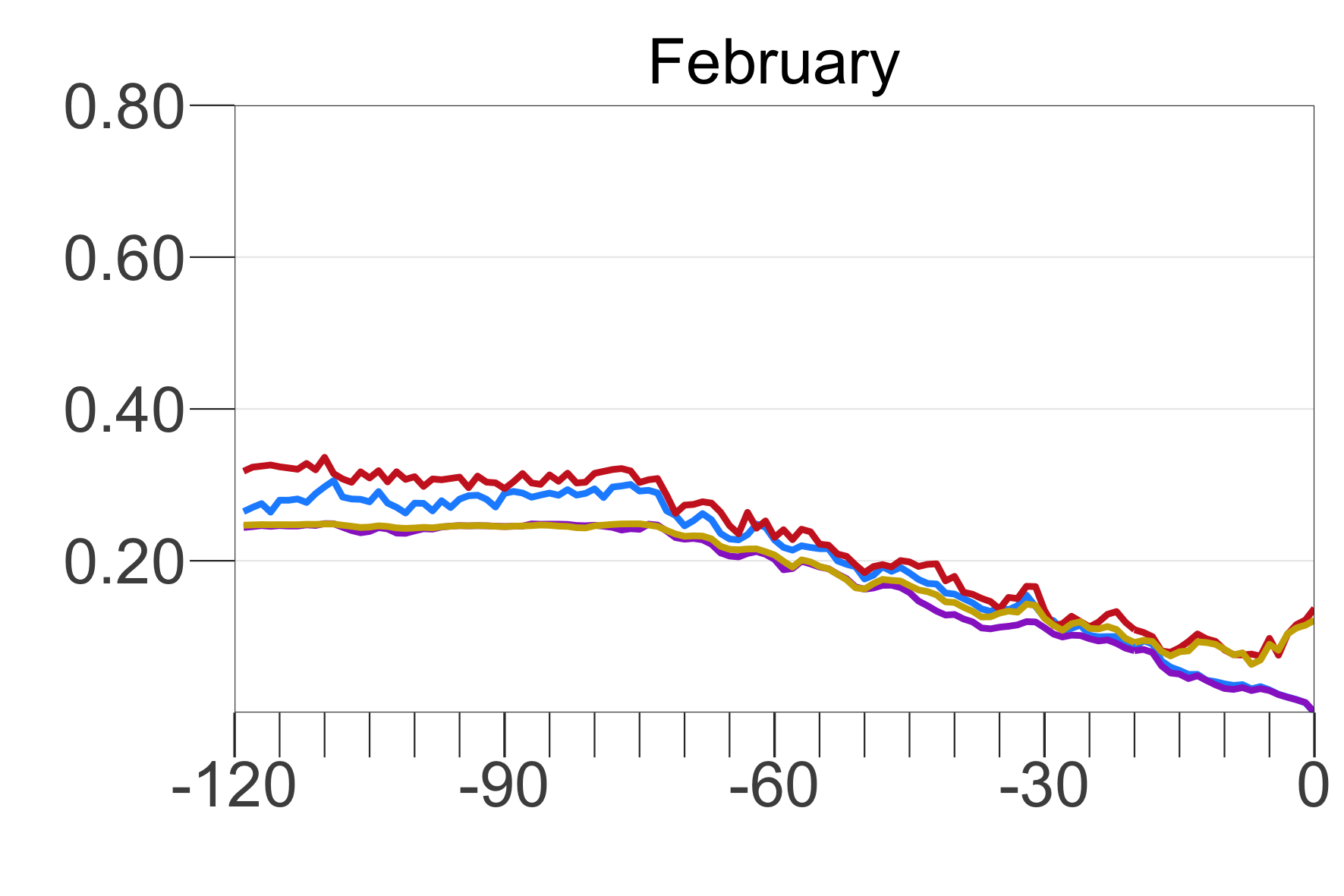}}}
		\scalebox{1}[1.25]{\includegraphics[trim={0mm 10mm 0mm 10mm},clip,width=.325\textwidth]{{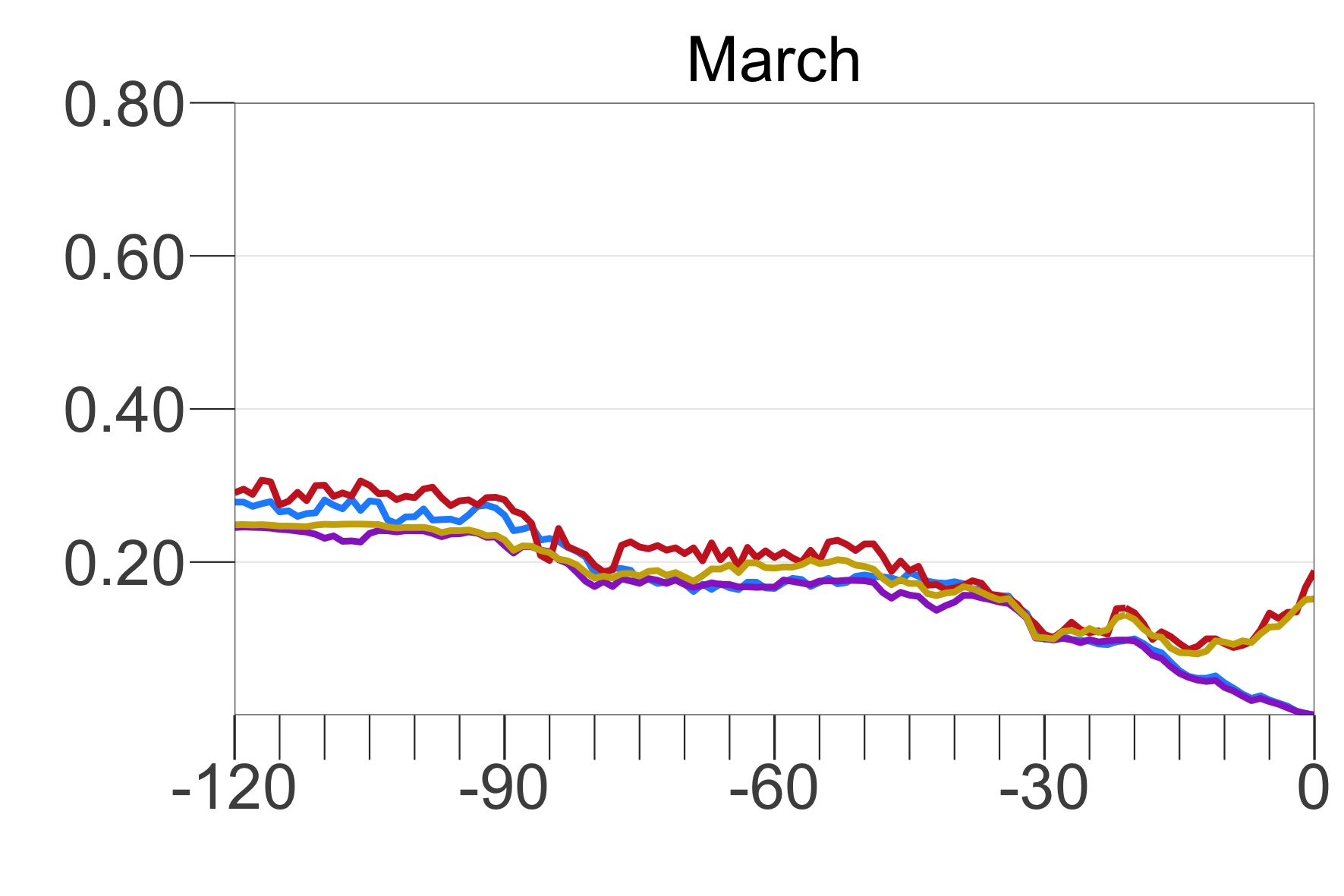}}}
		\scalebox{1}[1.25]{\includegraphics[trim={0mm 10mm 0mm 10mm},clip,width=.325\textwidth]{{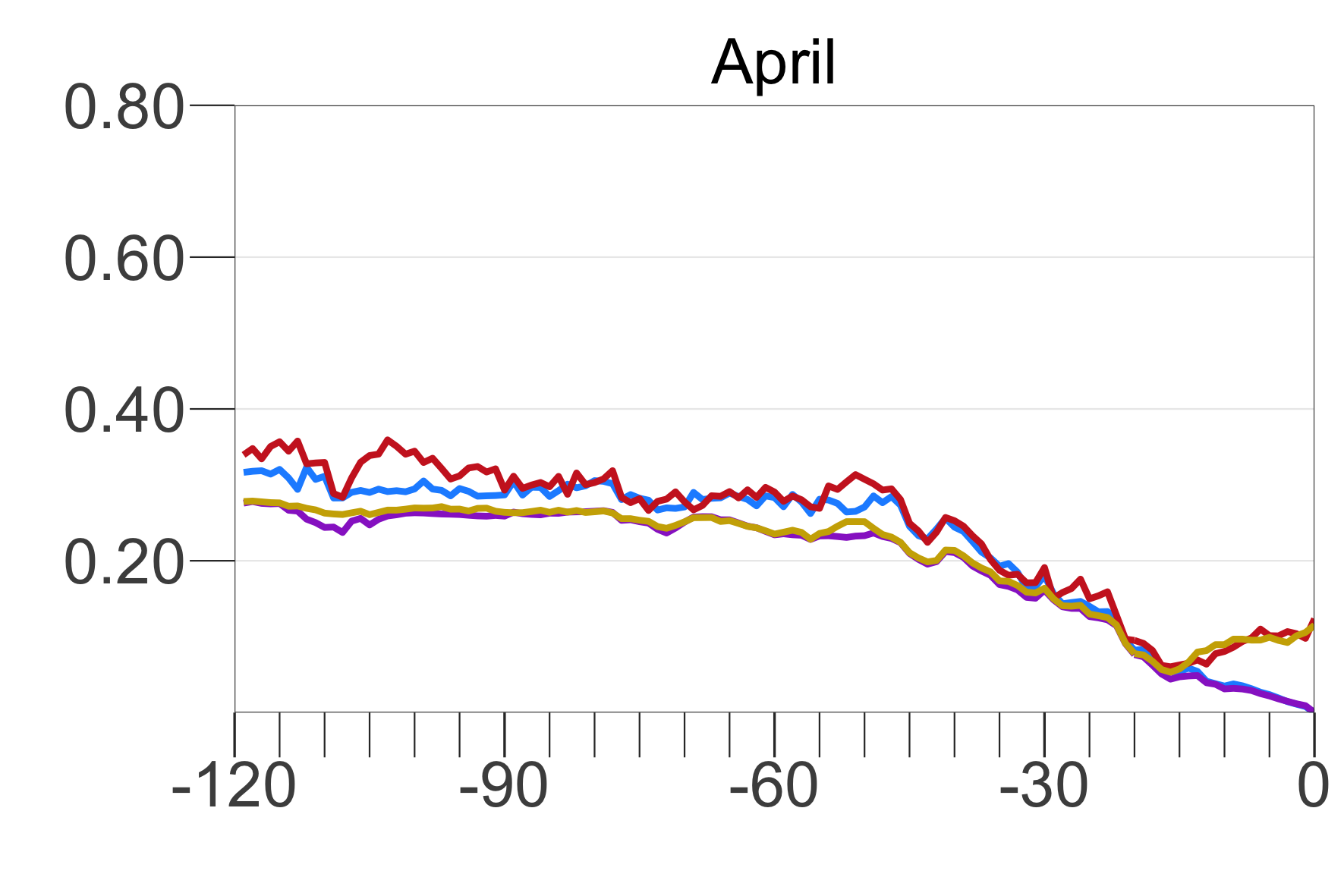}}}
		\scalebox{1}[1.25]{\includegraphics[trim={0mm 10mm 0mm 10mm},clip,width=.325\textwidth]{{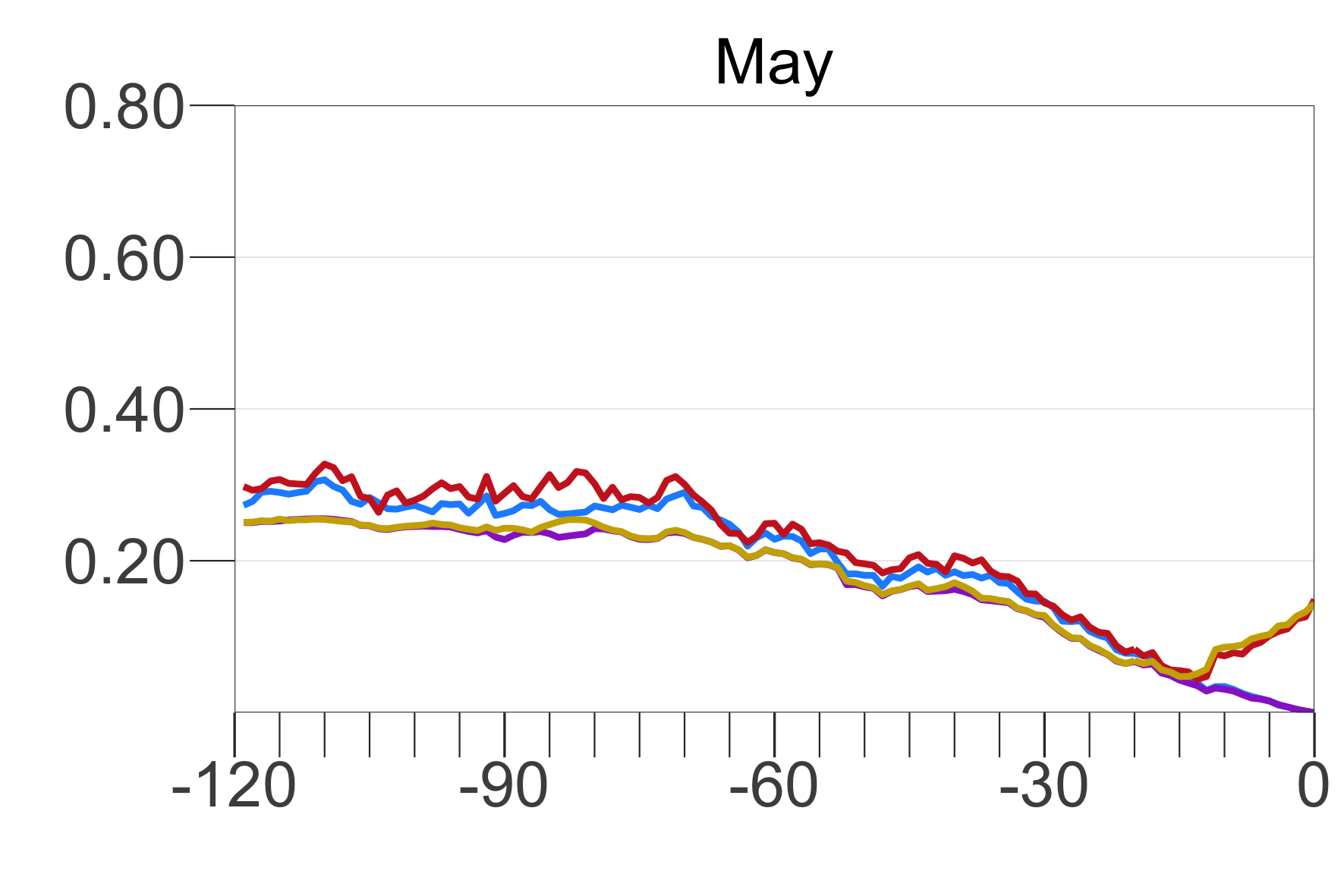}}}
		\scalebox{1}[1.25]{\includegraphics[trim={0mm 10mm 0mm 10mm},clip,width=.325\textwidth]{{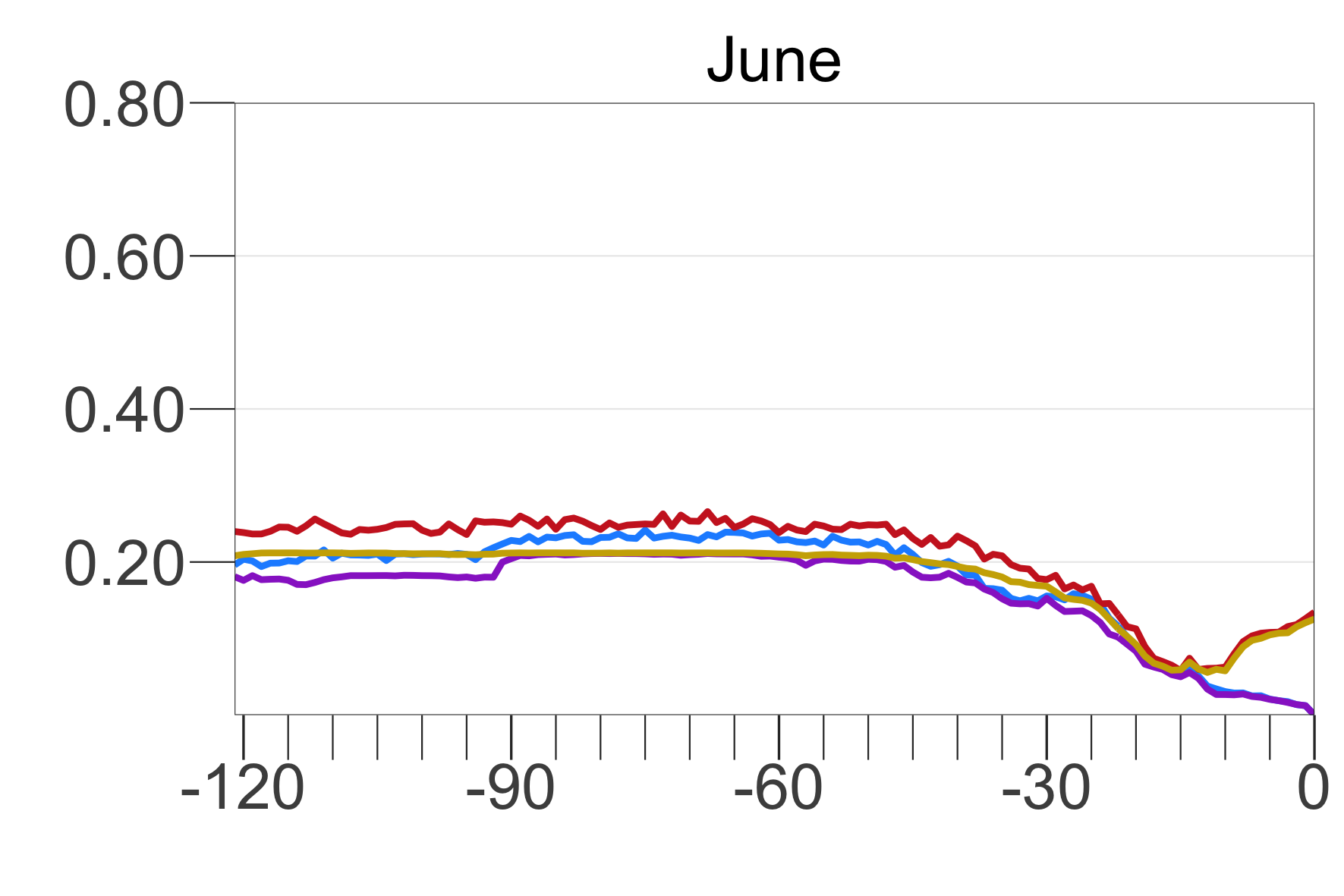}}}
		\scalebox{1}[1.25]{\includegraphics[trim={0mm 10mm 0mm 10mm},clip,width=.325\textwidth]{{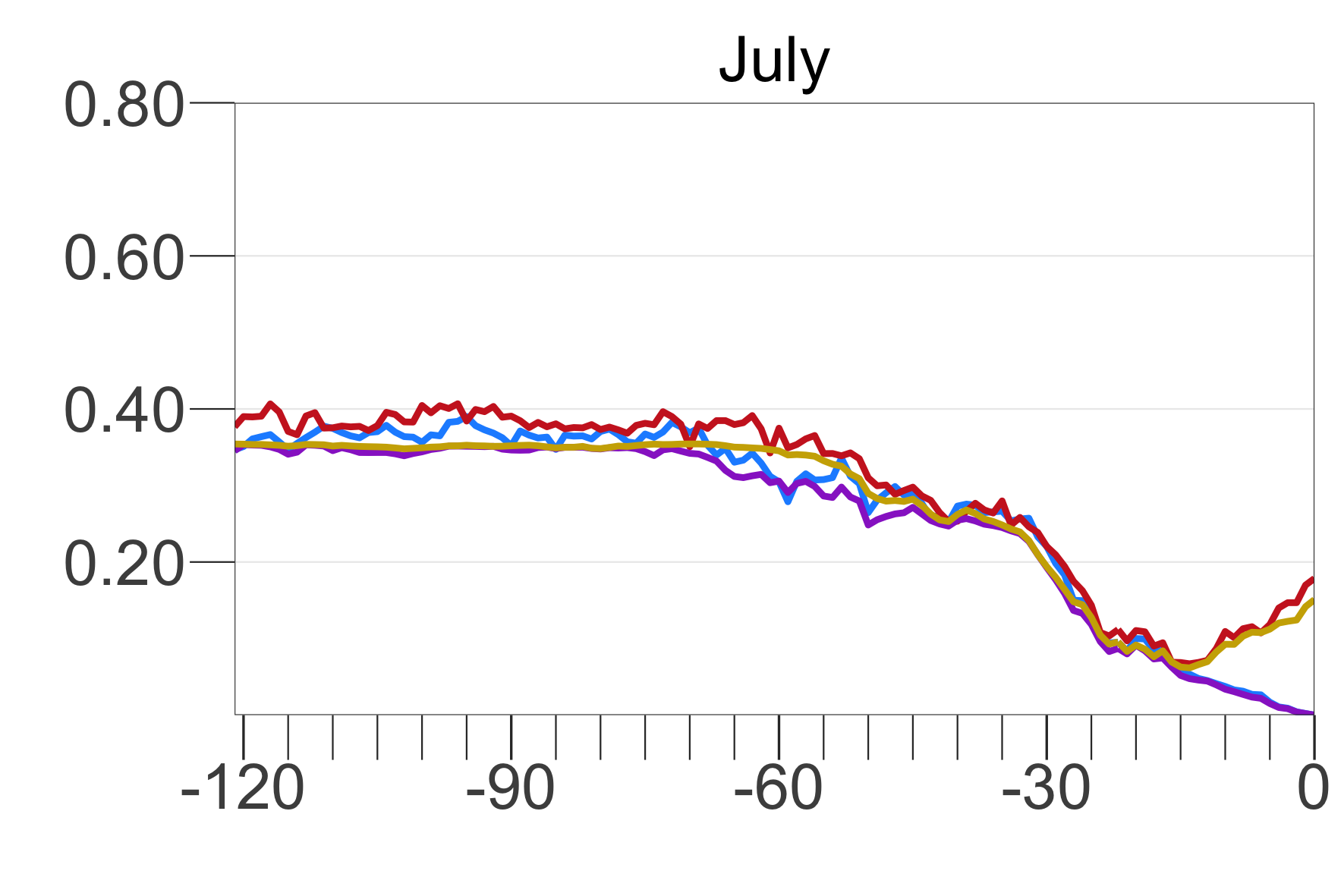}}}
		\scalebox{1}[1.25]{\includegraphics[trim={0mm 10mm 0mm 10mm},clip,width=.325\textwidth]{{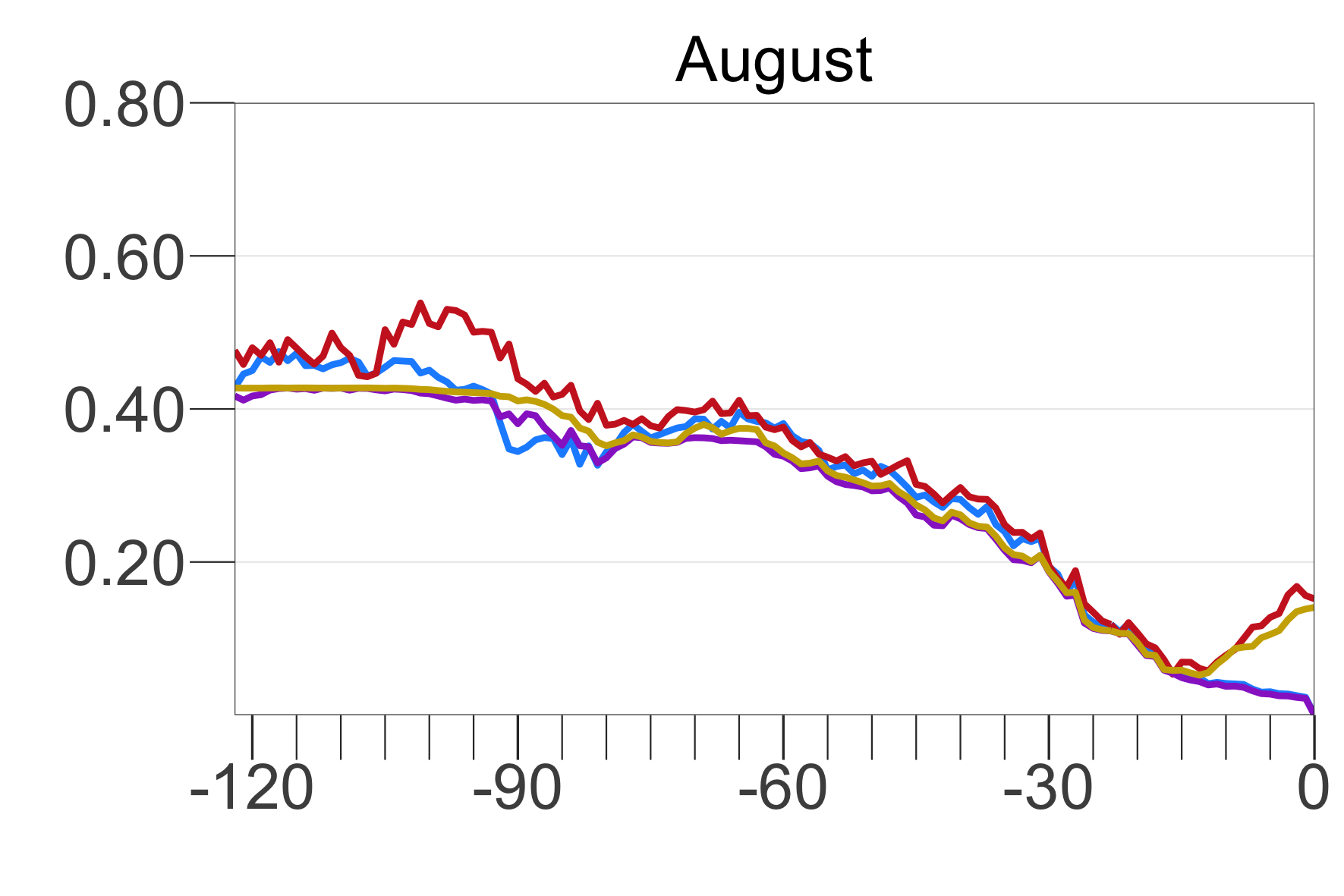}}}
		\scalebox{1}[1.25]{\includegraphics[trim={0mm 10mm 0mm 10mm},clip,width=.325\textwidth]{{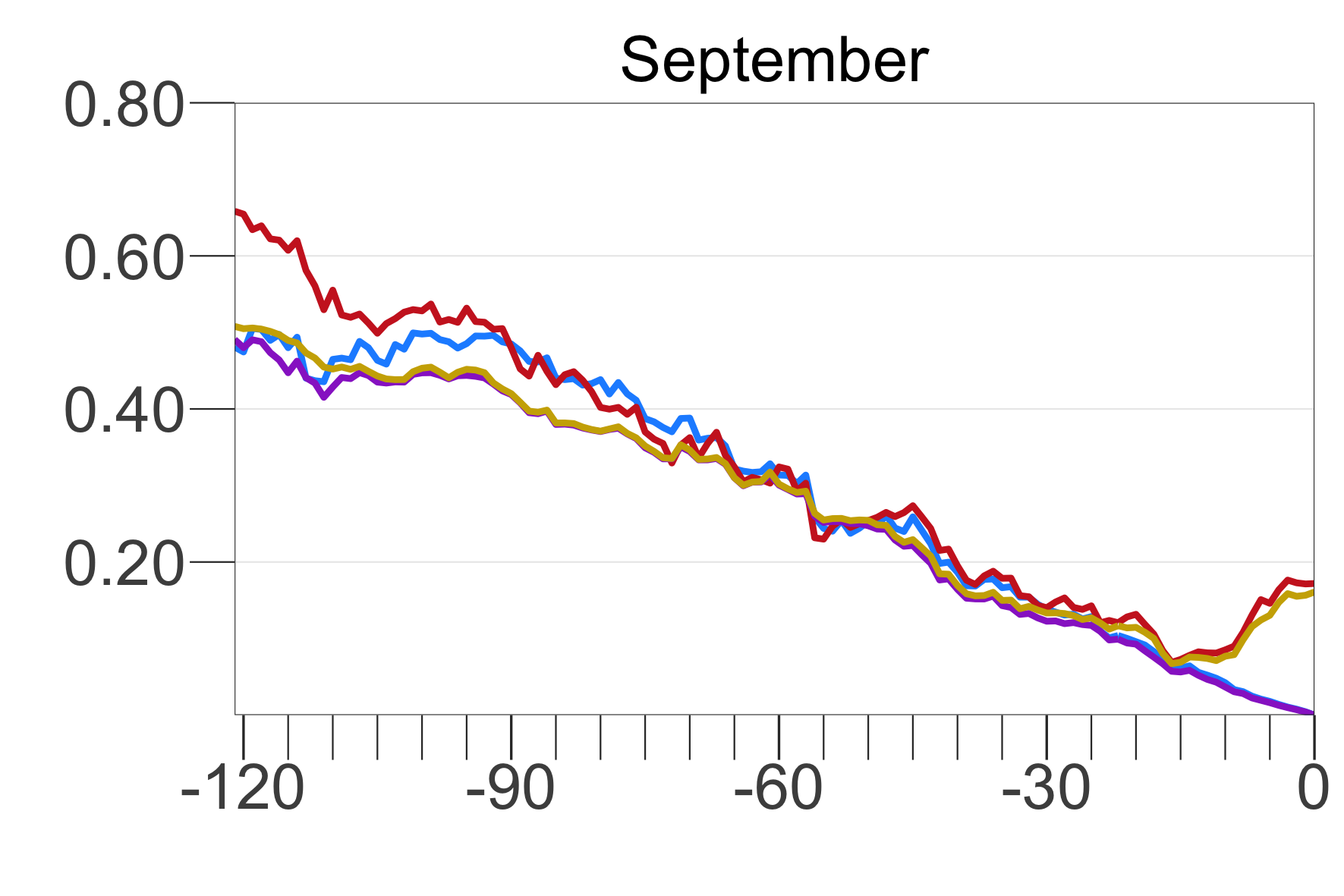}}}
		\scalebox{1}[1.25]{\includegraphics[trim={0mm 10mm 0mm 10mm},clip,width=.325\textwidth]{{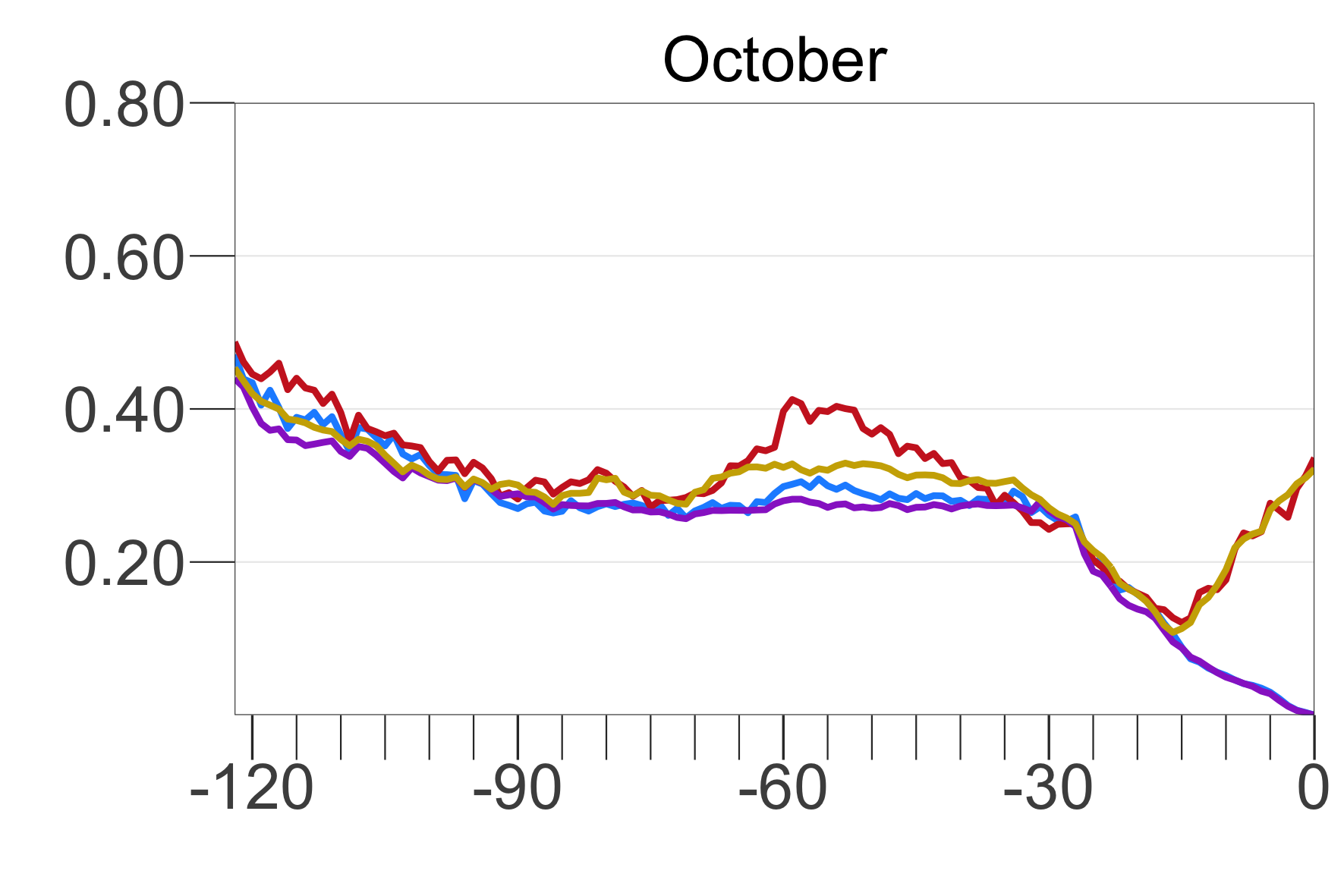}}}
		\scalebox{1}[1.25]{\includegraphics[trim={0mm 10mm 0mm 10mm},clip,width=.325\textwidth]{{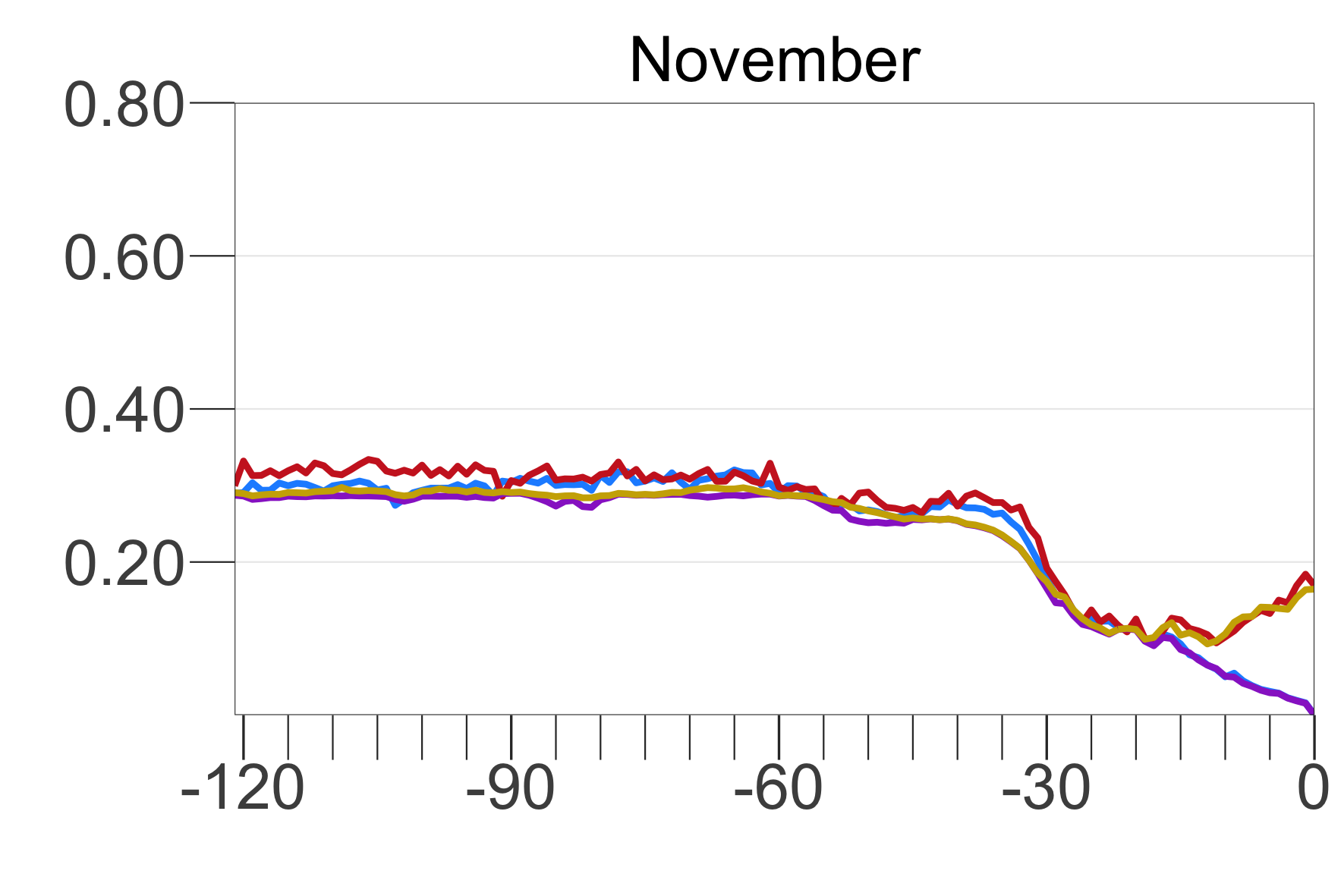}}}
		\scalebox{1}[1.25]{\includegraphics[trim={0mm 10mm 0mm 10mm},clip,width=.325\textwidth]{{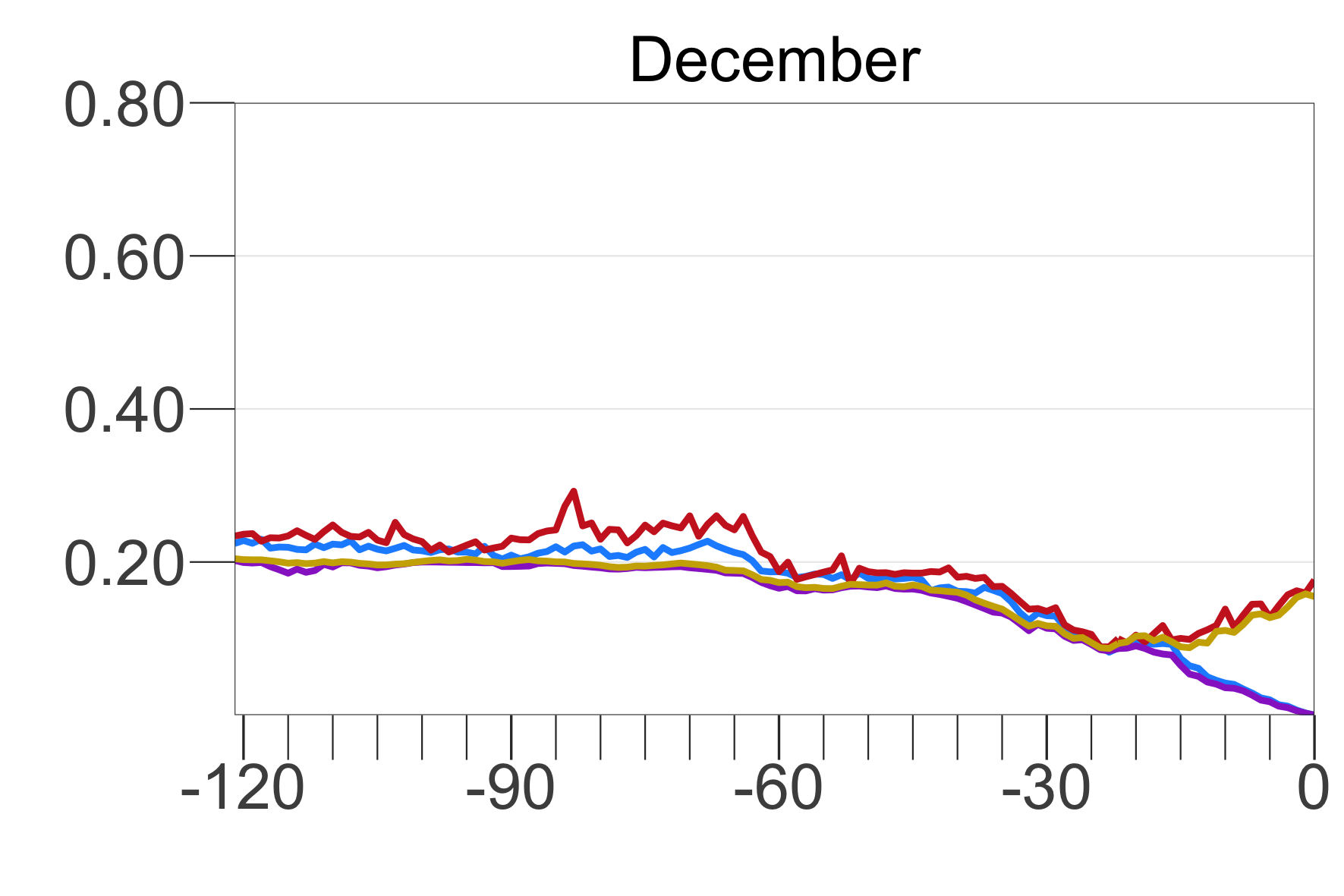}}}
		\includegraphics[trim={0mm 0mm 0mm 0mm},clip,width=0.5\textwidth]{{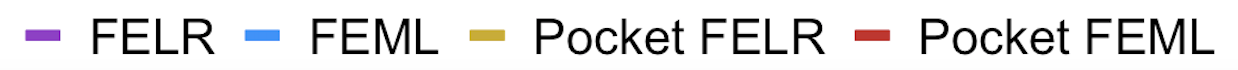}}
	\end{center}
	\label{glideRMSE_FELR_FEML}
	\begin{spacing}{1.0}  \noindent \footnotesize Notes: We show \textbf{in-sample} RMSFE glide charts for several FELR and FEML models. The estimation period is 1979-2020.  		The horizontal axes show  the number of days until the end of the target month.   In some instances $SIE_{LastMonth} = SIE_{Last30Days}$, so some models would suffer from perfect multicollinearity. In such cases, we drop $SIE_{Last30Days}$. See text for details.  
	\end{spacing}
\end{figure}

In Figure \ref{glideRMSE_FELR_FEML} we show day-by-day in-sample RMSFEs of selected FELR and FEML models for each month.  Calculation of  RMSFE requires  training set residuals.  While those are perfectly fine to use for linear regressions,  they are not for random forest-based models.  It is customary that successful (as per test set performance) RFs vastly overfit the training data,  reducing residuals to dust \citep{TBTP} -- i.e., a form of ``benign overfitting".  Consequently, it is typical to rely on the so-called out-of-bag error to internally evaluate the goodness-of-fit from such models \citep{breiman2001random}. We do so using block subsampling as in \cite{MRF} which is more adequate in the context of time series data.  \textcolor{black}{Here, we set the block size to two years.}

Mechanically, we observe that among the FELRs, the model with the smallest degrees of freedom (FELR, with 5 parameters) is the lower envelope.  It is noted that the differences between FELR and Pocket FELR are often small, except for the longer horizons of spring and summer months.  This indicates that the cost of forgoing certain regressors in Pocket FEML's linear part may not be too \textcolor{black}{wise a choice} in the forthcoming out-of-sample evaluation.  In this in-sample evaluation,  FEML RMSFEs are higher than those of FELRs in almost every instance. However, it is worth remembering that RMSFEs are computed differently (by necessity) and that FEMLs' calculations account for degrees of freedom while FELRs' do not.  To provide an apples-to-apples comparison of the competing FE approaches, we switch to a uniform recursive out-of-sample evaluation metric.

\subsubsection{A Pseudo-Out-of-Sample Analysis of the Last Decade}

\begin{figure}[tp]  
	\caption{Out-of-Sample Glide Charts: FELR and FEML  \\ Average Over 2012-2021 }
	\begin{center}
		\scalebox{1}[1.25]{\includegraphics[trim={0mm 10mm 0mm 10mm},clip,width=.325\textwidth]{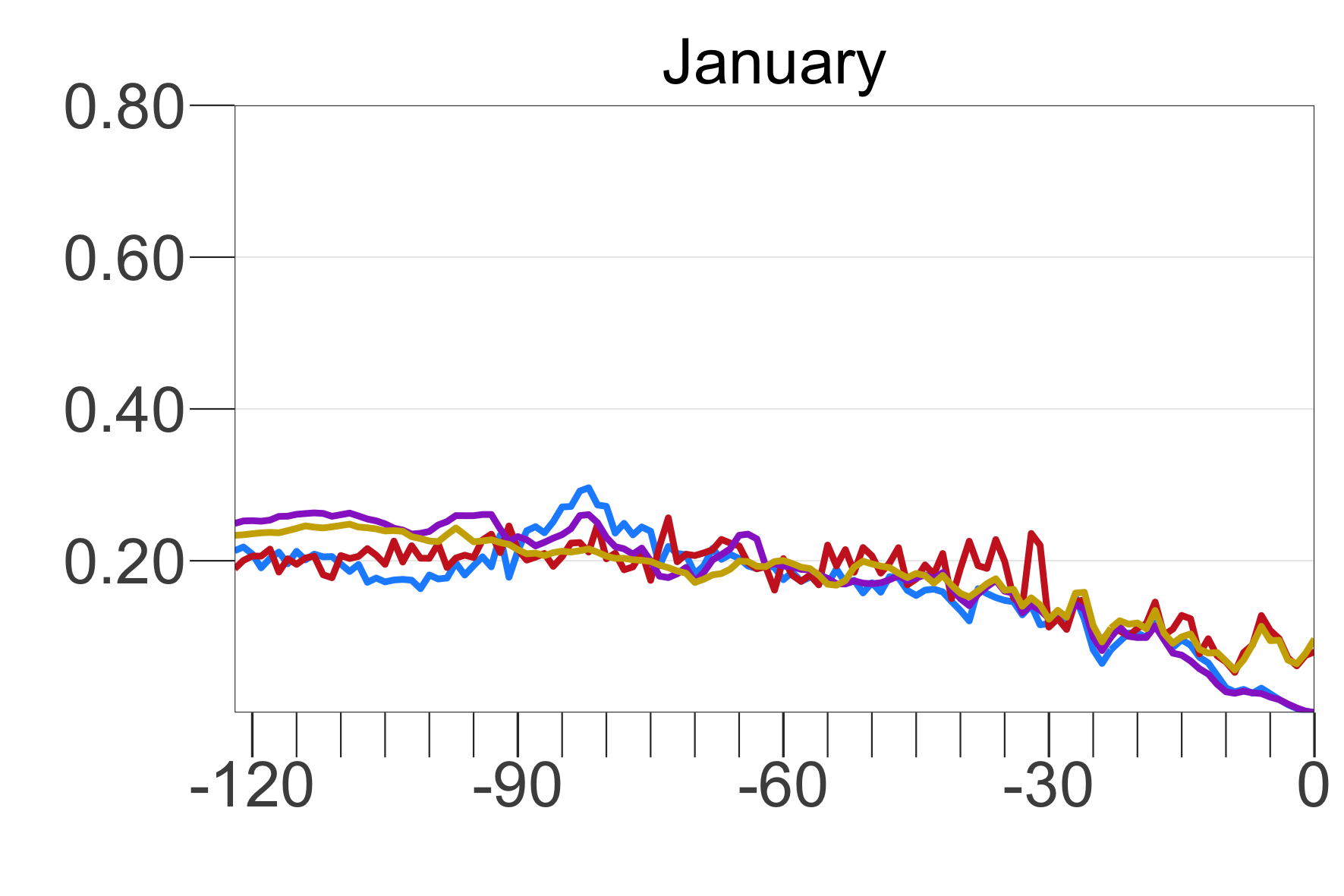}}
		\scalebox{1}[1.25]{\includegraphics[trim={0mm 10mm 0mm 10mm},clip,width=.325\textwidth]{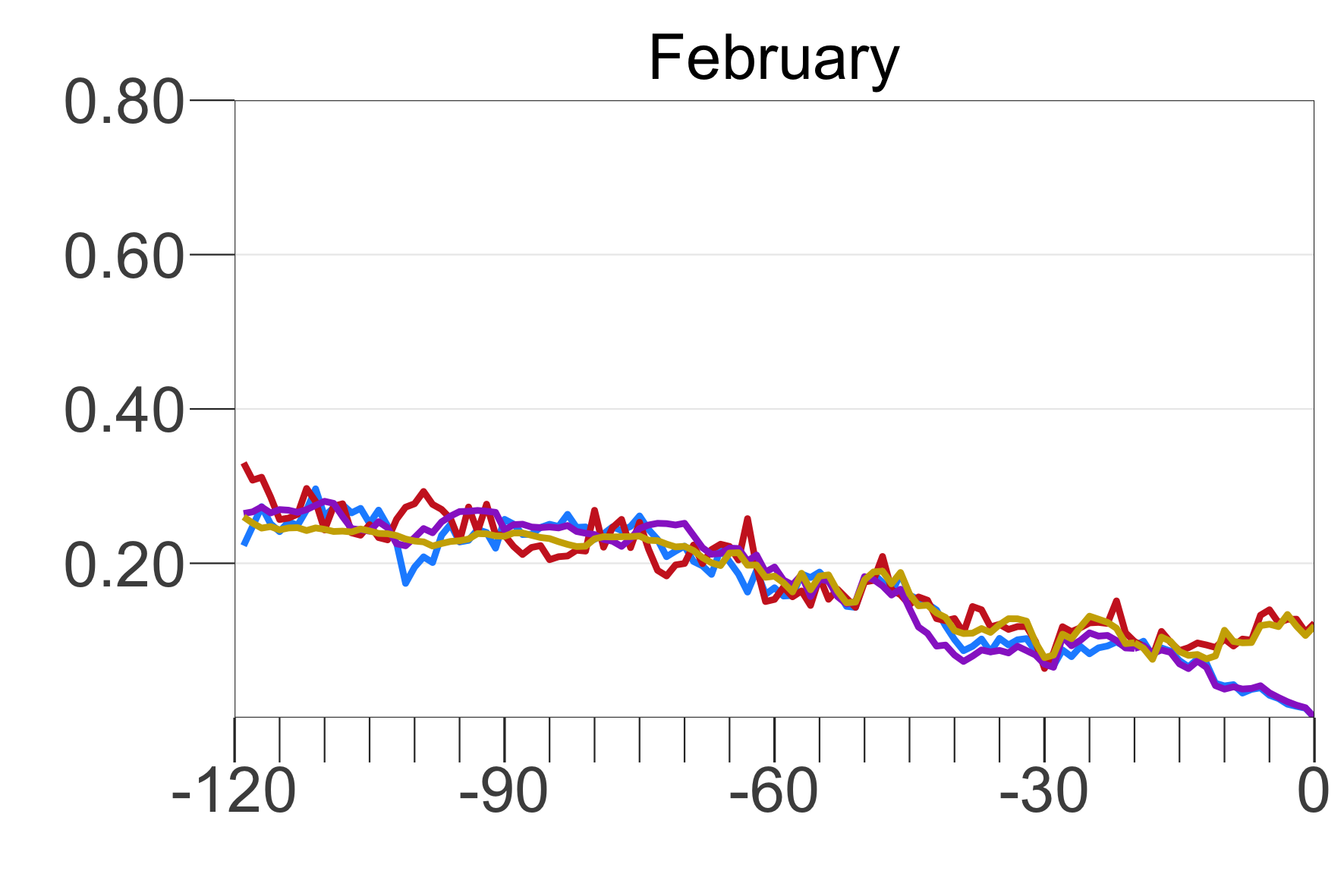}}
		\scalebox{1}[1.25]{\includegraphics[trim={0mm 10mm 0mm 10mm},clip,width=.325\textwidth]{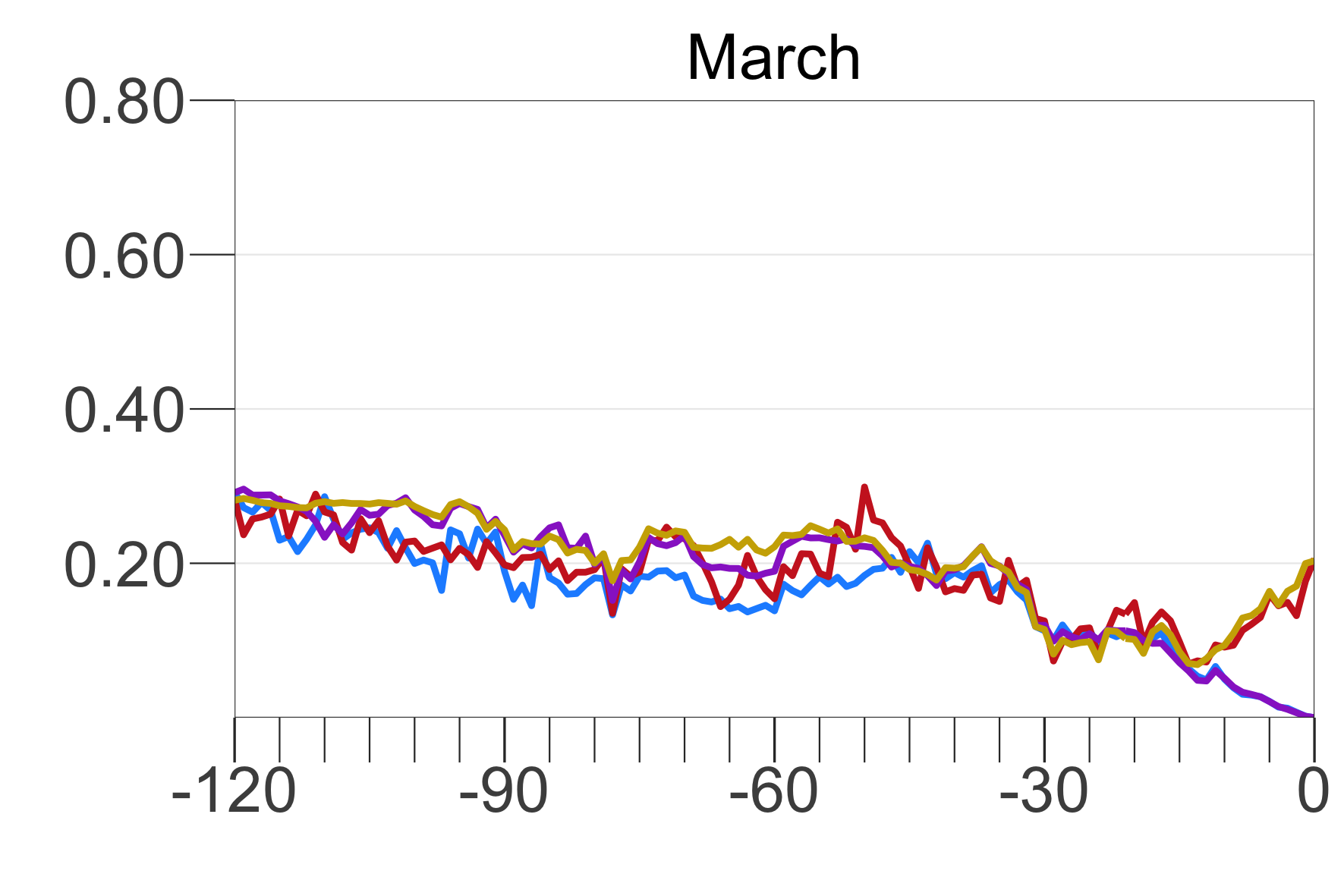}}
		\scalebox{1}[1.25]{\includegraphics[trim={0mm 10mm 0mm 10mm},clip,width=.325\textwidth]{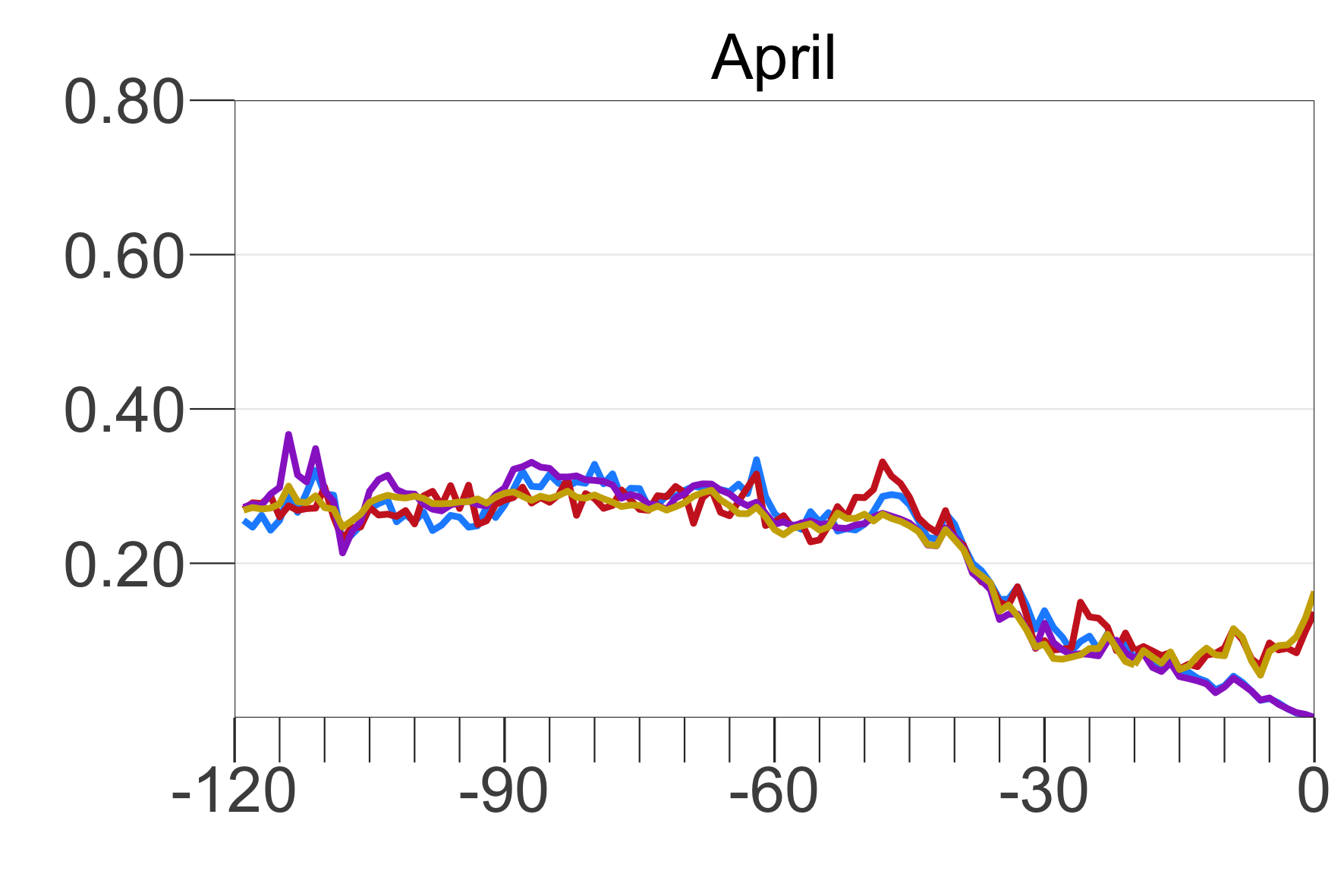}}
		\scalebox{1}[1.25]{\includegraphics[trim={0mm 10mm 0mm 10mm},clip,width=.325\textwidth]{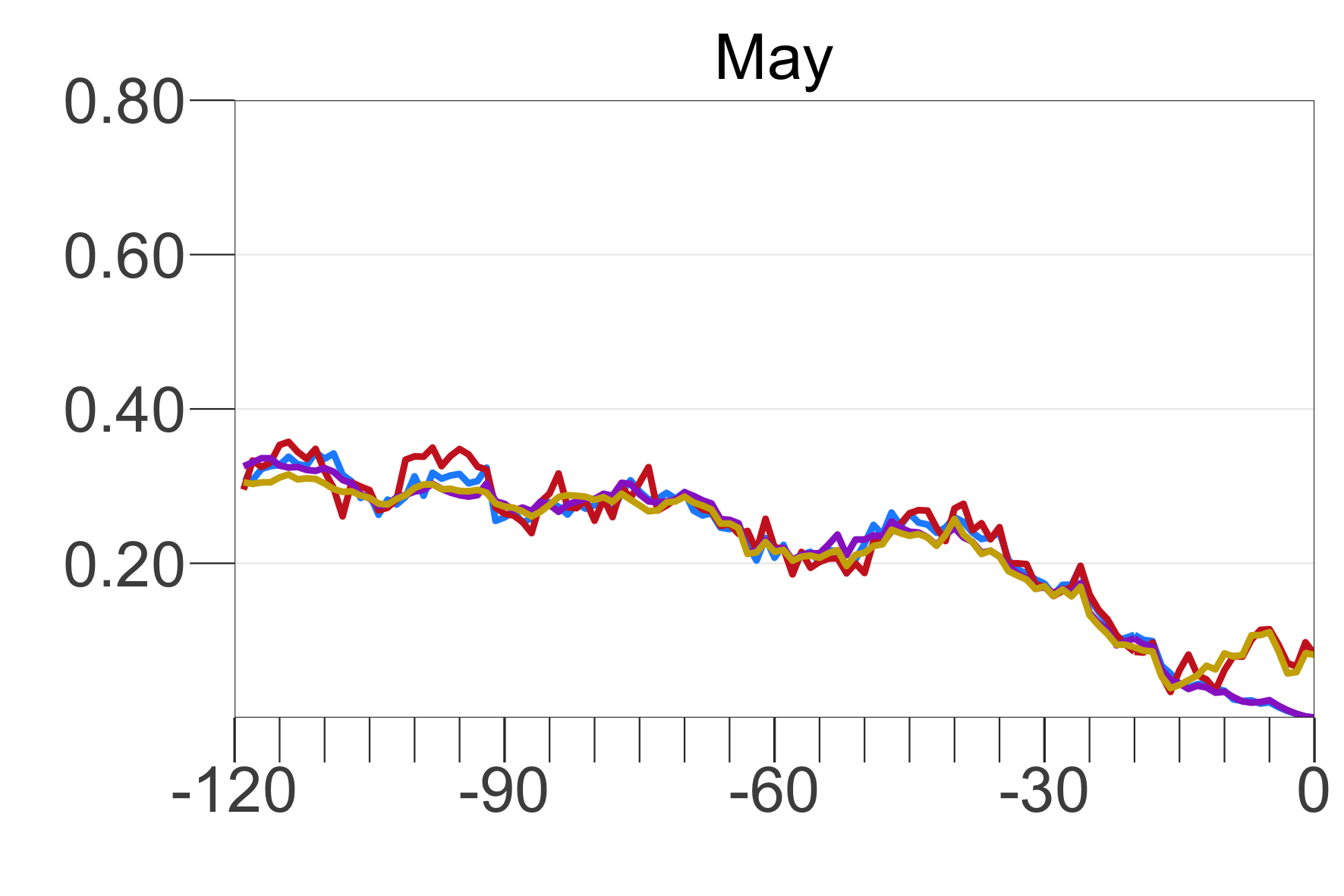}}
		\scalebox{1}[1.25]{\includegraphics[trim={0mm 10mm 0mm 10mm},clip,width=.325\textwidth]{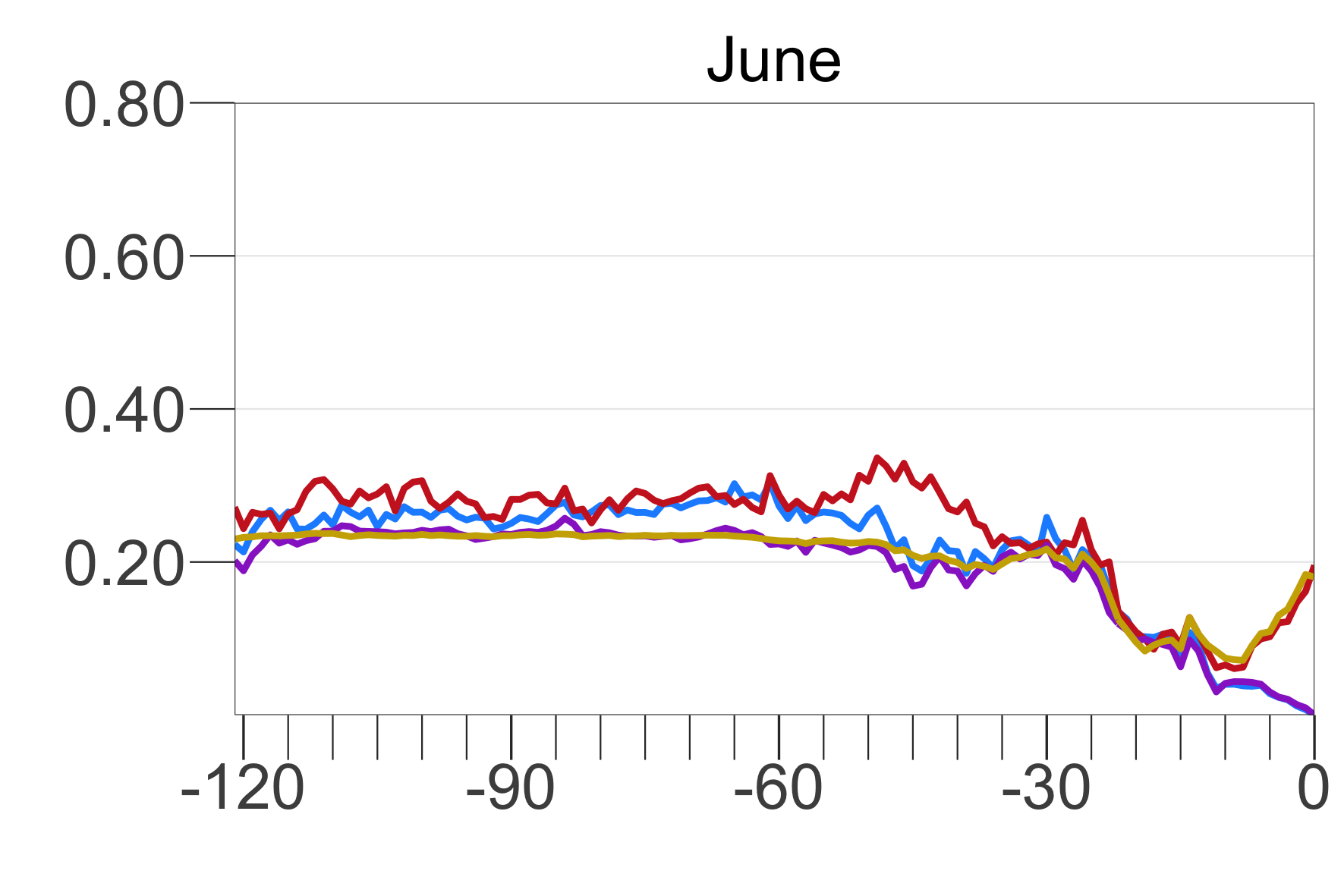}}
		\scalebox{1}[1.25]{\includegraphics[trim={0mm 10mm 0mm 10mm},clip,width=.325\textwidth]{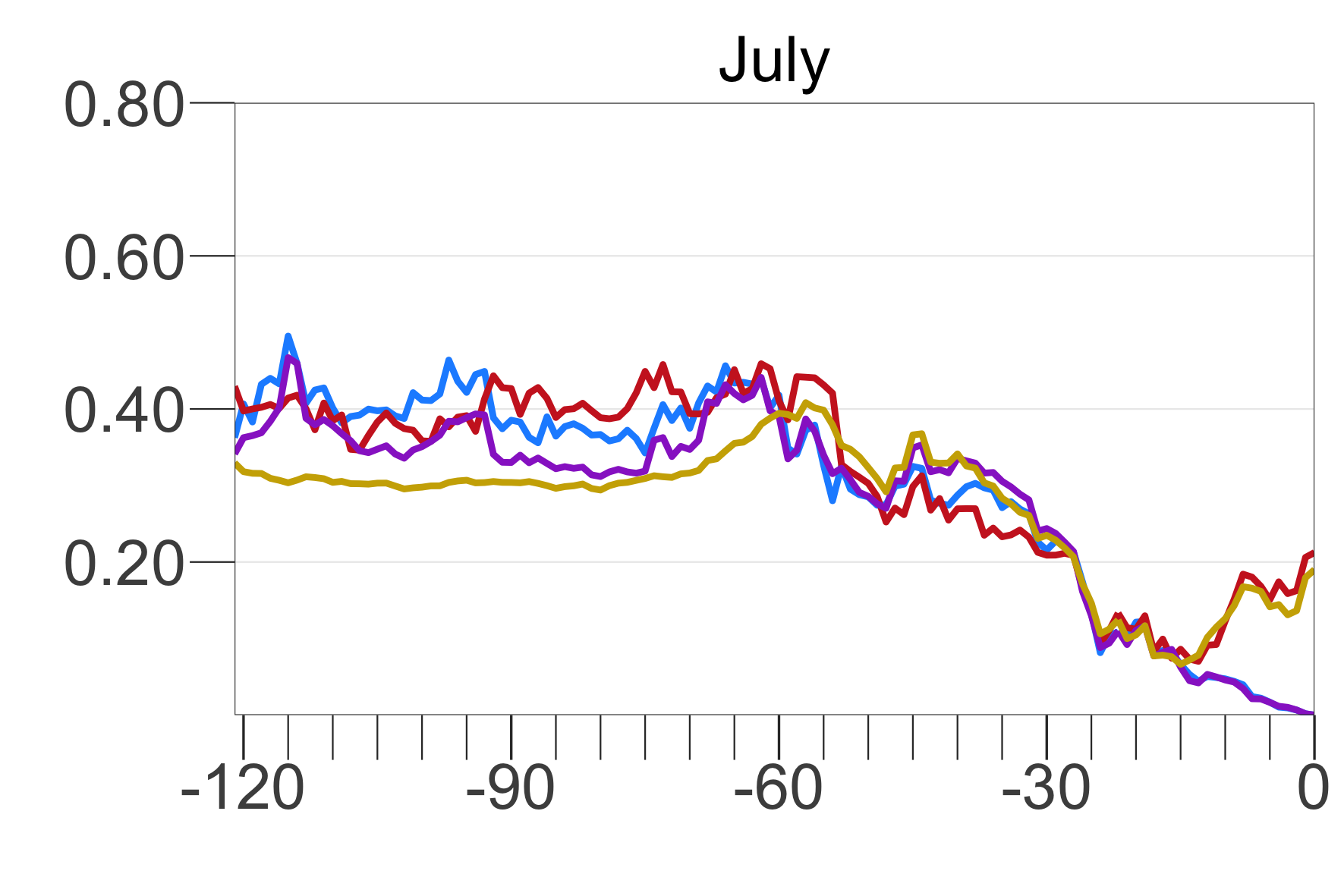}}
		\scalebox{1}[1.25]{\includegraphics[trim={0mm 10mm 0mm 10mm},clip,width=.325\textwidth]{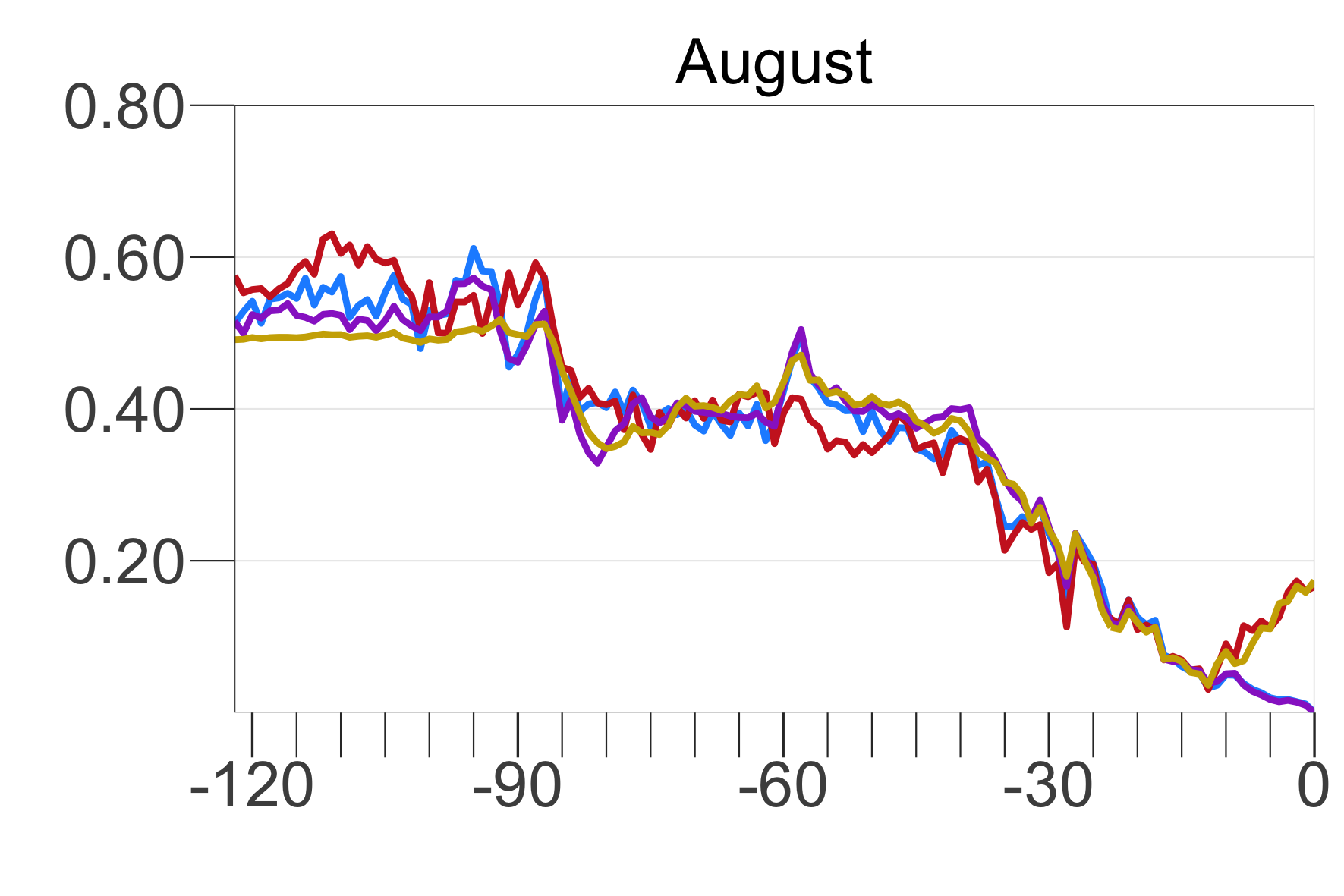}}
		\scalebox{1}[1.25]{\includegraphics[trim={0mm 5mm 0mm 0mm},clip,width=.325\textwidth]{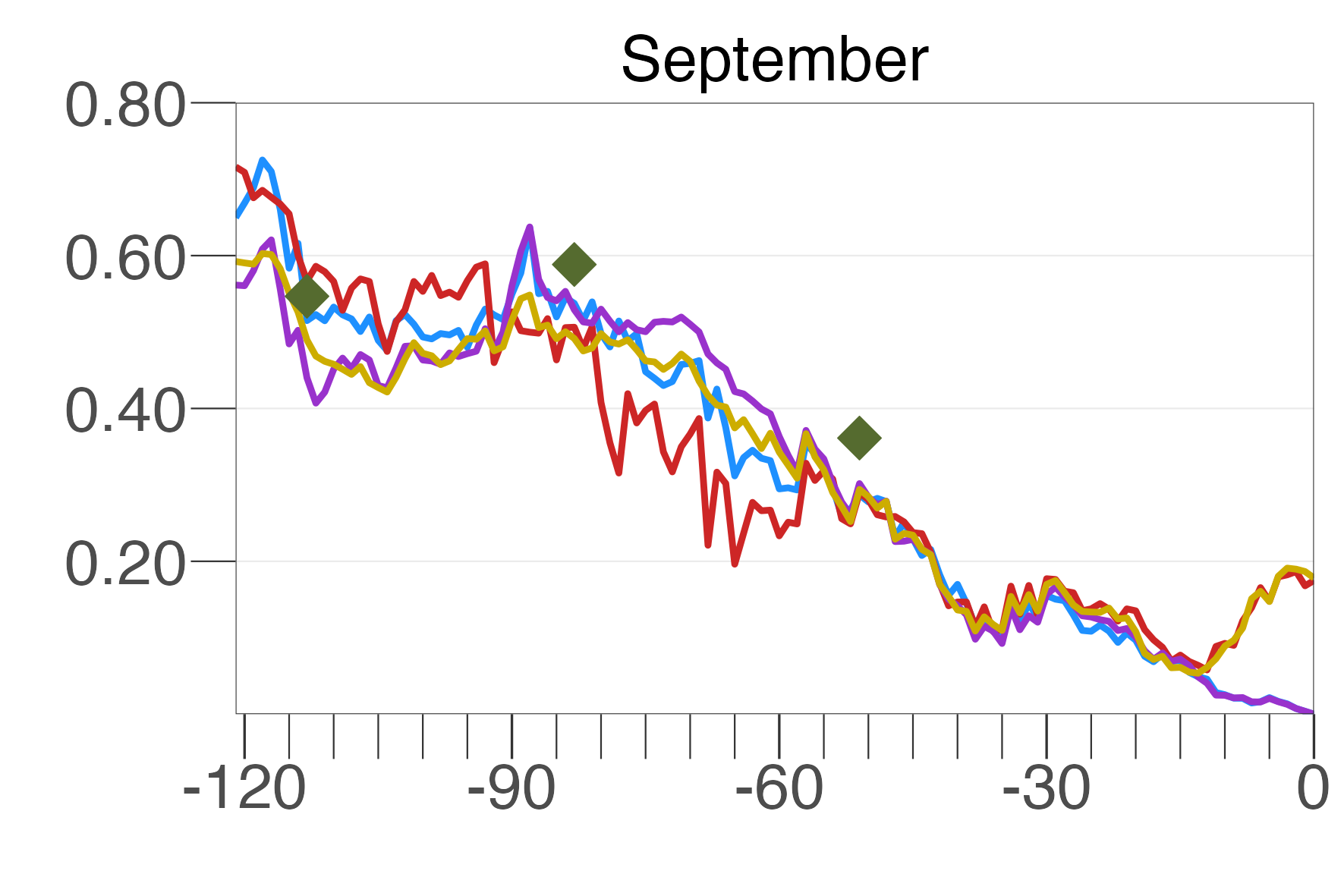}}
		\scalebox{1}[1.25]{\includegraphics[trim={0mm 10mm 0mm 10mm},clip,width=.325\textwidth]{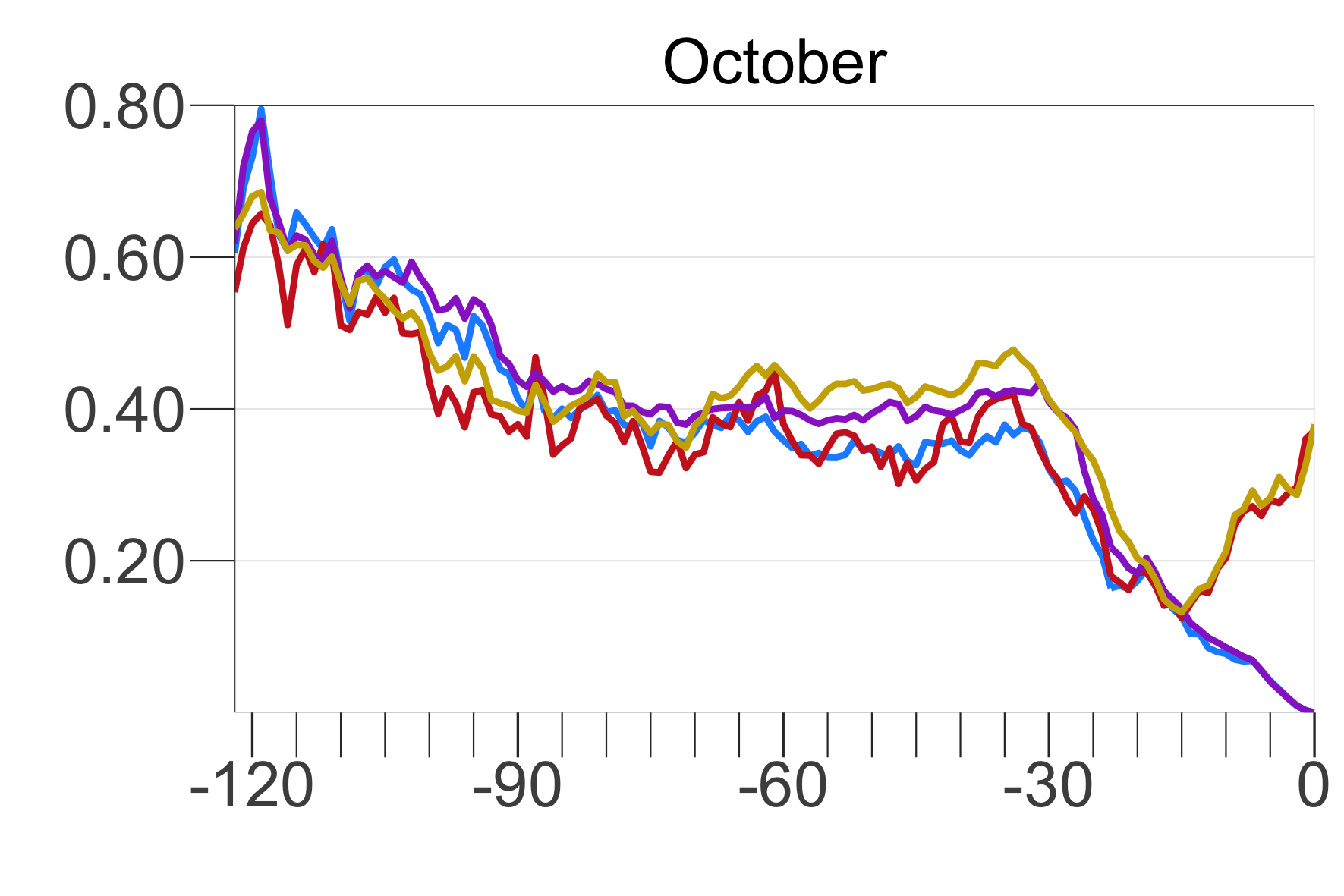}}
		\scalebox{1}[1.25]{\includegraphics[trim={0mm 10mm 0mm 10mm},clip,width=.325\textwidth]{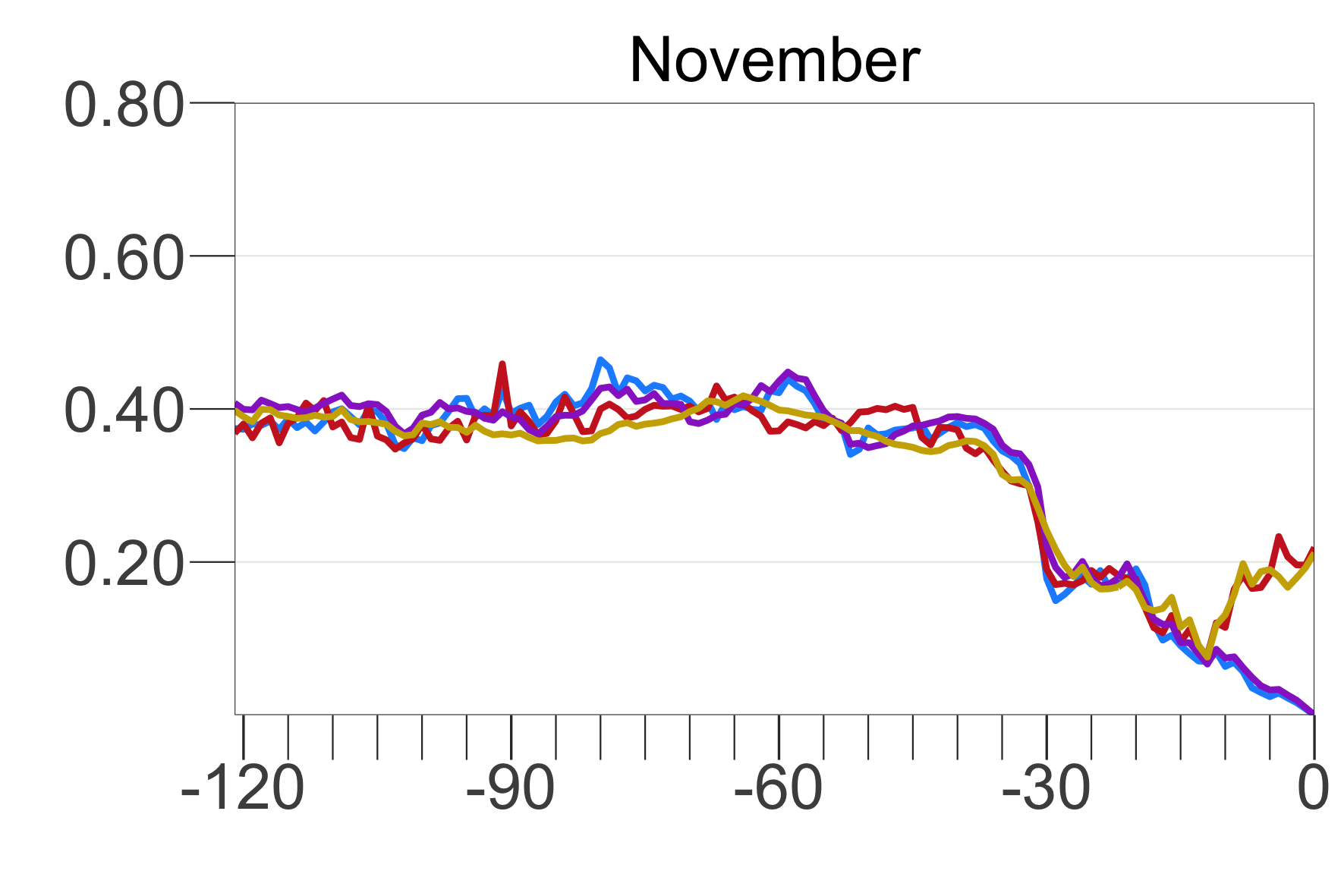}}
		\scalebox{1}[1.25]{\includegraphics[trim={0mm 10mm 0mm 10mm},clip,width=.325\textwidth]{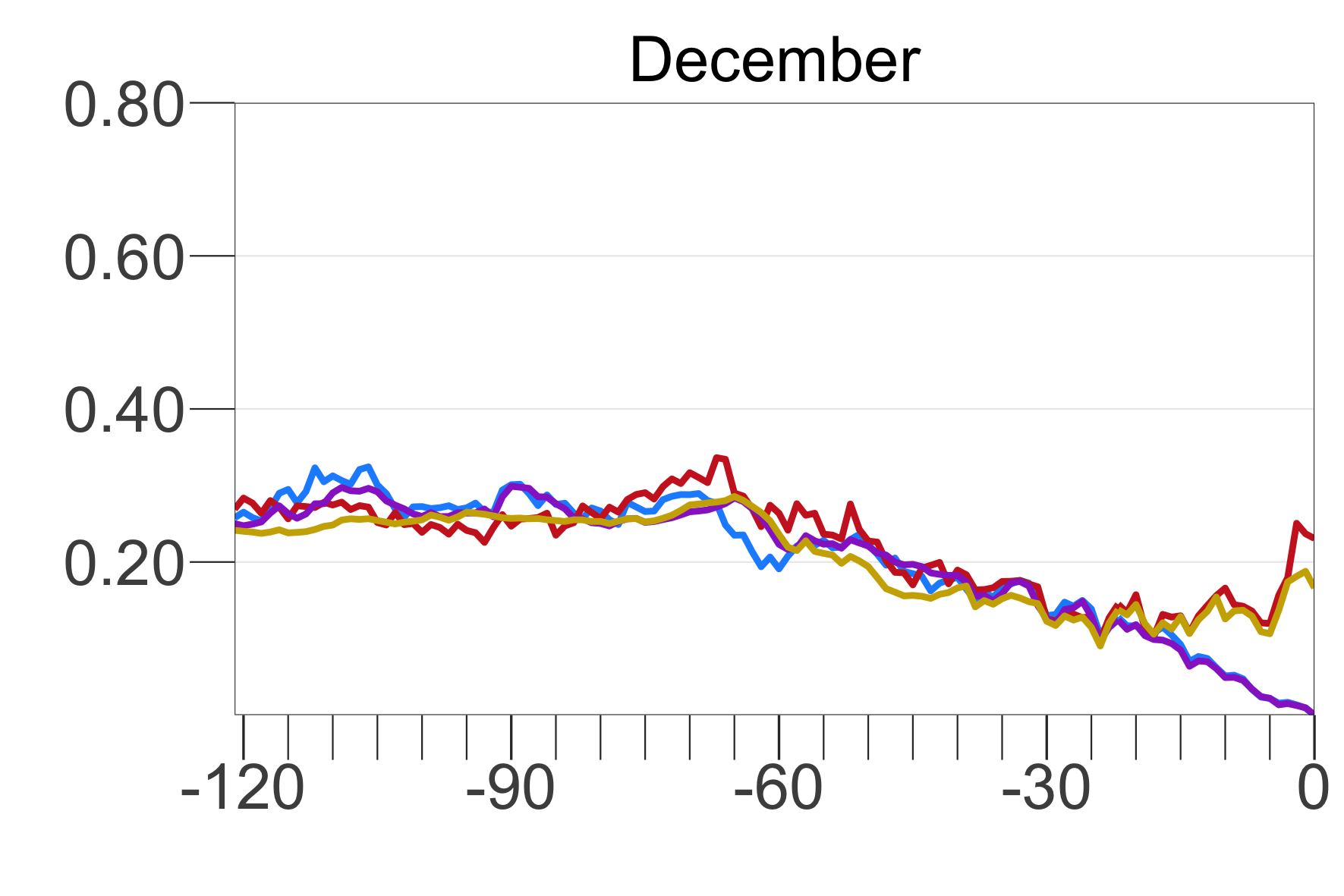}}
		\includegraphics[trim={0mm 0mm 0mm 0mm},clip,width=0.5\textwidth]{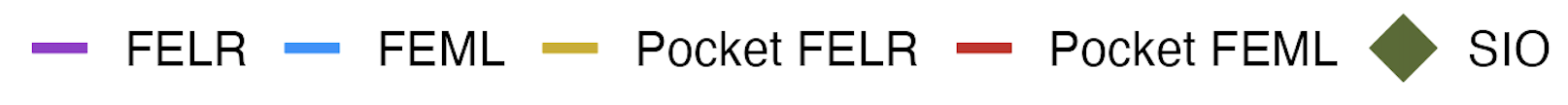}
	\end{center}
	\label{MSEOOS_FELR_FEML}
	\begin{spacing}{1.0}  \noindent \footnotesize Notes: We show \textbf{out-of-sample} RMSFE glide charts. We make forecasts each day through the end of the target month in each year from 2012-2021. We then plot the square root of the 10-year average of each day's squared forecast error.  	The horizontal axes show  the number of days until the end of the target month.  In some instances $SIE_{LastMonth} = SIE_{Last30Days}$, so some models would suffer from perfect multicollinearity. In such cases, we drop $SIE_{Last30Days}$.  See text for details. 
	\end{spacing}
\end{figure}

Glide charts need not necessarily be used in conjunction  with RMSFE based on in-sample residuals or variants of them.  In fact,  any loss can be used.   In Figure \ref{MSEOOS_FELR_FEML},  we remain within the realm of squared errors,  but those are computed from a recursive expanding-window pseudo-out-of-sample experiment.  This sort of exercise is standard in the modern macroeconomic forecasting literature comparing econometric and machine learning models \citep{GCLSS2019}.  The choice of the 2012-2021 window for the "test" set is inspired by \cite{andersson2021seasonal}.\footnote{They conduct an evaluation of SIE predictability for their convolution neural network trained directly on satellite imagery data.  The benchmarks they consider are a climate model and a linear time-trend model.} Given data limitations,  it is a fair balance between avoiding training models on too small of a sample size and calculating RMSFE based on too few out-of-sample errors.  Models are re-estimated every year to leverage the gradually incoming new data points.  

{\color{black}  There are benefits and costs to this alternative evaluation setup.  The obvious cost is that the test set RMSFE is the average of 10 errors rather than 40 (as considered in the previous section for the in-sample analysis), which inevitably increases the variance of the evaluation metric. } The benefits are threefold.  First,  it is not unthinkable that the magnitude of the last 10 years' forecasting errors is more informative about the near future than that of those in the 1980s and 1990s.  Second,  semi-flexible trend models (like FELR) will have a built-in advantage for in-sample evaluation over what  prevails when one uses such models to \textit{really} forecast next year's SIE.  The reason for this is that in-sample evaluation uses a residual at time $t$ from a model that is trained on both $t-1$ and $t+1$.  Information from $t+1$ is extremely useful \textit{when estimating the parameters} of a time trend,  but such information about $t+1$ is not available when one is truly forecasting $t$ from $t-1$.  The recursive evaluation addresses this potential bias by mimicking directly the reality of forecasting every year using a model estimated only on available data at that particular point in time.  Lastly,  an advantage of recursive estimation in our setting is that OLS-based and RF-based models are now evaluated using an identical metric and differences between performances cannot be attributed to various choices on how to account for degrees of freedom (like setting up the out-of-bag metric).

\textcolor{black}{In this out-of-sample evaluation, a couple of observations are worth mentioning}. First,  FELR and its pocket counterpart stand out as solid benchmarks,  by routinely yielding the smallest RMSFE,  which is especially clear for longer horizons of early summer months.  Given the small estimation sample limitations, it is not entirely surprising that FEML's reductions in RMSFE are limited in size.  There are, however,  notable and important exceptions.  The first is Pocket FEML's performance from 90 to 45 days ahead for September.  
\textcolor{black}{Needless to say, if there are any $SIE$ forecasts of superior interest, it is exactly those lead-times prior to the end of September (hence period of the annual SIO forecasting competition).}
Improvements of this particular FEML over FELRs are over 0.1 $\times$ 10$^6$ km$^2$  for the whole period.  For October,  the dominance of nonlinear models, albeit  quantitatively smaller, is present for almost all horizons up to 15 days ahead.  Finally,  FEMLs (with FEML leading among them) also outperform FELRs for the vast majority of horizons for March -- sometimes offering reduction up to 50\% in RMSFE.

\begin{figure}[tp]  
	\caption{Fraction of Days in which any FEML outperforms any other FELR or FEML \\ Years 2012-2021}
	\begin{center}
			\includegraphics[trim={0mm 0mm 0mm 0mm},clip,width=0.45\textwidth]{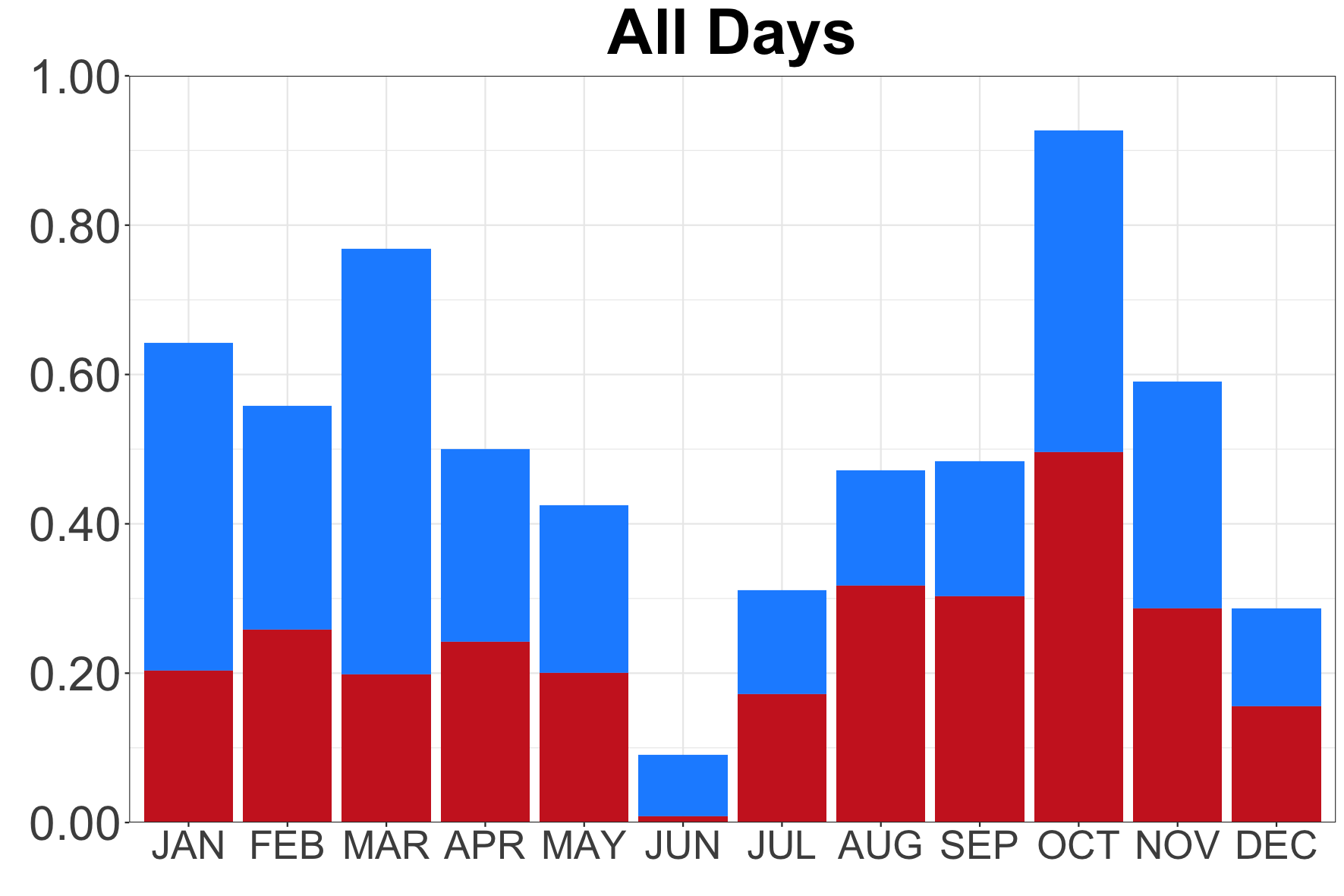}
		\includegraphics[trim={0mm 0mm 0mm 0mm},clip,width=0.45\textwidth]{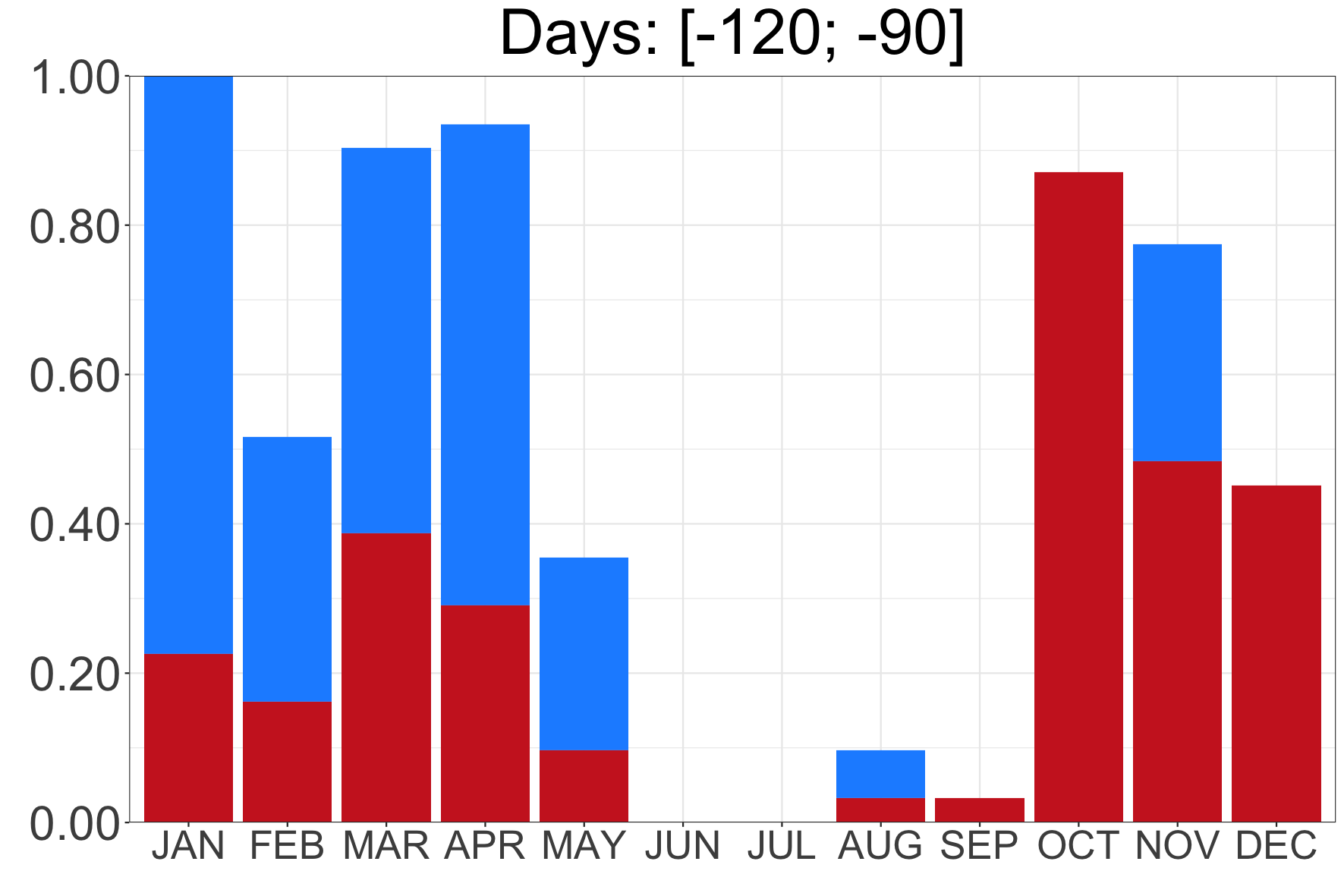}
				\vskip 5pt
		\includegraphics[trim={0mm 0mm 0mm 0mm},clip,width=0.45\textwidth]{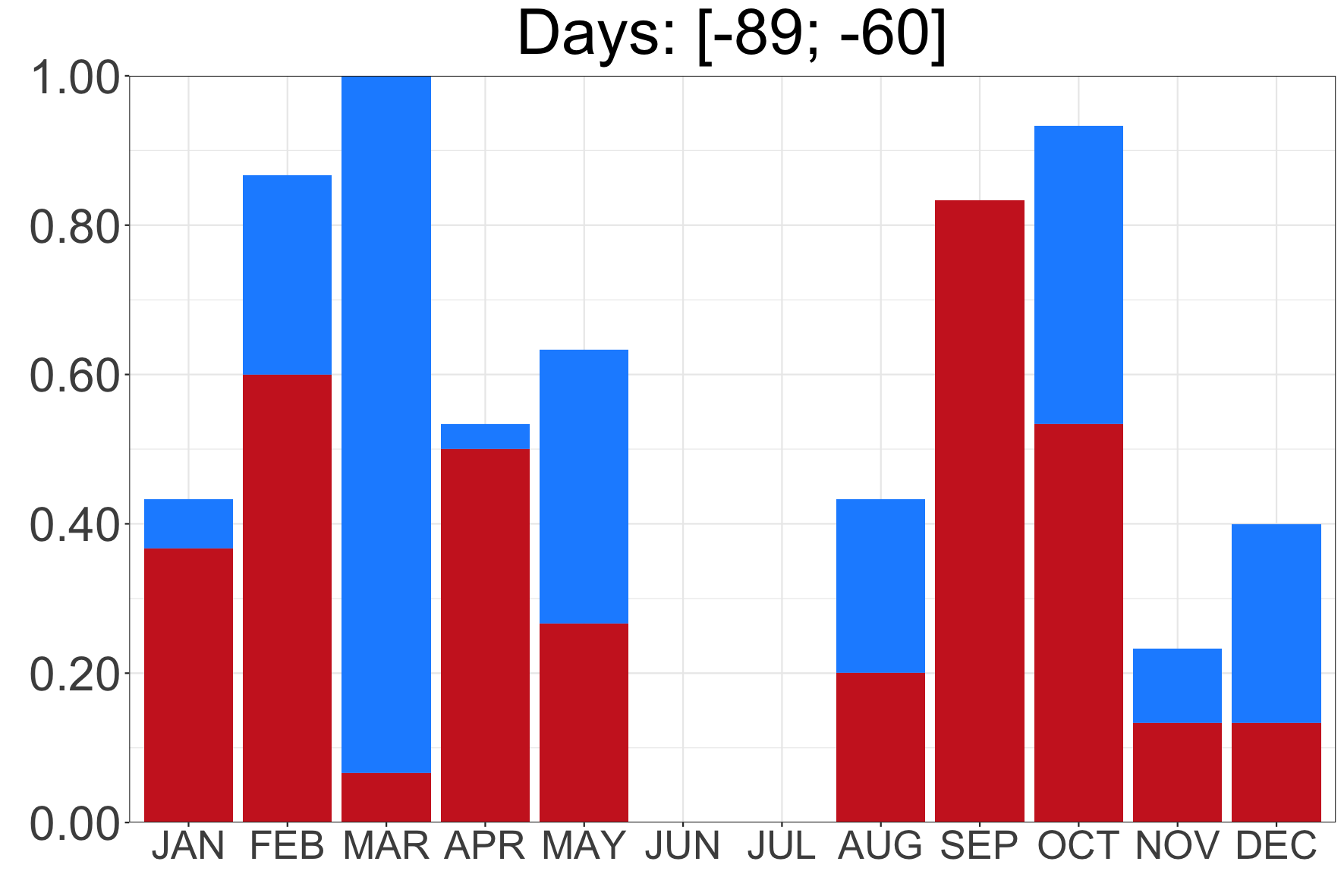}
		\includegraphics[trim={0mm 0mm 0mm 0mm},clip,width=0.45\textwidth]{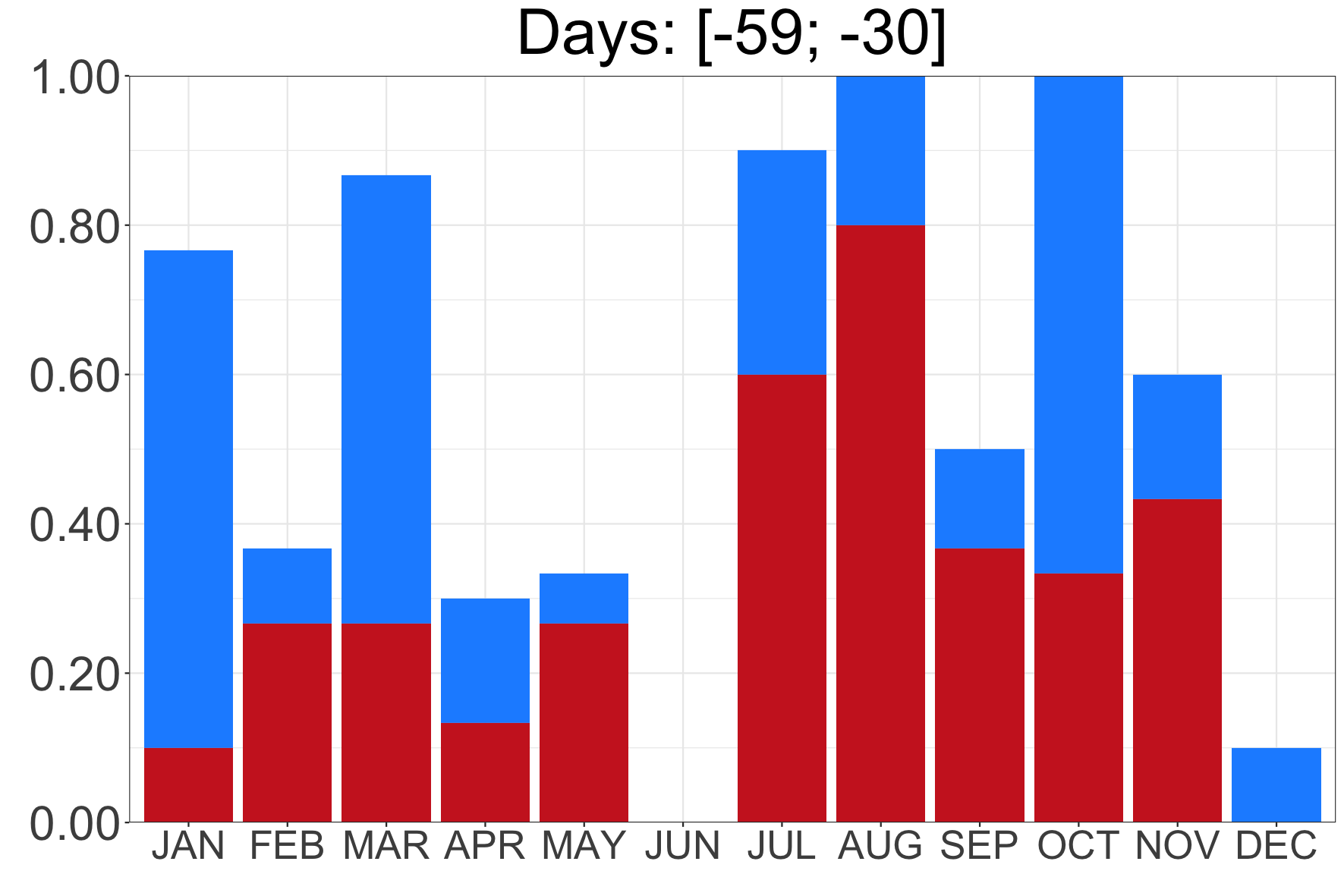}
		\vskip 5pt
		\includegraphics[trim={0mm 0mm 0mm 0mm},clip,width=0.3\textwidth]{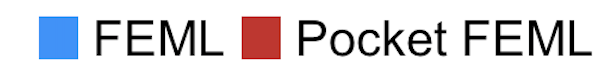}
\end{center}
	\label{fig:FractionBestDays_FEML_byMonth_All_3}
	\begin{spacing}{1.0}  \noindent  \footnotesize  Notes: We show the fraction of days for which each FEML model showed the lowest root-mean-squared forecast errors of out-of-sample predictions between 2012-2021. That is, in each year from 2012-2021 we calculate the squared point-forecast error made on each day through the end of the target month. We calculate the square-root of the 10-year average squared forecast error for each day (see Figure \ref{MSEOOS_FELR_FEML}). We then plot the fraction of days, for which each FEML model achieved the lowest 10-year RMSFE. The set of all models includes ``FELR'',  ``Pocket FELR'', ``FEML'', and ``Pocket FEML''.
		The horizontal axes show  the number of days until the end of the target month $m$.  See text for details. Notice that in some instances $SIE_{LastMonth} = SIE_{Last30Days}$ and some models would suffer from perfect multicollinearity. In these cases, $SIE_{Last30Days}$ is dropped. This causes the red and green line to sometimes coincide.  
	\end{spacing}
\end{figure}

In Figure \ref{fig:FractionBestDays_FEML_byMonth_All_3},  we report the fraction of days for which any FEML in Figure \ref{MSEOOS_FELR_FEML} offers the lowest RMSFEs.  This helps \textcolor{black}{summarizing and synthesizing} the abundant information in glide charts.  It is clear that October, March, and to a lesser extent, January, are all months where gains (albeit small for certain horizons) are generalized over the whole 120 days.  Their respective shares of optimal forecasts are above 90\%, 80\%, and  70\% respectively.  In contrast,  September reductions in RMSFE are substantial in size but are localized within a specific forecasting range.  Accordingly,  the fraction of days in which any FEML outperforms FELR for September is around 50\%.  Many months exhibit fractions in such a range, but unlike September, they are typically due to FEML and FELR forecasts being roughly similar.  Overall, \textcolor{black}{the} early summer months June and July are better predicted using FELRs, and this is mostly attributable to long-range forecasts made in the first months of the melting phase.   As can be seen in the bottom quadrants of Figure \ref{fig:FractionBestDays_FEML_byMonth_All_3},  the opposite can be said for mid-range horizons where FEMLs are very frequently the best option for many months (excluding June, July and August).

The fact that FEMLs offer clear gains for September, October, \textit{as well as March} forecasts,  and much less so for other months suggests that nonlinearities (and \textcolor{black}{an} expanded data set) are particularly beneficial for \textcolor{black}{detecting(?)} turning points in the annual SIE cycle,  that is, when SIE stops expanding or stops retracting.  A working hypothesis is that nonlinearities and additional information helps in avoiding either too low or too large $SIE$ predictions around the trough based on slowly evolving physical limits of the seasonal component.  In the case of September and October,  this could be \textcolor{black}{due to FEML's} moderate downward pressures on the prediction from very low  readings of $SIE_{Today}$ and $SIE_{LastMonth}$ during early Arctic summer to account for the fact that as ice melts,  perhaps more than previous summers,  the weighting of multi-year thicker and older ice increases,  ultimately slowing the melting process in late summer  \citep{maslanik2007younger}.   Another potential source of nonlinearity,  now in favor of accelerated melting beyond what a linear dynamic relationship suggests, is the presence of feedback loops, like the ice-albedo effect, which can manifest even within short time spans \citep{VARCTIC}. 

\begin{figure}[h!]  
	\caption{Annual Out-Of-Sample Forecasts on Different Days} \label{hist_oos}
	\vspace*{-0.5cm}
	\begin{center}
		
		\begin{subfigure}[t]{0.9\textwidth}
		\centering
			\includegraphics[trim={0mm 50mm 0mm 40mm},clip,width=\textwidth]{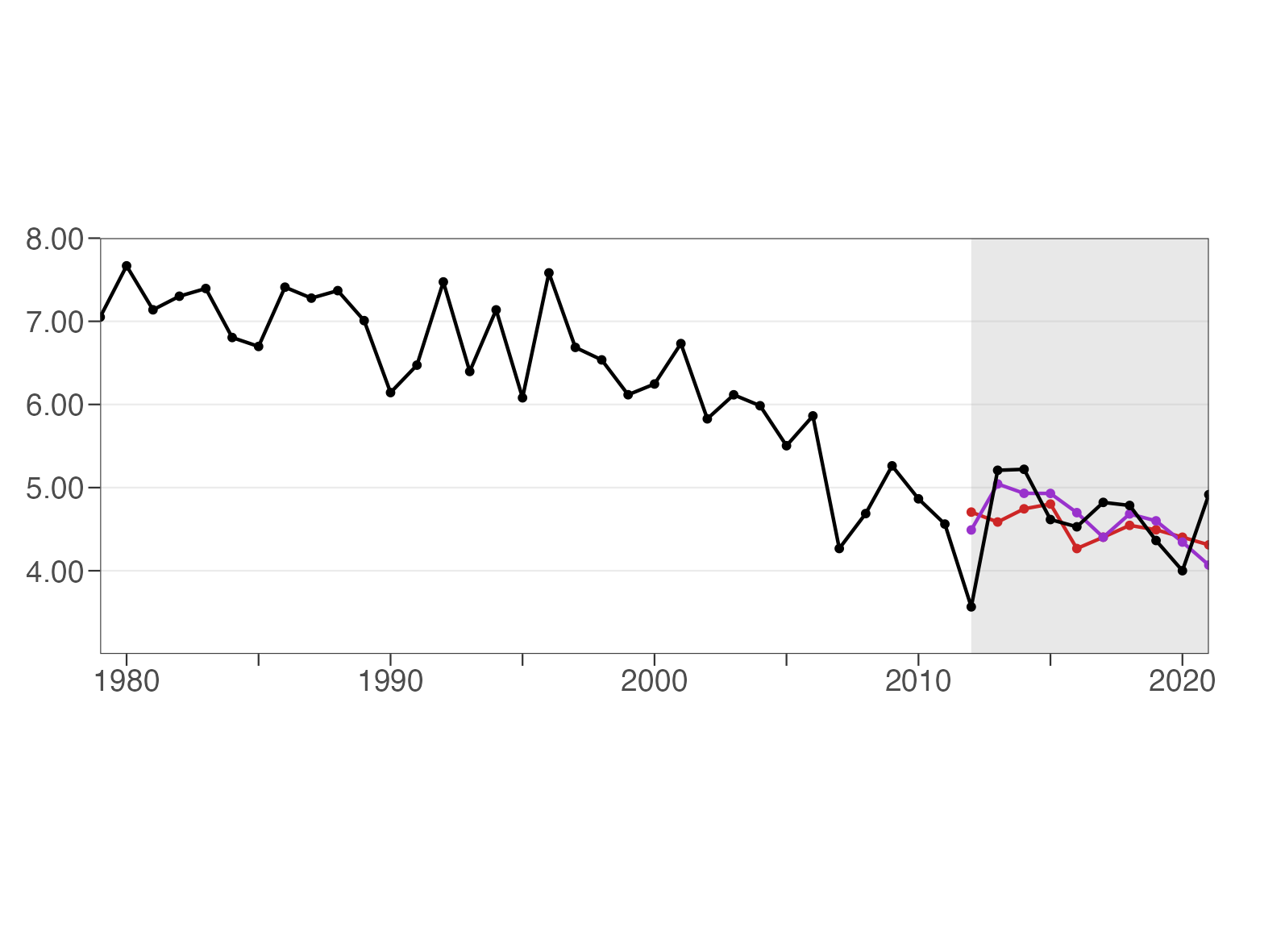}
			\caption{Forecast-Day: June 14$^\text{th}$}
		\end{subfigure}
		\vskip 10pt
		\begin{subfigure}[t]{0.45\textwidth}
		\centering
			\includegraphics[trim={0mm 20mm 0mm 20mm},clip,width=\textwidth]{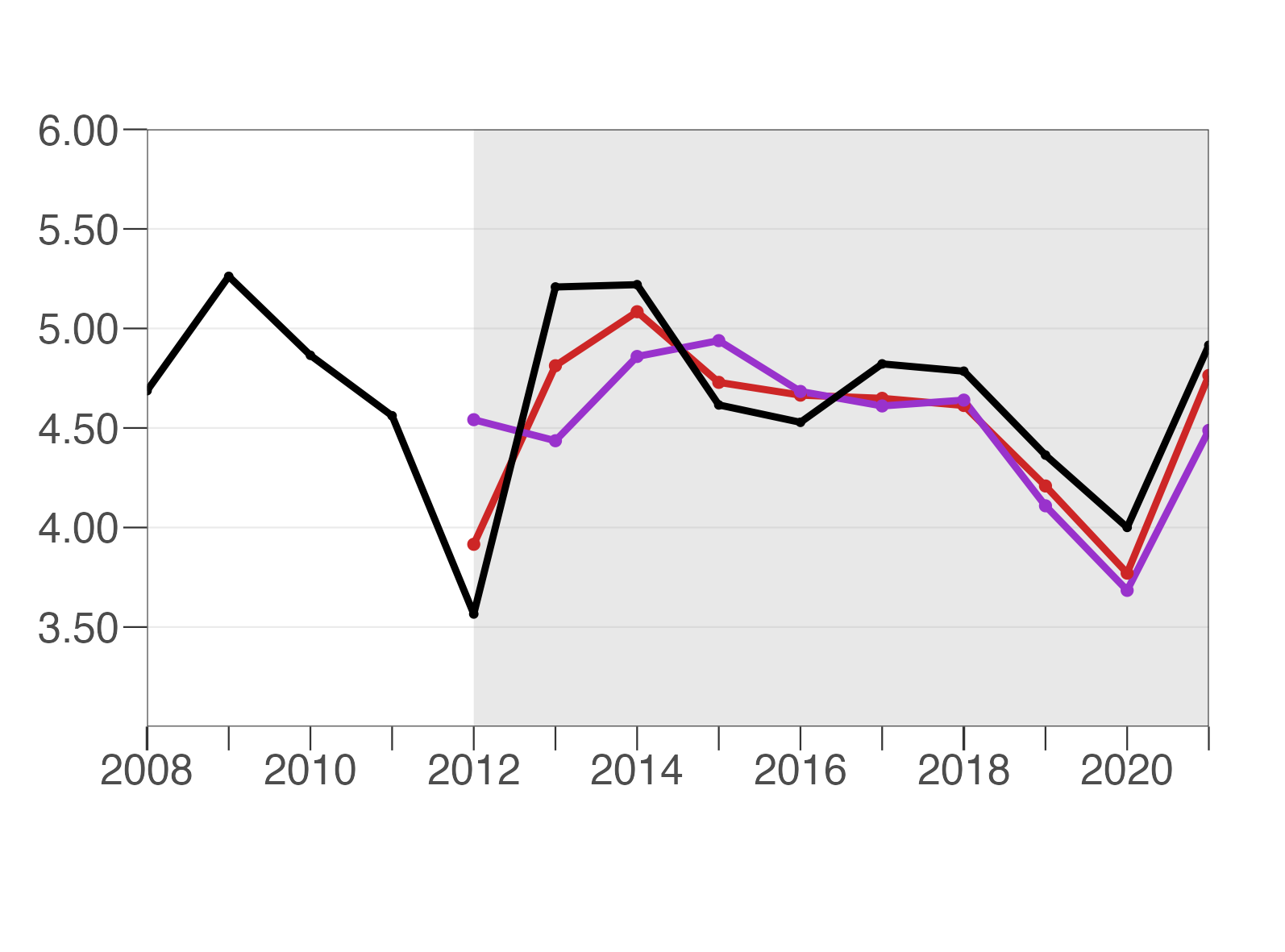}
			\caption{Forecast-Day: July 25$^\text{th}$}
		\end{subfigure}%
		\begin{subfigure}[t]{0.45\textwidth}
		\centering
			\includegraphics[trim={0mm 20mm 0mm 20mm},clip,width=\textwidth]{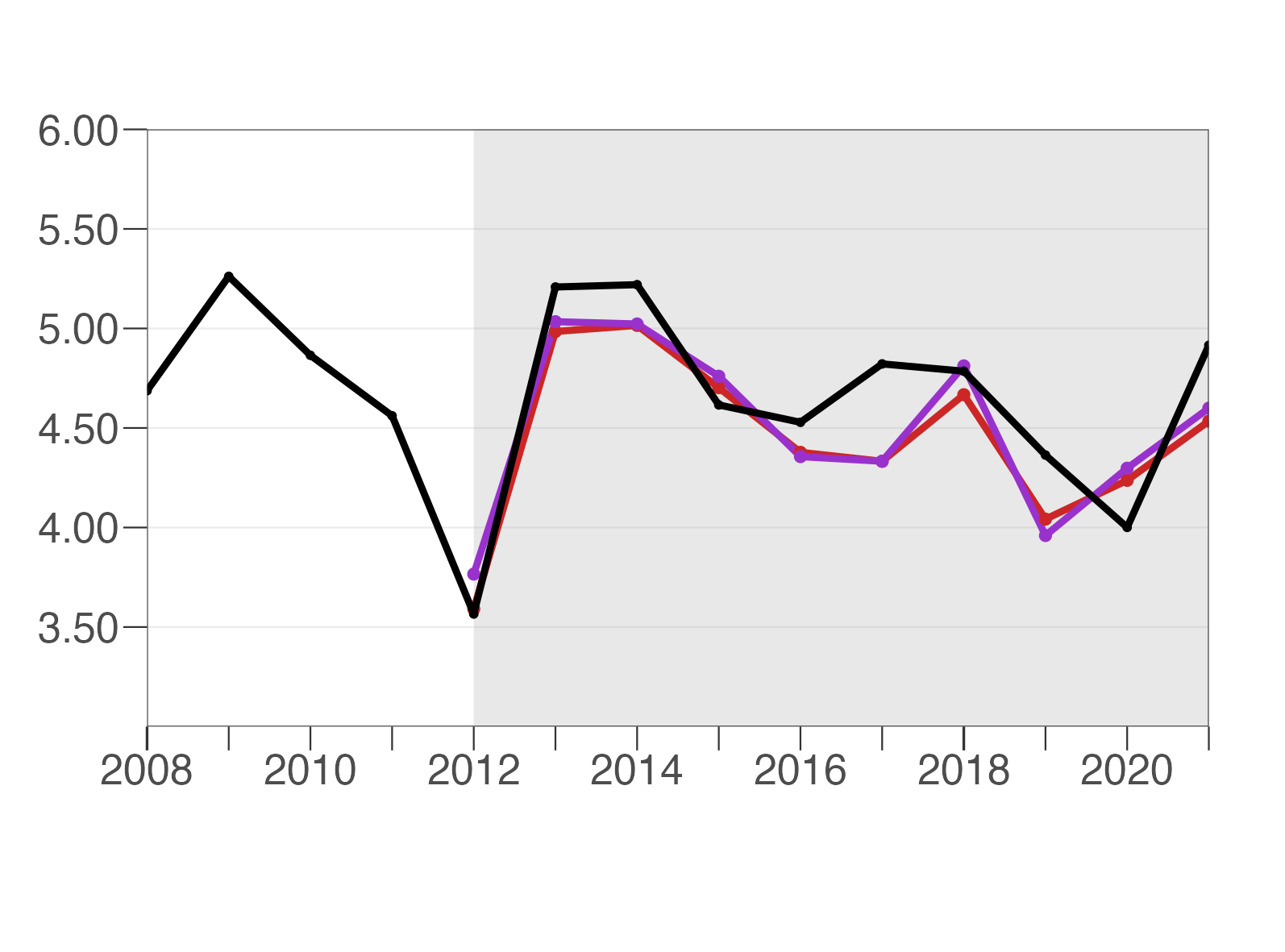}
			\caption{Forecast-Day: August 13$^\text{th}$}
		\end{subfigure}

		\vskip 10pt
		\includegraphics[trim={0mm 0mm 0mm 0mm},clip,width=0.5\textwidth]{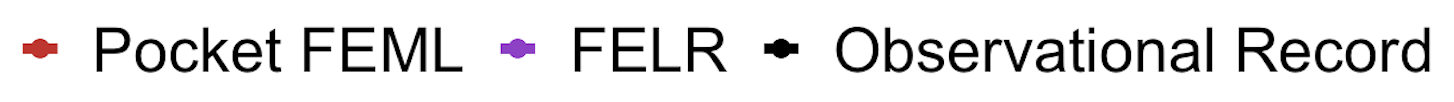}
		
	\end{center}
	\begin{spacing}{1.0}  \noindent \scriptsize Notes: We show out-of-sample forecasts for September $SIE$ for the years 2012-2021. The black line is the realized $SIE$. The scale of the Y-axis is in 10$^6$ km$^2$.
	\end{spacing}
\end{figure}

{\color{black}  Lastly,  it can be informative to look at the raw series and corresponding forecasts themselves for the key month of September. Figure \ref{hist_oos} \textcolor{black}{shows} three horizons where disagreement between linear and nonlinear models can be substantial (June 14$^\text{th}$,  July 25$^\text{th}$,  August 13$^\text{th}$).   We see FELR $\succ$ Pocket-FEML in June is due to the latter being overly pessimistic in the first half of the out-of-sample \textcolor{black}{period}.  Disagreement inevitably shrinks in July as the target date approaches.   Nonetheless,  Pocket FEML clearly gets the upper hand in the early 2010s by better capturing the large deviations from trend starting in 2012 (the lowest SIE on record).  Finally,  forecasts converge to near-identical values by mid-August. }

\subsubsection{Which Model(s) to Use and When}

As mentioned earlier,  \textit{Pocket} models are mechanically handicapped for horizons less than 30 days by excluding the slowly accumulating September data.  Fortunately,  the glide chart's vocation is to recommend a model to use,  and that recommendation may depend on the horizons of interest.  In the case of September, the outcome is clear: one should use FELR or Pocket FELR up to 90 days ahead, then switch to Pocket FEML for the next 60 days, and then revert back to FELR for the remaining 30 (short-run) horizons.  Glide charts prove to be \textcolor{black}{a} particularly useful analytic\textcolor{black}{al} tool in this exercise  given that the optimal model choice for various months is clearly horizon-dependent.  

In unreported results,  we considered  FEML ($S_t=X_t$),  a MRF with the linear part $X_t$ still being FELR,  but with the set of ``forest" variables $S_t$ being restricted to only include the elements in $X_t$.  Naturally, this helps in gauging how much of FEML gains/losses are attributable to the use of a larger information set vs plain nonlinearities.  

In the overwhelming majority of cases,  FEML supplants or performs equally well as  FEML ($S_t=X_t$).  This suggests that focused nonlinearities \textit{and} an expanded data set can provide the largest gains over FELR.  Thus,  in line with \cite{MRF}\textcolor{black}{'s} observations in macroeconomic forecasting applications, a larger $S_t$ is almost always preferable  to a restricted one.  Moreover,  as noted in \cite{TBTP},  a larger $S_t$ spurs diversification of the underlying trees which helps \textcolor{black}{to keep} overfitting in check.  Given the short length of our time series,  potential for tree diversification  is \textcolor{black}{more} easily obtained from feature randomization than \textcolor{black}{from} bagging.  

In sum,  FEMLs can provide timely forecasting improvements over FELRs for important months in the SIE annual cycle.  Given the data limitation, these are not extremely large and are not observed for every horizon, even in successful months.  In such a context,  glide charts are particularly useful to provide guidance on which feature-engineered model to use and when.   Our results unequivocally indicate that FELRs and FEMLs are more adequate benchmarks for out-of-sample predictive accuracy than the oversimplistic linear trend model -- which is nonetheless widely used for such purposes \citep{andersson2021seasonal}.

\section{Concluding Remarks  and  Directions for Future Research}  \label{concl}

 We have used glide charts -- plots of sequences of root mean squared forecast errors as the target date is approached  --  to evaluate and compare  fixed-target forecasts of Arctic sea ice.  We first used them   to evaluate  the feature-engineered linear regression (FELR) forecasts of \cite{DieboldGoebel2021}, and to compare them to naive pure-trend benchmark forecasts.   Then we introduced a much more sophisticated feature-engineered \textit{machine learning} (FEML) model, and we used glide charts to evaluate its forecasts  and compare them to a FELR benchmark.  Our substantive results include the frequent appearance of predictability thresholds, which differ across months, meaning that accuracy initially fails to improve as the target date is approached but then increases progressively once a threshold lead time is crossed.  We also compared FELR and FEML, finding that FEML can improve on FEML for turning point months in the annual SIE cycle,  namely September, October, and March.  Those gains are  particularly evident for forecasts made 90 to 30 days before the target date.

In addition, we have built a  \href{https://chairemacro.esg.uqam.ca/arctic-sea-ice-forecasting/?lang=en}{ web site} that expands on the analysis of this paper, providing weekly updates of forecasts for target date September 2022.\footnote{See \url{https://chairemacro.esg.uqam.ca/arctic-sea-ice-forecasting/?lang=en}.} The forecasts are based on FELR models, FEML models, and the  VARCTIC model of \cite{VARCTIC}.  The user can explore the 2022 forecasts and those of previous years through a series of interactive plots.  Among other things,  the site features a glide chart of the key models as well as a continuously-updated rolling history of 2022 point forecasts and  associated prediction intervals.  This provides publicly available real-time SIE predictions from four competitive statistical/machine learning   models.  It therefore complements the Sea Ice Outlook, which is of much larger scope in terms of included models,  but which publishes the results of the survey only on a monthly basis and  with a lag of two to three weeks.

 Several directions for future research are apparent, all of which are related to this paper's 
use of  glide charts  for comparing different fixed-target SIE forecasting models, and the related idea that a ``better" model should have a ``better" glide chart.  First, although 
in this paper we focused exclusively on comparing glide charts of statistical/econometric sea ice forecasting models, one could also (a) include glide charts of structural global climate models (GCMs) (e.g., ``How well does a particular GCM's glide chart match the FEML glide chart?"), or (b) use glide charts to help calibrate/estimate GCMs  (e.g., ``For a particular GCM, what parameter configuration minimizes the divergence between \textcolor{black}{the} GCM glide chart and the FEML glide chart?").  

Second, one could consider glide-chart loss functions, or accuracy measures, other than the ubiquitous quadratic loss underlying RMSFE. In particular, one may want to entertain  asymmetric loss functions.  Consider, for example, a firm contemplating in June whether to attempt a September trans-Arctic shipment, using September fixed-target sea ice forecasts to guide  the decision, and consider positive vs. negative forecast errors:
\begin{enumerate}

\item  Positive errors (realized September ice greater than forecast): the overly-optimistic forecasts may produce a decision to undertake shipping, which may be regretted as the shipping will be  more risky and costly than expected, or even impossible.
	
\item  Negative errors (realized September ice less than forecast): the overly-pessimistic forecasts may produce a decision not to undertake shipping, which may be regretted as business is lost unnecessarily.

\end{enumerate}
Both positive and negative errors are of course costly, but there is no reason why the loss associated with a given positive error should necessarily match that of a negative error of the same absolute magnitude.  Asymmetric loss functions capture such effects.

\clearpage

\bibliographystyle{Diebold}
\addcontentsline{toc}{section}{References}
\bibliography{Bibliography}

@article{RohdeHausfather2020,
title = {The Berkeley Earth Land/Ocean Temperature Record},
journal = {Earth Syst. Sci. Data},
volume = {12},
number = {},
pages = {3469-3479},
year = {2020},
doi = {https://doi.org/10.5194/essd-12-3469-2020},
author = {Rohde, R.A. and Hausfather, Z.},
}

@article{iceplus,
  title={Optimal Combination of Arctic Sea Ice Extent Measures: A Dynamic Factor Modeling Approach},
  author={Diebold, F.X.  and G{\"o}bel, M.  and  Goulet Coulombe, P.  and Rudebusch, G.D. and Zhang, B.},
  journal={International Journal of Forecasting},
  volume={37},
  number={4},
  pages={1509--1519},
  year={2021},
  publisher={Elsevier}
}

@article{SeaIceBookCh4,
	title={Changes in Arctic Sea Ice Cover in the Twentieth and Twenty-First Centuries},
	author={Shalina, E.V. and Johannessen, O.M.  and Sandven, S.},
	note={in Johannessen, O.M., Bohylev, L.P., Shalina, E.V., and Sandven, S., eds., \emph{Sea Ice in the Arctic: Past Present and Future}, 93-166, Springer Nature},
	year={2020},
}

@article{Zellner1992,
	title={Statistics, Science and Public Policy},
	author={Zellner, A.},
	journal={Journal of the
American Statistical Association},
volume={87},
pages={1-6},	
year={1992}
}

@article{StroeveEtAl2014,
author = {Stroeve, J. and Hamilton, L.C. and Bitz, C.M. and Blanchard-Wrigglesworth, E.},
title = {Predicting September Sea Ice: Ensemble Skill of the SEARCH Sea Ice Outlook 2008–2013},
journal = {Geophysical Research Letters},
volume = {41},
number = {7},
pages = {2411-2418},
doi = {https://doi.org/10.1002/2014GL059388},
year = {2014}
}

@article{DieboldGoebel2021,
  title={A Benchmark Model for Fixed-Target Arctic Sea Ice Forecasting},
  author={Diebold, F.X. and G\"obel, M.},
  year={2022},
journal={Economics Letters},
volume={215},
pages={110478}
%note={Manuscript, University of Pennsylvania, \href{https://arxiv.org/abs/1907.06303}{arXiv:2101.10359}}

@article{DRice,
  title={Probability Assessments of an Ice-Free Arctic: Comparing Statistical and Climate Model Projections},
  author={Diebold, F.X. and Rudebusch, G.D.},
  year={2022},
journal={Journal of Econometrics},
volume={231},
pages={520-534}

@article{Ing2003, title={Multistep Prediction in Autoregressive Processes}, volume={19}, DOI={10.1017/S0266466603192031}, number={2}, journal={Econometric Theory}, publisher={Cambridge University Press}, author={Ing, C.-K.}, year={2003}, pages={254–279}}

@article{VARCTIC,
  title={Arctic Amplification of Anthropogenic Forcing: A Vector Autoregressive Analysis},
  author={Goulet Coulombe, P. and G{\"o}bel, M.},
  journal={Journal of Climate},
pages={5523–5541},
volume={34},
  year={2021}
}

@article{SIO_postseason2022,
        author = {Bhatt, U.S. and Meier, W. and Blanchard-Wrigglesworth, E. and Massonnet, F. and Goessling, H. and Ludwig V. and Bieniek, P. and Eicken, H. and Fisher, M. and Hamilton, L.C. and Little, J. and Overland, J.E. and Serreze, M. and Steele, M. and Stroeve, J. and Walsh, J. and Wang, M. and Wiggins, H.V.},
        title = {},
        year = {2022},
      url = { https://www.arcus.org/sipn/sea-ice-outlook/2022/postseason},
        note = {\emph{Sea Ice Outlook: 2022 Post Season Report.} (Edited by  B. Turner-Bogren, and H.V. Wiggins), \url{ https://www.arcus.org/sipn/sea-ice-outlook/2022/postseason}}
}

@article{BushukEtAl2019,
    author = {Bushuk, M. and Msadek, R. and Winton, M. and Vecchi, G. and Yang, X. and Rosati, A. and Gudgel, R.},
    title = "{Regional Arctic Sea–Ice Prediction: Potential versus Operational Seasonal Forecast Skill}",
    journal = {Climate Dynamics},
    volume = {52},
    number = {5},
    pages = {2721-2743},
    year = {2019}
}

@article{DayEtAl2014,
    author = {Day, J.J. and Tietsche, S. and Hawkins, E.},
    title = "{Pan-Arctic and Regional Sea Ice Predictability: Initialization Month Dependence}",
    journal = {Journal of Climate},
    volume = {27},
    number = {12},
    pages = {4371-4390},
    year = {2014}
}

@article{HawkinsEtAl2016,
author = {Hawkins, E. and Tietsche, S. and Day, J. J. and Melia, N. and Haines, K. and Keeley, S.},
title = {Aspects of Designing and Evaluating Seasonal-to-Interannual Arctic Sea-Ice Prediction Systems},
journal = {Quarterly Journal of the Royal Meteorological Society},
volume = {142},
number = {695},
pages = {672-683},
doi = {10.1002/qj.2643},
year = {2016}
}

@article{ChevallierEtAl2013,
    author = {Chevallier, M. and Salas y Mélia, D. and Voldoire, A. and Déqué, M. and Garric, G.},
    title = "{Seasonal Forecasts of the Pan-Arctic Sea Ice Extent Using a GCM-Based Seasonal Prediction System}",
    journal = {Journal of Climate},
    volume = {26},
    number = {16},
    pages = {6092-6104},
    year = {2013},
    doi = {10.1175/JCLI-D-12-00612.1}
}

@article{Bekkers2016,
	author = {Bekkers, E. and Francois, J.F. and Rojas-Romagosa, H.},
	title = {Melting Ice Caps and the Economic Impact of Opening the Northern Sea Route},
	journal = {Economic Journal},
	volume = {128},
	pages = {1095-1127},
	year = {2016},
}

@article{Petrick2017,
	author = {Petrick, S. and Riemann-Campe, K. and Hoog, S. and Growitsch, C. and Schwind, H. and Gerdes, R. and Rehdanz, K.},
	title = {Climate Change, Future Arctic Sea Ice, and the Competitiveness of European Arctic Offshore Oil and Gas Production on World Markets},
	journal = {Ambio},
	volume = {46},
	number = {3},
	pages = {410-422},
	year = {2017},
}

@article{HamiltonStroeve2016,
author = {Hamilton, L.C. and Stroeve, J.},
title = {400 Predictions: the SEARCH Sea Ice Outlook 2008-2015},
journal = {Polar Geography},
volume = {39},
number = {4},
pages = {274-287},
year = {2016},
}

@misc{NSIDC_SII,
  author = {Fetterer, F. and Knowles, K. and Meier, W.N. and Savoie, M. and Windnagel, A.K.},
  title = {Sea Ice Index, Version 3},
  howpublished = {\url{https://doi.org/10.7265/N5K072F8}},
  year = {2017},
}

@article{ebinger2009,
  title={The Geopolitics of Arctic Melt},
  author={Ebinger, C.K. and Zambetakis, E.},
  journal={International Affairs},
  volume={85},
  number={6},
  pages={1215-1232},
  year={2009},
  publisher={Oxford University Press}
}

@Article{Hamilton2020,
	author = {Hamilton, L.},
	title = {1000 Predictions: What's New and What's Old in a Retrospective Analysis of the Sea Ice Outlook, 2008-2020},
	journal = {},
	year = 2020,
	volume = {},
pages={},
note={Presentation at American Geophysical Union Annual Meeting}
}

@ARTICLE{Fettereretal2017,
  author = {Fetterer, F. and Knowles, K. and Meier, W. and Savoie, M. and Windnagel, A.K.},
  title = {Sea Ice Index, Version 3, Dataset ID G02135},
note={Boulder, Colorado, USA. NSIDC: National Snow and Ice Data Center. \url{https://doi.org/10.7265/N5K072F8}, updated daily},
  year = {2017}
}

@article{MRF,
  title={The Macroeconomy as a Random Forest},
  author={Goulet Coulombe, P.},
  journal={Available at SSRN 3633110},
  year={2020}
}

@article{breiman2001random,
  title={Random Forests},
  author={Breiman, Leo},
  journal={Machine learning},
  volume={45},
  number={1},
  pages={5--32},
  year={2001},
  publisher={Springer}
}

@article{GCLSS2019,
  title={How is Machine Learning Useful for Macroeconomic Forecasting?},
  author={Goulet Coulombe, P.  and Leroux, M.  and Stevanovic, D.  and Surprenant, S.},
  journal={Journal of Applied Econometrics},
  volume = {37},
 number = {5},
 pages = {920-964},
  year={2022}
}

@article{TBTP,
  title={To Bag is to Prune},
  author={Goulet Coulombe, P.},
  journal={arXiv e-prints},
  pages={arXiv--2008},
  year={2020}
}

@article{andersson2021seasonal,
  title={Seasonal Arctic Sea Ice Forecasting with Probabilistic Deep Learning},
  author={Andersson, T.R. and Hosking, J.S. and P{\'e}rez-Ortiz, M. and Paige, B. and Elliott, A. and Russell, C. and Law, S. and Jones, D.C. and Wilkinson, J. and Phillips, T. and others},
  journal={Nature Communications},
  volume={12},
  number={1},
  pages={1--12},
  year={2021},
  publisher={Nature Publishing Group}
}

@article{maslanik2007younger,
  title={A Younger, Thinner Arctic Ice Cover: Increased Potential for Rapid, Extensive Sea-Ice Loss},
  author={Maslanik, J.A. and Fowler, C. and Stroeve, J. and Drobot, S. and Zwally, J. and Yi, D. and Emery, W.},
  journal={Geophysical Research Letters},
  volume={34},
  number={24},
  year={2007},
  publisher={Wiley Online Library}
}

\clearpage
\appendix
\appendixpage
\addappheadtotoc
\newcounter{saveeqn}
\setcounter{saveeqn}{\value{section}}
\renewcommand{\theequation}{\mbox{\Alph{saveeqn}.\arabic{equation}}} \setcounter{saveeqn}{1}
\setcounter{equation}{0}
%
%\clearpage
%
\section{Data}\label{data_appendix}

Daily SIE data are from  {Sea Ice Index, Version 3} \citep{NSIDC_SII} provided by the {National Snow and Ice Data Center} (NSIDC).\footnote{See  \url{https://doi.org/10.7265/N5K072F8}.}  Until August 1986, data are reported only every other day. For model estimation we fill missing $SIE_{Today,t}$ observations with the average of the two adjacent days, $SIE_{Yesterday,t}$ and $SIE_{Tomorrow,t}$.

Daily SIT data are from PIOMAS provided by the Polar Science Center.\footnote{See \url{http://psc.apl.uw.edu/wordpress/wp-content/uploads/schweiger/ice_volume/PIOMAS.thick.daily.1979.2022.Current.v2.1.dat.gz}.}  Data for month $m$ are not known until the end of month $m+1$. Hence, on any day prior to the end of month $m+1$, information on SIT is only available through the end of month $m - 1$. This results in a one- to two-month lag.

Daily  AT data are based on \cite{RohdeHausfather2020}.\footnote{See \url{http://berkeleyearth.lbl.gov/auto/Global/Land_and_Ocean_complete.txt}.} The monthly measurements are reported as anomalies relative to the January 1951 - December 1980 average.

Monthly CO2 concentration data are from Mauna Loa, provided by the NOAA Global Monitoring Laboratory.\footnote{See the ``deseasonalized" column in   \url{https://gml.noaa.gov/webdata/ccgg/trends/co2/co2_mm_mlo.csv}.} The data for month $m$ are made available during the first days of month $m+1$. 

\end{document}